%% file: main.tex
\documentclass[runningheads,a4paper]{svjour3}

\usepackage{sosy-paper}

\usepackage{times}
\usepackage{geometry}
\newcommand{\blkf}{\sigma} 
\newcommand{\imc}{{interpolation-based model checking}\xspace}
\newcommand{\Imc}{{Interpolation-based model checking}\xspace}
\newcommand{\IMC}{{Interpolation-Based Model Checking}\xspace}

\definetool{\duality}{Duality}
\definetool{\spacer}{Spacer}
\definetool{\safari}{Safari}

\usepackage[firstpage]{draftwatermark}
\SetWatermarkText{\hspace*{10.5cm}\raisebox{130mm}{\href{https://doi.org/10.5281/zenodo.8245824}{\includegraphics[width=17mm]{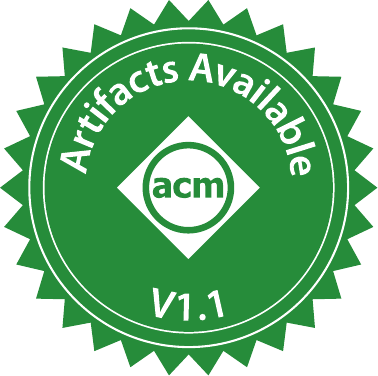}}}}
\SetWatermarkAngle{0}

\newcommand{\mypaperkeywords}{
Software verification \and
Program analysis \and
Model checking \and
Interpolation \and
Interpolation-based model checking \and
CPAchecker \and
SMT \and
SAT
}

\tikzset{
  abstraction state/.style={
    fill=black!10!white,
    rounded corners=3pt,
    inner sep=3pt,
  },
  covering edge/.style={
    draw,
    ->,
    dashed,
  },
  arg edge/.style={
    draw,
    ->,
  },
  multi arg edge/.style={
    arg edge,
    decorate, decoration={snake, amplitude=.5pt, segment length=6pt,},
    -LaTeX,
  },
}

\newcommand{\intstate}[7]{$\astate_{#1}$: $(#2, (#3, #4, #5, #6), \{\pc_4\mapsto#7\})$}%

\title{Interpolation and SAT-Based Model Checking Revisited: \\ Adoption to Software Verification
}
\titlerunning{Interpolation and SAT-Based Model Checking Revisited}

\begin{document}

\author{Dirk Beyer\orcidID{0000-0003-4832-7662}
  \and
  Nian-Ze Lee\orcidID{0000-0002-8096-5595}
  \and
  Philipp Wendler\orcidID{0000-0002-5139-341X}}

\authorrunning{D.~Beyer, N.-Z.~Lee, and P.~Wendler}
\institute{LMU Munich, Germany}
\date{Received: 2022-08-09}

\def\makeheadbox{}

\maketitle

\begin{abstract}
  \input{abstract}
\end{abstract}

\input{introduction}
\input{related-work}
\input{background}
\input{encoding}
\input{approach}
\input{implementation}
\input{evaluation}
\input{conclusion}

\section*{Data-Availability Statement}
To ensure verifiability and transparency of the results reported in this paper,
all used software, input programs, and raw experimental results are
available in a supplemental reproduction package~\cite{IMC-artifact-JAR-final}.
For convenient browsing through the results,
interactive tables are available at \url{https://www.sosy-lab.org/research/cpa-imc}.
Current versions of \cpachecker are also available at \url{https://cpachecker.sosy-lab.org}.

\section*{Funding Statement}
This project was funded in part by the LMU Postdoc Support Fund.

\interlinepenalty=10000
\bibliography{sw,dbeyer,artifacts,svcomp,svcomp-artifacts}


\vfill
\noindent
{\bf Open Access.} This article is licensed under the terms of the Creative Commons\break Attribution 4.0 International License (\url{http://creativecommons.org/licenses/by/4.0/}), which permits use, sharing, adaptation, distribution, and reproduction in any medium or format, as long as you give appropriate credit to the original author(s) and the source, provide a link to the Creative Commons license and indicate if changes were made.

The images or other third party material in this chapter are included in the chapter's Creative Commons license, unless indicated otherwise in a credit line to the material.~If material is not included in the chapter's Creative Commons license and your intended\break use is not permitted by statutory regulation or exceeds the permitted use, you will need to obtain permission directly from the copyright holder.

\medskip\noindent\href{http://creativecommons.org/licenses/by/4.0/}{\includegraphics{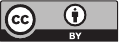}}

\end{document}

%% file: abstract.tex
The article \emph{Interpolation and SAT-Based Model Checking} \href{https://doi.org/10.1007/978-3-540-45069-6_1}{(McMillan, 2003)}
describes a formal-verification algorithm,
which was originally devised to verify safety properties of finite-state transition systems.
It derives interpolants from unsatisfiable \bmc queries
and collects them to construct an overapproximation of the set of reachable states.
Although 20~years old, the algorithm is still state-of-the-art in hardware model checking.
Unlike other formal-verification algorithms, such as \kinduction or \pdr,
which have been extended to handle infinite-state systems and investigated for program analysis,
McMillan's interpolation-based model-checking algorithm from 2003 has not been used to verify programs so far.
%
Our contribution is to close this significant, two decades old gap in knowledge by
adopting the algorithm to software verification.
We implemented it in the verification framework \cpachecker
and evaluated the implementation against other state-of-the-art software-verification techniques
on the largest publicly available benchmark suite of C~safety-verification tasks.
The evaluation demonstrates that McMillan’s interpolation-based model-checking algorithm from 2003
is competitive among other algorithms in terms of both the number of solved verification tasks and the run-time efficiency.
%
Our results are important for the area of software verification,
because researchers and developers now have one more approach to choose from.

%% file: introduction.tex
\section{Introduction}
Automatic software verification~\cite{SoftwareModelChecking} is an active research field in which automated solutions of the following problem are studied:
Given a program and a specification, decide whether the program satisfies the specification or not.
In this paper, we focus on the verification of reachability-safety properties, asserting that some error location in the program should never be reached by the control flow.
Other specifications, including termination, memory safety, concurrency safety, and overflows, are also investigated in the literature.
Although the problem of software verification is in general undecidable,
many important concepts, including various predicate-abstraction techniques~\cite{GrafSaidi97,FlanaganQadeer02,LazyAbstraction,AutomaticPredicateAbstraction2001},
counterexample-guided abstraction refinement~(CEGAR)~\cite{ClarkeCEGAR},
large-block encoding~\cite{LBE,ABE},
interpolation~\cite{AbstractionsFromProofs,IMPACT},
together with the advances in SMT solving~\cite{HBMC-SMT} and combinations with data-flow analysis~\cite{HBMC-dataflow},
make it feasible to apply verification technology to industry-scale software, such as
device drivers~\cite{LDV,LDV12,SLAM,SLAMtransfer},
web services~\cite{INFER,AWS},
and operating systems~\cite{LDV-Toolset}.

To illustrate the reachability-safety verification of a program,
consider the C~program in~\cref{fig:even-code}.
The program first initializes the variable~\texttt{x} to $0$ and keeps incrementing \texttt{x} by $2$ while the nondeterministic value returned from the function~\texttt{nondet()} is nonzero.
Once the nondeterministic value equals zero, the control flow exits the loop and tests whether \texttt{x} is odd.
If \texttt{x} is odd, the control flow reaches the error location at line~\texttt{8};
otherwise the program terminates without errors.
The goal of the reachability-safety verification is to either prove that the error location is unreachable by the control flow or find an execution path of the program reaching the error location.

\newsavebox{\exampleCode}
\begin{lrbox}{\exampleCode}
  \begin{minipage}[b]{0.48\textwidth}
    \centering
\begin{lstlisting}[
      style=C,
      basicstyle=\ttfamilywithbold,
      numberstyle=\scriptsize,
      lineskip=2pt,
      aboveskip=0pt,
      belowskip=0pt,
    ]
extern int nondet();
int main(void) {
  unsigned int x = 0;
  while (nondet()) {
    x += 2;
  }
  if (x % 2) {
    ERROR: return 1;
  }
  return 0;
}
\end{lstlisting}
  \end{minipage}
\end{lrbox}

\newsavebox{\exampleCFA}
\begin{lrbox}{\exampleCFA}
  \begin{minipage}[b]{0.48\textwidth}
    \centering
    \scalebox{0.9}{\input{figures/even-cfa.tex}}
  \end{minipage}
\end{lrbox}

\begin{figure*}[t]
  \centering
  \subfloat[C program]{\usebox{\exampleCode}\label{fig:even-code}}
  \subfloat[Control-flow automaton]{\usebox{\exampleCFA}\label{fig:even-cfa}}
  \caption{
    An example C program (a) and its CFA (b)
    (adopted from \href{https://gitlab.com/sosy-lab/benchmarking/sv-benchmarks/-/blob/svcomp22/c/loop-invariants/even.c}{\texttt{loop-invariants/even.c}} in the benchmark set of the 2022 Competition on Software Verification (SV-COMP\,'22)~\cite{SVCOMP22})
  }
  \label{fig:example}
\end{figure*}

As the verification of finite-state and infinite-state transition systems share much similarity,
some classic model-checking algorithms for software (infinite-state systems), such as bounded model checking (BMC)~\cite{BMC,CBMC} or \kinduction~\cite{kInduction,k-Induction,PKind},
were originally developed for hardware (finite-state systems).
A well-known example of such technology transfer is property-directed reachability (\pdr)~\cite{IC3}.
After it obtained huge success in hardware model checking,
many research efforts have been invested for its software-verification adoption~\cite{SoftwareIC3,IC3-CFA,CTIGAR,PDR-kInduction,PDR}.

\subsection{Interpolation-Based Verification Approaches}
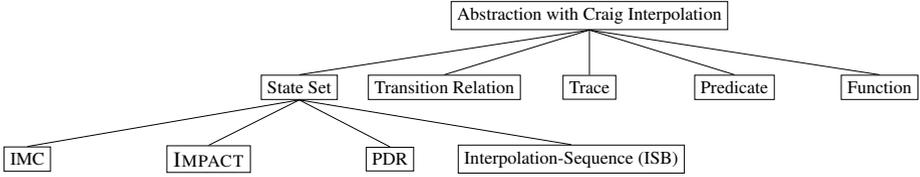
\begin{figure*}[t]
  \centering
  \scalebox{0.85}{\input{figures/overview-tree}}
  \caption{Classification of different abstractions using Craig interpolation}
  \label{fig:classification}
\end{figure*}
McMillan's algorithm~\cite{McMillanCraig} from 2003 is another state-of-the-art approach for hardware model checking, prior to the invention of \pdr.
It utilizes Craig interpolation~\cite{Craig57} to derive interpolants from unsatisfiable \bmc queries
and computes an overapproximation of the set of reachable states as the union of the interpolants.
Its idea of abstracting objects with interpolants has been extended beyond state sets and underpinned various interpolation-based verification approaches and tools.
Abstractions of
transition relations~\cite{TRApproximation},
traces~\cite{TraceAbstraction},
predicates over program variables~\cite{AbstractionsFromProofs,AutomaticPredicateAbstraction2001},
and function calls~\cite{SeryFunFrog11}
have been studied in the literature.
We classify in~\cref{fig:classification} different usages
of Craig interpolation and highlight some important algorithms regarding state-set abstraction.
An overview of several representative interpolation-based formal-verification approaches is provided in~\cref{sect:related-work}.

Despite its success in hardware model checking and profound theoretical impact on program analysis,
McMillan's algorithm~\cite{McMillanCraig} from 2003 has not been investigated for software verification.
We emphasize that
McMillan's interpolation-based algorithm for model checking from 2003 should not be mistaken for other, later interpolation-based verification approaches.
In the following, we refer to the algorithm proposed by McMillan from 2003 as \emph{\imc} and abbreviate it as IMC.

One potential concern to apply IMC to software,
raised by its inventor McMillan in his later paper~\cite{IMPACT} presenting the algorithm \impact,
was the scalability of the underlying decision procedure to handle the entire unrolled program.
Compared to IMC, \impact derives interpolants only for individual execution paths, reducing the workload of the solver.
Fortunately, due to the advancements in SMT solving,
delegating formulas encoding the entire unrolled program to the solvers has become feasible.
Therefore, it is time to revisit IMC and evaluate its performance against the state of the art.
Other SMT-based approaches have been thoroughly compared already in the literature~\cite{AlgorithmComparison-JAR}.

\subsection{Our Research Questions and Contributions}
\label{sect:intro-research-question}
In this paper, we explore the applicability of the IMC algorithm to software verification.
Specifically, we answer the following two research questions.
First, we investigate \emph{how to efficiently adopt IMC to software verification}.
As mentioned earlier, IMC was originally invented to verify sequential Boolean-logic circuits (hardware),
whose transition relations, required to perform IMC, are easy to derive:
The downstream circuitry of the memory elements (i.e., registers) encodes the next-state function of the system,
which can then be naturally expressed as a transition relation between system states.
It is less straightforward, by contrast, to extract a transition relation from a program (software).
Although representing a program as a transition relation with the program counter is in principle possible,
such conversion mixes the reasoning of the control-flow structure and the program semantics,
and hence, does not work for IMC in practice as we show in \cref{sect:symbolic-pc}.
To address this research question,
we propose an efficient software adoption of IMC via large-block encoding,
separating the analysis of the control flow and the program semantics
by exploring the analogy between the execution paths of a sequential Boolean-logic circuit and a program.
We also present the first implementation of the IMC algorithm for software verification
and make it available in the open-source framework \cpachecker~\cite{CPA,CPACHECKER}.
The details of the proposed adoption and implementation will be discussed in~\cref{sect:approach,sect:implementation}, respectively.
Our second research question focuses on \emph{evaluating the performance of the IMC adoption against the state of the art}.
To address this research question,
we compare the proposed implementation against other state-of-the-art software-verification algorithms,
including \pdr, BMC, \kinduction, predicate abstraction, and \impact,
on the largest benchmark suite of C safety-verification tasks in~\cref{sect:evaluation}.
Our experimental results show that IMC is competitive in terms of both effectiveness (the number of solved tasks)
and efficiency (the elapsed CPU time).

\inlineheadingbf{Novelty}
(1)~This paper closes the two decades old gap of knowledge by investigating the applicability of IMC to software verification.
We analyze the characteristics of IMC in the context of software verification,
and our empirical evaluation indicates its competitiveness against the state-of-the-art approaches.
(2)~Our replication of the IMC algorithm as open-source implementation broadens
the spectrum of available software-verification techniques,
which is important in practice
because researchers, developers, and tool users now have more choices at their disposal.
(3)~While the application of large-block encoding to program analysis has a long history,
to the best of our knowledge, using large-block encoding to represent an algorithm that originated from a different research community for software is a new idea, which may shed light on the efficient adoptions of other algorithms.

\inlineheadingbf{Significance}
IMC is an important verification algorithm in hardware verification.
It is a risk to leave the potential of it unexplored for the verification of software.
Therefore, we believe that the knowledge about the algorithm's adoption to software is
a significant improvement of the state of the art and has the potential to inspire other
works in the area of software verification.

\inlineheadingbf{Correctness}
We show the correctness of our algorithms in~\cref{thm:imc-soundness}.
Our implementation is based on components from the \cpachecker framework~\cite{CPACHECKER},
which is a well-maintained software project with lots of evidence that the components work well.
Large-block encoding is a sound component from the literature~\cite{LBE,ABE}.

The effectiveness and efficiency of our implementation is empirically evaluated with
experiments on a large benchmark set in~\cref{sect:evaluation}.
We discuss possible threats to validity that might affect the soundness of our conclusions from the experimental results
in \cref{sec:threats}.

%% file: figures/even-cfa.tex
\begin{tikzpicture}
\centering
\node (s) [inner sep=0pt] {};
\node (3) [below of = s, node distance=0.7cm, draw, circle, inner sep=0.03cm]{$l_3$};
\node (4) [below of = 3, node distance=1cm, draw, circle, inner sep=0.03cm]{$l_4$};
\node (5) [right of = 4, node distance=2.5cm, draw, circle, inner sep=0.03cm]{$l_5$};
\node (7) [below of = 4, node distance=1cm, draw, circle, inner sep=0.03cm]{$l_7$};
\node (8) [below of = 7, node distance=1cm, draw, circle, inner sep=0.03cm]{$l_8$};
\node (10) [below of = 8, node distance=1cm, draw, circle, inner sep=0.03cm]{$l_{10}$};
\node (11) [below of = 10, node distance=1cm, draw, circle, inner sep=0.03cm]{$l_{11}$};

\draw[->] (s) --(3);
\draw[->] (3) to node[left] {\texttt{unsigned int x = 0;}}(4);
\draw[->] (4) to node[above] {\texttt{[nondet()]}}(5);
\draw[->] (5) edge[bend left] node[below] {\texttt{x += 2;}} (4);
\draw[->] (4) to node[left] {\texttt{[!nondet()]}}(7);
\draw[->] (7) to node[left] {\texttt{[x \% 2]}}(8);
\draw[->] (7) edge[bend left] node[right] {\texttt{[!(x \% 2)]}}(10);
\draw[->] (8) edge[bend right] node[left] {\texttt{ERROR: return 1;}}(11);
\draw[->] (10) to node[right] {\texttt{return 0;}}(11);
\end{tikzpicture}

%% file: figures/overview-tree.tex
\begin{tikzpicture}[
        level 1/.style={sibling distance=8em},
        level 2/.style={sibling distance=10em},
        level distance=4em,
        every node/.style={
                draw,
                align=center,
            },
    ]
    \node {Abstraction with Craig Interpolation}
    child { node {State Set}
            child { node {IMC} }
            child { node {\larger\impact} }
            child { node {PDR} }
            child { node {Interpolation-Sequence (ISB)} }
        }
    child { node {Transition Relation} }
    child { node {Trace} }
    child { node {Predicate} }
    child { node {Function} }
    ;
\end{tikzpicture}

%% file: related-work.tex
\section{Related Work}
\label{sect:related-work}

IMC has popularized the idea of using interpolation for verification,
and although IMC itself has not been applied to software so far,
there are many approaches for software verification that make use of interpolation.
Based on the classification in~\cref{fig:classification},
we will discuss several representative interpolation-based approaches and tools,
as summarized in~\cref{tab:related-work}.
Interested readers are referred to the chapter~\cite{HBMC-interpolation} by McMillan
in the Handbook of Model Checking for a broader survey.
Of course, there exist many techniques for computing interpolants.
We do not discuss them here as interpolant computation is typically orthogonal
to the used verification algorithm.
In our implementation, we use an off-the-shelf SMT solver for interpolation (\mathsat~\cite{MATHSAT5}).

\begin{table*}[t]
    \centering
    \caption{Important interpolation-based formal-verification approaches and tools}
    \label{tab:related-work}
    \newcommand\newapproach{\\[0.1cm]}
    \begin{tabular}{l|ccl}
        \toprule
        Approach                               & Year                  & Publication                                    & Contribution                                        \\
        \midrule
        IMC                                    & 2003                  & \cite{McMillanCraig}                           & First interpolation-based model-checking algorithm
        \newapproach
        \multirow{2}{*}{Predicate abstraction} & \multirow{2}{*}{2004} & \multirow{2}{*}{\cite{AbstractionsFromProofs}} & Discovering relevant predicates                     \\
                                               &                       &                                                & from interpolants to refute false alarms
        \newapproach
        \multirow{2}{*}{TR approximation}      & \multirow{2}{*}{2005} & \multirow{2}{*}{\cite{TRApproximation}}        & Refining an abstract TR with interpolants           \\
                                               &                       &                                                & to avoid exact image computation
        \newapproach
        \multirow{2}{*}{\impact}               & \multirow{2}{*}{2006} & \multirow{2}{*}{\cite{IMPACT}}
                                  & Performing lazy abstraction by computing \\
                                               &                       &                                                & sequences of interpolants on program paths
        \newapproach
        Slicing abstraction                    & 2007                  & \cite{SlicingAbstractions}                     & Splitting abstract states with interpolants
        \newapproach
        \multirow{2}{*}{ISB}                   & \multirow{2}{*}{2009} & \multirow{2}{*}{\cite{VizelFMCAD09}}           & Imitating BDD-based model checking by               \\
                                               &                       &                                                & abstracting states with interpolants
        \newapproach
        \multirow{2}{*}{Trace abstraction}     & \multirow{2}{*}{2009} & \multirow{2}{*}{\cite{TraceAbstraction}}       & Refining an overapproximation of                    \\
                                               &                       &                                                & possible traces with interpolant automata
        \newapproach
        \multirow{2}{*}{Lazy annotation}       & \multirow{2}{*}{2010} & \multirow{2}{*}{\cite{LazyAnnotation}}         & Annotating a program with interpolants              \\
                                               &                       &                                                & derived from Hoare triples
        \newapproach
        \multirow{2}{*}{Function summaries}    & \multirow{2}{*}{2011} & \multirow{2}{*}{\cite{SeryFunFrog11}}          & Summarizing function calls with interpolants        \\
                                               &                       &                                                & to reduce future analysis effort
        \newapproach
        \multirow{2}{*}{Software PDR}          & \multirow{2}{*}{2012} & \multirow{2}{*}{\cite{SoftwareIC3}}            & Combining {\impact}-like proof-based interpolants   \\
                                               &                       &                                                & and PDR clause generation
        \newapproach
        \multirow{2}{*}{\ctigar}               & \multirow{2}{*}{2014} & \multirow{2}{*}{\cite{CTIGAR}}                 & Refining abstraction failures                       \\
                                               &                       &                                                & relative to single steps with interpolants          \\
        \midrule
        {\blast}                               & 2004                  & \cite{AbstractionsFromProofs,BLAST}            & First software model checker using interpolation
        \newapproach
        {\tool{CSIsat}}                        & 2008                  & \cite{CSIsat}                                  & First open-source interpolation engine
        \newapproach
        {\cpachecker}                          & 2009                  & \cite{LBE,CPACHECKER}                          & Large-block encoding and interpolation
        \newapproach
        {\wolverine}                           & 2011                  & \cite{Wolverine}                               & First public implementation of {\impact}
        \newapproach
        {\ufo}                                 & 2012                  & \cite{UFO}                                     & Combining predicate and interpolation methods
        \newapproach
        {\duality}                             & 2013                  & \cite{DUALITY}                                 & Solving constrained Horn clauses with interpolation
        \newapproach
        {\spacer}                              & 2013                  & \cite{SPACER}                                  & Combining proof-based approaches and CEGAR
        \newapproach
        {\safari}                              & 2014                  & \cite{LazyAbstractionWithArrays-JOURNAL}       & Backward {\impact}-like analysis with arrays        \\
        \bottomrule
    \end{tabular}
    \vspace{-1mm}
\end{table*}

\subsection{State Sets}
The most closely related algorithm is \impact~\cite{IMPACT} from the same author,
which is also based on the idea of computing a fixed point from interpolants.
\impact applies interpolation to formulas of single program paths instead of the whole program
and generates a \textit{sequence of interpolants} for a spurious counterexample,
one interpolant after each program statement on the execution path.
It also computes fixed points of reachable states per program location
instead of globally.
One adaptation~\cite{SoftwareIC3} of property-directed reachability~(PDR)~\cite{IC3} to software
computes sequences of sets of clauses for refuting spurious counterexamples,
and these sequences also form valid sequences of interpolants.
Under this view, the approach is similar to \impact, only differing in how the interpolants are computed.
A hybrid approach with a combination of proof-based interpolation (as in \impact)
and PDR-based clause generation has also been suggested~\cite{SoftwareIC3}.
\ctigar~\cite{CTIGAR} is another attempt to extend PDR to software.
It combines Cartesian predicate abstraction with PDR and considers an abstract state as a conjunction of the predicates satisfied by the corresponding concrete state.
Different from other adaptations of PDR,
\ctigar avoids expensive pre-image computation by focusing on refinement relative to single steps of the transition relation.

A related approach for hardware model checking
is \emph{interpolation-sequence based model checking} (ISB)~\cite{VizelFMCAD09}.
In contrast to IMC, which computes only one interpolant at a time
that overapproximates states reachable within a certain number of steps,
ISB derives a sequence of interpolants from an unsatisfiable BMC query,
and each interpolant is an overapproximation of the states reachable within
an increasing number of steps.
This is similar to \impact,
just with ISB computing sequences of interpolants for an unrolling of the whole transition relation
instead of single program paths like \impact.
In ISB, the fixed point is found if the interpolant derived at the last unrolled loop head
implies the disjunction of all previous interpolants.

The approach of \emph{lazy annotation}~\cite{LazyAnnotation} combines symbolic execution and interpolation to generate Hoare-style annotations for a program in a similar way as a conflict-driven clause-learning SAT solver.
An annotation on a program edge is a condition that will block any future execution from this edge to an error location.
The method symbolically executes the input program along some chosen path to search for an error location.
If the execution is blocked by an edge,
it backtracks and produces an annotation by interpolation,
which is a valid precondition of the edge's Hoare triple.
This method is also applicable to program testing because it explores only feasible traces.

\subsection{Predicates, Transition Relations, Traces, and Functions}
Another popular use of interpolation for software verification
is to derive predicates from interpolants for predicate abstraction~\cite{AbstractionsFromProofs,AutomaticPredicateAbstraction2001}
in the refinement step of CEGAR,
typically by breaking up the interpolants into atomic predicates.
In contrast to IMC and \impact,
which both create the final abstract model of the program
(the overapproximation of the set of reachable states)
directly from interpolants,
predicate abstraction uses Boolean or Cartesian abstraction
over the set of derived predicates
and may generalize better.
Interpolation has also been used to avoid the expensive exact image computation in predicate abstraction~\cite{TRApproximation}, refining an abstract transition relation to guarantee convergence given adequate predicates.
Slicing abstraction~\cite{SlicingAbstractions} is another technique related to predicate abstraction.
It splits abstract states using predicates obtained from Craig interpolants to refine the abstraction.

Trace abstraction~\cite{TraceAbstraction,UAUTOMIZER2013} extends the concept of abstracting information by Craig interpolation to representing program paths with interpolants.
Given an unsatisfiable BMC query,
it derives a sequence of interpolants and constructs an interpolant automaton out of them.
This interpolant automaton excludes spurious traces that share the same reason of infeasibility with the current one.
A novel counterexample-guided abstraction refinement scheme is proposed for trace abstraction to prove the correctness of a program.

Interpolants are also applied to summarize function calls in a program~\cite{SeryFunFrog11}.
This approach replaces function calls with interpolants obtained in a previous analysis to reduce the subsequent verification effort.
Given an unsatisfiable BMC query involving a function call,
a summary of the function is computed as an interpolant between the function's corresponding formula and the rest of the BMC formula.
Recently, Craig interpolation is also used to abstract sequences of transition relations
to find deep counterexamples~\cite{BlichaTACAS2022}.

\subsection{Tools Based on Craig Interpolation}
Several software-verification tools are developed on top of Craig interpolation.
The tool \blast~\cite{AbstractionsFromProofs,BLAST} provides the first implementation
of a software-verification tool that uses interpolants for computing abstractions.
The tool \tool{CSIsat}~\cite{CSIsat} was the first freely available SMT solver with interpolation support.
The verification framework \cpachecker~\cite{CPACHECKER} applies Craig interpolation to large-block encodings of program code.
The tool \wolverine~\cite{Wolverine} provides the first publicly available implementation of \impact,
featuring a built-in interpolation procedure and some support for bit-vector operations.
The framework \ufo~\cite{UFO} is parameterized by definable components of abstract post, refinement, and expansion,
allowing various verification techniques based on overapproximation and underapproximation.
Craig interpolation has also been applied to solve constrained Horn clauses (CHC).
The tool \duality~\cite{DUALITY} generalizes \impact to gradually unroll a program
and solves the corresponding CHC formulas with interpolation until it yields valid inductive invariants.
The tool \spacer~\cite{SPACER} combines proof-based techniques with CEGAR,
maintaining both an overapproximation and an underapproximation of the input program.
The tool \safari~\cite{LazyAbstractionWithArrays-JOURNAL} implements a backward reachability analysis with lazy abstraction based on the MCMT framework~\cite{MCMT}, which can be understood as a backward variant of \impact, to support reasoning of arrays with unknown length.

%% file: background.tex
\section{Background}
\label{sect:background}
In the following, Boolean connectives $\lnot, \lor, \land, \rightarrow, \equiv$ are used in their conventional semantics.
A first-order logical formula is also interpreted as a set of (program) states that satisfy the formula, and we use the two terms interchangeably when it is clear from the context.

\subsection{\IMC}
\label{sect:background-imc}
\Imc (IMC)~\cite{McMillanCraig} is an algorithm for unbounded model checking to verify safety properties of state-transition systems.
It can be considered as an extension of BMC, which is well-known for bug hunting.
In order to describe IMC, we first define the notation to formalize a state-transition system.
Second, we review Craig's interpolation theorem~\cite{Craig57}, which is the core concept to extend BMC to unbounded model checking.

\subsubsection{State-Transition System}
Let $s$ and $s'$ be two arbitrary states in the state space of a state-transition system.
We formalize the state-transition system by three predicates over states.
Predicate $I(s)$ evaluates to $\true$ if state $s$ is an initial state of the system.
Predicate $T(s,s')$ evaluates to $\true$ if the system can transit from state $s$ to state $s'$.
It is also called the \textit{transition relation} of the system.
Predicate $P(s)$ evaluates to $\true$ if state $s$ satisfies the safety property to be verified.

In the above formulation of a state-transition system, we do not assume the state space to be finite or infinite.
The working of IMC is similar in both cases, provided that the underlying constraint solver (SAT/SMT solver) supports the reasoning over the corresponding logical formulas.

\subsubsection{Craig's Interpolation Theorem}
Given two first-order logical formulas~$\alpha$ and~$\beta$, if $\alpha \implies \beta$,
Craig's interpolation theorem~\cite{Craig57} guarantees the existence of a logical formula $\gamma$
such that $\alpha \implies \gamma$ and $\gamma \implies \beta$ hold,
and $\gamma$ only refers to the common variables of $\alpha$ and $\beta$.
Formula~$\gamma$ is called an \emph{interpolant} of~$\alpha$ and~$\beta$
as it is \emph{between}~$\alpha$ and~$\beta$.
In the model-checking community,
Craig's interpolation theorem is usually stated in an equivalent form based on unsatisfiability:
Given an unsatisfiable formula $A \land B$, $C$ is an interpolant of this formula if
(1)~$A \implies C$,
(2)~$C \land B$ is unsatisfiable, and
(3)~$C$ only refers to the common variables of~$A$ and~$B$.

\subsubsection{Algorithm Description}
The overall procedure of IMC~\cite{McMillanCraig} can be decomposed into two phases.
The first phase poses a BMC query by unrolling the transition relation $k$~times
and constructing a formula representing all possible execution paths from an initial state to a bad state
(a~state that violates the safety property) with $k$~transitions.\footnote{%
    The original BMC query in McMillan's 2003 paper~\cite{McMillanCraig}
    encodes all possible execution paths violating the safety property with \emph{at most} $k$~transitions.
    In this work, we use an optimization discussed in Section~3.2 of the~2003 paper
    to perform IMC incrementally and consider the property violation only after the last transition.
}
We use variable $s_i$ to denote the state after the $i^{th}$ transition.
Furthermore, to facilitate Craig interpolation in the second phase, the BMC query is partitioned into two formulas $A$ and $B$
(we omit $\land$ for brevity):
\begin{align}
    \underbrace{I(s_0)T(s_0,s_1)}_{\text{$A(s_0,s_1)$}}\underbrace{T(s_1,s_2)\ldots T(s_{k-1},s_k)\neg P(s_k)}_{\text{$B(s_1,s_2,\ldots,s_k)$}}
    \label{eq:BMC}
\end{align}

If this formula is satisfiable, a violation is found,
and we conclude that the system does not fulfill the safety property.
Otherwise, instead of simply increasing the unrolling upper bound, IMC tries to prove the safety property from the unsatisfiable BMC query in its second phase.
According to Craig's interpolation theorem, there exists an interpolant $C(s_1)$ referring to the common variable $s_1$, such that the following two conditions hold:
\begin{align*}
     & I(s_0)T(s_0,s_1) \rightarrow C(s_1)\text{ ~and}                           \\
     & C(s_1)T(s_1,s_2)\ldots T(s_{k-1},s_k)\neg P(s_k)\text{ is unsatisfiable.}
\end{align*}
The above two conditions indicate that $C(s_1)$ is an overapproximation of the set of states reachable from the initial states with one transition, and that states in $C(s_1)$ will not violate the safety property after $(k-1)$ transitions.

An overapproximation of the set of reachable states can be built by iteratively computing these interpolants.
Suppose the interpolant contains some noninitial states.
Changing the variable used in the interpolant from $s_1$ to $s_0$,
we pose another BMC query starting from the interpolant, that is, with $I(s_0)$ replaced by $C(s_0)$:
\begin{align*}
    \underbrace{C(s_0)T(s_0,s_1)}_{\text{$A'(s_0,s_1)$}}\underbrace{T(s_1,s_2)\ldots T(s_{k-1},s_k)\neg P(s_k)}_{\text{$B'(s_1,s_2,\ldots,s_k)$}}
\end{align*}
If the formula is again unsatisfiable, another interpolant $C'(s_1)$ exists, which is an overapproximation of the set of states reachable from the initial states with two transitions.
Such computation is repeated until the newly derived interpolant is contained in the union of the initial states and all previous interpolants.
In other words, the procedure stops when the union of the initial states and all previous interpolants grows to a \textit{fixed point}, i.e.,
a set of states that is inductive with respect to the transition relation and hence contains all reachable states.
From the second condition of Craig's interpolation theorem, it is guaranteed that this fixed point implies the safety property, and hence the safety property is proved.

If any BMC query is satisfiable during the iteration in the second phase, we cannot conclude that the property is violated.
The violation could be a wrong alarm, as some starting states in the interpolants might not be reachable.
Therefore, we have to return back to the first phase, increase the unrolling upper bound, and precisely check the existence of a violation starting from the initial states.

\subsubsection{Towards an Efficient Adoption}
\label{sect:background-efficient-adoption}
While IMC is described in terms of logical formulas in the above discussion, the adoption of this algorithm to a concrete state-transition system, such as a sequential Boolean-logic circuit (hardware) or a program (software), requires a conversion from the system under verification to the three predicates $I(s)$, $T(s,s')$, and $P(s)$.
The conversion is simple for sequential Boolean-logic circuits,
which IMC originally focused on,
as the input wires to the registers of the circuit encode the function to compute the next state
(i.e., the state after transition)
via the downstream circuitry in terms of the output wires of the registers (i.e., the current state).
This state-transition function can be naturally expressed as a transition relation.
It is less straightforward, by contrast, to extract a transition relation from a program.
Although a brute-force conversion is available,
representing a program via a transition relation with symbolic program counters
conceals the structural information of the program from the analysis.
In~\cref{sect:symbolic-pc},
we examine why encoding a program as a transition relation with symbolic program counters
is not suitable for adopting IMC to software verification.
The main challenge towards an efficient adoption to software verification
thus lies in obtaining all required predicates
while taking the program's structure into consideration.

\subsection{Program Representation}
To facilitate the subsequent discussion of program analysis,
here we provide some fundamental definitions for program representation from the literature~\cite{BLAST,HBMC-dataflow}.
We consider an imperative programming language whose variables are all integers.
The operations are either variable assignment or Boolean-expression evaluation.
We represent such a program as a \textit{control-flow automaton} (CFA) $A = (\locs, \pci, G)$.
A CFA is a directed graph with a set $L$ of nodes being program locations,
an initial location $l_0\in\locs$ indicating the entry point of the program,
and a set $G \subseteq (\locs \times Ops \times \locs)$
being control-flow edges annotated with program operations.

A \textit{reachability-safety verification task} consists of a CFA and an error location of the CFA.
The task is to either prove that the error location is unreachable from the initial location
or find a feasible error path to the error location otherwise.
For instance, the CFA of the example C program in~\cref{fig:even-code} is shown in~\cref{fig:even-cfa}.
The initial location of this CFA is $l_3$, and the error location is $l_8$.

\subsection{Configurable Program Analysis}
\label{sec:cpa}
A configurable program analysis (CPA)~\cite{CPA,CPAplus,HBMC-dataflow} defines the abstract domain used for a program analysis.
As we implemented the proposed adoption of IMC in the framework \cpachecker~\cite{CPACHECKER},
which utilizes CPA as the core concept,
we provide necessary background knowledge about CPA as follows.
To simplify the presentation,
we omit the dynamic precision adjustment of CPA
because it is irrelevant for this paper.
Please refer to the literature~\cite{CPAplus,AlgorithmComparison-JAR} for further details.

\subsubsection{Definition}
A CPA $\cpa = (D,\transabs{}{},\mergeop,\stopop)$ consists of
an abstract domain~$D$,
a transfer relation~$\transabs{}{}$,
and the operators~$\mergeop$ and $\stopop$.
The abstract domain $D = (C, \lattice, \sem{\cdot})$ consists of
a set~$C$ of concrete program states,
a semilattice~$\lattice = (\astates, \sqsubseteq)$
over a set~$\astates$ of abstract states and a partial order~$\sqsubseteq$,
and a concretization function~$\sem{\cdot}$
to map an abstract state to the represented set of concrete program states.
The transfer relation $\transabs{}{} \subseteq \astates \times \astates$
computes abstract successor states.
The merge operator $\mergeop: \astates\times\astates \to \astates$
specifies how to merge two abstract states when the control flow meets.
The stop operator $\stopop: \astates\times2^\astates \to \Bools$
determines whether an abstract state is covered by a given set of abstract states.
The operators $\mergeop$ and $\stopop$ can be chosen appropriately
to influence the abstraction level of the analysis.
Common choices include $\mergeop^\sep(\astate,\astate') = \astate'$
(which does not merge abstract states) and
$\stopop^\sep(\astate,R) = (\exists e' \in R : e \sqsubseteq e')$
(which determines coverage by checking whether the given abstract state is less than or equal to
any other reachable abstract state according to the semilattice).

\subsubsection{Fundamental CPAs and Composite CPA}
Several fundamental CPAs are used in this paper:
The \emph{Location CPA}~$\loccpa$~\cite{CPAplus} uses a flat lattice over all program locations to track the program counter explicitly;
the \emph{Loop-Bound CPA}~$\boundscpa$~\cite{kInduction,AlgorithmComparison-JAR} tracks in its abstract states for every loop of the program how often the loop body has been traversed on the current program path.
Another important CPA,
namely the \emph{Predicate~CPA}~$\predcpa$~\cite{AlgorithmComparison-JAR},
serves as the core data structure underlying the proposed IMC adoption.
The \emph{Predicate~CPA}~$\predcpa$ for \emph{adjustable-block encoding} (ABE)~\cite{ABE} uses a triple~$(\as,\labs{}{},\pf)$ of an abstraction formula~$\as$, an abstraction location~$\labs{}{}$, and a path formula~$\pf$ as an abstract state.
The abstraction formula~$\as$ stores the abstraction of the program state computed at the program location~$\labs{}{}$.
The path formula~$\pf$ syntactically encodes the program behavior from the abstraction location~$\labs{}{}$ to the current program location.
Abstract states where the path formula~$\pf$ is $\true$ are called \emph{abstraction states}; other abstract states are \emph{intermediate states}.

Several CPAs can be combined using a \emph{Composite CPA}~\cite{CPA} to achieve synergy.
The abstract states of the Composite CPA are tuples of one abstract state from each component CPA
and the operators of the Composite CPA can delegate to the component CPAs' operators accordingly.
We also use the \emph{ARG CPA}~$\argcpa$ to store the predecessor-successor relationship
between abstract states to track the \emph{abstract reachability graph}~(ARG).

\subsubsection{CPA Algorithm}
CPAs can be used by the CPA~algorithm~\cite{AlgorithmComparison-JAR,HBMC-dataflow,CPA},
which gets as input a CPA and an initial abstract state, for reachability analysis.
The algorithm performs a classic fixed-point iteration by looping
until all abstract states have been completely processed and returns the set of reachable abstract states.
The proposed adoption of IMC relies on an extension of the CPA~algorithm,
named \cpapa~\cite{AlgorithmComparison-JAR}.
Instead of an initial abstract state,
the \cpapa algorithm takes a set of reached abstract states
and a set of frontier abstract states awaiting processing.
It additionally receives as input a function $\mathsf{abort}$ to determine
whether it should abort early for some abstract state.
Upon completion, the \cpapa algorithm returns
the updated reached set and waiting list of abstract states.

%% file: encoding.tex
\newcommand{\nlooptasks}{\num{723}}
\newcommand{\nlooptruetasks}{\num{505}}
\newcommand{\nloopcorrect}{\num{36}}
\newcommand{\nloopproofs}{\num{11}}
\newcommand{\nloopalarms}{\num{25}}

\section{A Straightforward Adoption with Symbolic Program Counters}
\label{sect:symbolic-pc}
Before presenting the proposed efficient adoption of IMC for software verification,
we take a step back by first discussing a straightforward method
to encode IMC with symbolic program counters,
as mentioned in~\cref{sect:background-efficient-adoption}.
We will also demonstrate, both conceptually and empirically,
why such a brute-force encoding is not suitable for IMC on software.

\subsection{Encoding Transition Relations with Symbolic Program Counters}
The straightforward method derives the three predicates, namely,
the initial condition,
transition relation,
and safety property,
via introducing a variable $pc$ to store the program counter.
Given a CFA whose initial location is $l_0$,
the initial condition can be encoded as $pc=l_0$.
The transition relation of the CFA is the disjunction of the formula
$pc=l_i \land op \land pc'=l_j$ for each edge $(l_i,op,l_j)$ of the CFA,
where $op$ is the operation annotated to the edge,
and the variable~$pc'$ stores the program counter after the operation is executed.
Suppose a location $l_E$ of the CFA is specified as the error location
for a reachability-safety verification task,
then the corresponding safety property of the task can be expressed as $pc\neq l_E$.

We use the example CFA in~\cref{fig:even-cfa} to illustrate the encoding.
Recall that a primed variable denotes the program variable after one transition.
With the symbolic program counter~$pc$,
the initial condition is encoded as $I(pc)=(pc=l_3)$;
the safety property is expressed as $P(pc)=(pc\neq l_8)$;
the transition relation $T(pc,x,pc',x')$ of the CFA is captured as:
\begin{align*}
     & (pc=l_3 \land x'=0 \land pc'=l_4) ~\lor                               \\
     & (pc=l_4 \land \texttt{nondet()}\neq 0 \land pc'=l_5 \land x'=x) ~\lor \\
     & (pc=l_4 \land \texttt{nondet()}=0 \land pc'=l_7 \land x'=x) ~\lor     \\
     & (pc=l_5 \land x'=x+2 \land pc'=l_4) ~\lor                             \\
     & (pc=l_7 \land x\%2\neq 0 \land pc'=l_8 \land x'=x) ~\lor              \\
     & (pc=l_7 \land x\%2=0 \land pc'=l_{10} \land x'=x) ~\lor               \\
     & (pc=l_8 \land pc'=l_{11} \land x'=x) ~\lor                            \\
     & (pc=l_{10} \land pc'=l_{11} \land x'=x)
\end{align*}
Note that for each program variable that is not assigned by an operation,
the primed variable must be set equal to the respective unprimed variable.
For example, $x$ is not assigned by the edge $(l_4, [\texttt{nondet()}], l_5)$,
so $x'=x$ needs to be added to the edge's disjunctive term.

Having derived the three predicates,
we can perform IMC as described in~\cref{sect:background-imc}.
The following BMC query unrolls the transition relation $k$ times
and tests if the property can be violated after $k$ transitions:
\begin{align*}
    I(pc_0) \land T(pc_0,x_0,pc_1,x_1) \land \ldots \land T(pc_{k-1},x_{k-1},pc_k,x_k) \land \neg P(pc_k).
\end{align*}
To ease the readability,
we abbreviate $T(pc_i,x_i,pc_j,x_j)$ as $T_{i,j}$ in the following.
Note that the correctness of the example program can be proved
if the invariant $x\%2=0$ is established at the program location $l_4$.

\subsection{Drawbacks of the Encoding}
The obstacle hindering IMC to work with the encoding is the weak interpolants derived from the queries.
In particular, when program-counter variables appear in unsatisfiable queries,
the resultant interpolants tend to concern themselves mainly with the program counter $pc$
but seldom mention the program variables.
The lack of strong interpolants arises from syntactically infeasible paths encoded in the BMC queries.
We will discuss this problem using the example CFA in \cref{fig:even-cfa}.
First, note that for this CFA,
a potential error path from $l_3$ to $l_8$ must have an odd number of edges and at least three edges.
In other words, a BMC query with an even number of transitions is trivially unsatisfiable,
and the derived interpolant does not need to (and usually will not) refer to the program variable $x$.

For the BMC queries with a syntactically feasible path ($k = 3,5,7,9,\ldots$),
the derived interpolants may involve the program variable $x$.
However, IMC could still fail to reach a fixed point in this situation
(and does so in practice).
As an example, we continue on the CFA in~\cref{fig:even-cfa},
unroll the transition relation with $k=5$,
and pose the BMC query
$I \land T_{0,1} \land T_{1,2} \land T_{2,3} \land T_{3,4} \land T_{4,5} \land \neg P$.
This unsatisfiable BMC query allows syntactically feasible paths,
and thus it is possible to yield useful interpolants.
Suppose, with luck, a good interpolant $\itp_1=(pc=l_4 \land x\%2=0)$ is derived,
which is exactly the required invariant to prove the correctness of the program.
In the following,
we demonstrate how this valuable information might be discarded, unfortunately,
if IMC is performed without knowledge of the program structure.

The IMC algorithm described in~\cref{sect:background-imc} aims at building a fixed point
by iteratively computing more interpolants to overapproximate reachable states.
After computing~$\itp_1$,
the next interpolant is derived from a BMC query
with the initial condition replaced by the previous interpolant with appropriately shifted indices:
$\itp_1 \land T_{0,1} \land T_{1,2} \land T_{2,3} \land T_{3,4} \land T_{4,5} \land \neg P$.
This query is trivially unsatisfiable because
no path can go from $l_4$ to $l_8$ with five transitions,
which are the start and end points enforced by $\itp_1$ and $\neg P$, respectively.
Therefore, the interpolant for this query usually concerns only the program counter
and loses the information about program variables.
For example, we might obtain $\itp_2=(pc=l_5 \lor pc=l_7)$.
Starting from~$\itp_2$,
the next BMC query is satisfiable
because there is a feasible path from $l_5$ to $l_8$ with five transitions.
As a result, the IMC algorithm fails to reach a fixed point for $k=5$ and
has to go back to the BMC phase with an incremented unrolling bound.
A better interpolant without the program counter,
e.g., $(x\%2=0)$ instead of $\itp_2$,
could have prevented the loss of the information,
but it is rare to get such high-quality interpolants even from state-of-the-art interpolation engines.

We conducted an experiment on \nlooptasks~tasks
from the category \textit{ReachSafety-Loops} used in
the 2022 Competition on Software Verification~\cite{SVCOMP22-SVBENCHMARKS-artifact}
to support our conceptual reasoning above.
Among the \nlooptruetasks~tasks without property violation,
IMC with symbolic program counters only proved \nloopproofs~tasks.
Moreover, this encoding of IMC can prove the tasks because
the numbers of loop iterations in these \nloopproofs~tasks are bounded,
not because it constructed a fixed point from interpolants.
In our experiment,
IMC with symbolic program counters never
found a fixed point by interpolation for any task
and usually got trapped in the suboptimal situation explained above.

\subsection{Lessons Learned}
To avoid weak interpolants that only concern the program counter
and thus prevent reaching a fixed point,
we must not pose BMC queries about syntactically infeasible paths.
At the core of this issue is
mixing the information about the control flow and program semantics in the transition relation
when IMC is adopted with the brute-force conversion and symbolic program counters.
Next, we will present another adoption of IMC based on large-block encoding,
which separates the analysis of the control flow from the fixed-point computation of IMC
and hence improves the quality of derived interpolants.

%% file: approach.tex
\section{An Efficient Adoption with Large-Block Encoding}
\label{sect:approach}
In this section, we describe our proposed approach for adopting IMC to software verification.
In essence, we utilize \textit{large-block encoding} (LBE)~\cite{LBE} to draw an analogy between a program and the state-transition system discussed in~\cref{sect:background}.
The idea is not only helpful for this paper but might also shed light on the efficient adoptions of other algorithms.

As explained in~\cref{sect:symbolic-pc},
explicitly encoding symbolic program counters into the transition relation of a CFA
is not ideal for adopting IMC to software verification.
Sequential Boolean-logic circuits,
for which IMC was originally designed,
usually have only one feedback loop.
By contrast,
a CFA could have arbitrarily many loops.
To simplify the problem, we start by considering single-loop programs.
As a program with multiple loops can be converted into a single-loop program by a standard transformation~\cite{DragonBook,kIndForDMARaces},
this simplification will not hurt the generality of the proposed approach.
The effect of the single-loop transformation on the performance of IMC will be discussed in~\cref{sect:effect-SLT}.

To obtain the transition relation of a single-loop program, we take advantage of LBE~\cite{LBE}.
Given a CFA, LBE repeatedly rewrites the original CFA in order to summarize it.
In the summarized CFA, each loop-free subgraph of the original CFA is represented by a single control-flow edge.
The edge is annotated with a formula that encodes the program behavior of the represented subgraph of the original CFA.

For single-loop programs, applying LBE will always result in a summarized CFA
with a structure as shown in~\cref{fig:summarized-cfa}.
It has an initial location~$l_0$, a loop-head location~$l_H$, a loop-body location~$l_B$, a loop-tail location~$l_T$, and an error location~$l_E$.
These locations correspond to program locations in the original single-loop CFA before summarization.
The edges of the summarized single-loop CFA are labeled with the following formulas:
Formula~$\varphi_0$ summarizes the subgraph from~$l_0$ to~$l_H$,
formula~$C$ is the loop condition,
formula~$\varphi_L$ summarizes the subgraph from~$l_B$ back to~$l_H$, and
formulas~$\varphi_E, \varphi_E'$ summarize the subgraphs from~$l_T, l_B$ to~$l_E$, respectively.

\newsavebox{\summarizedCFA}
\begin{lrbox}{\summarizedCFA}
  \begin{minipage}[b]{0.4\textwidth}
    \centering
    \scalebox{0.8}{\input{figures/single-loop-cfa.tex}}
  \end{minipage}
\end{lrbox}

\newsavebox{\summarizedARG}
\begin{lrbox}{\summarizedARG}
  \begin{minipage}[b]{0.4\textwidth}
    \centering
    \scalebox{0.8}{\input{figures/single-loop-arg.tex}}
  \end{minipage}
\end{lrbox}

\begin{figure*}[t]
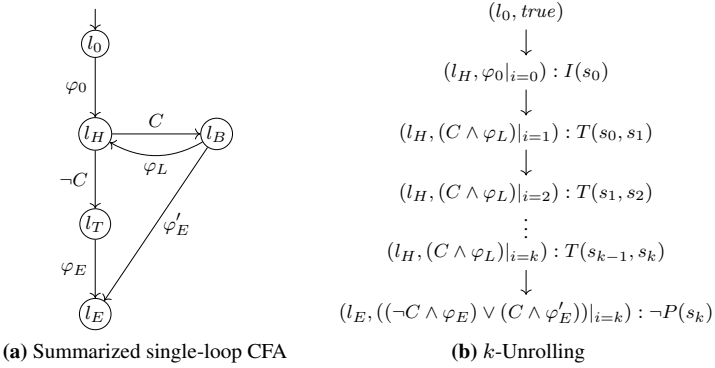

  \centering
  \subfloat[Summarized single-loop CFA]{\usebox{\summarizedCFA}\label{fig:summarized-cfa}}
  \subfloat[$k$-Unrolling]{\usebox{\summarizedARG}\label{fig:summarized-arg}}
  \caption{%
    A summarized single-loop CFA (a) and its $k$-unrolling (b)
  }
  \label{fig:summarized}
\end{figure*}

We notice that
the summarized single-loop CFA has a natural analogy to those predicates used in~\cref{sect:background-imc}:
The initial-state predicate $I(s)$ is analogous to $\varphi_0$,
the transition relation $T(s,s')$ is analogous to $C\land\varphi_L$, and
the negated safety property $\lnot P(s)$ is analogous to $(\lnot C\land\varphi_E)\lor (C\land\varphi_E')$.
Using LBE, we successfully obtain the required predicates
without explicitly encoding the program counter into the formulas.

Furthermore, in order to perform IMC, we have to unroll the summarized single-loop CFA and construct the BMC query~\cref{eq:BMC}.
In~\cref{fig:summarized-arg},
we unroll the CFA by drawing all possible paths starting from $l_0$,
iterating the loop $k$ times ($k+1$ visits to $l_H$),
and finally reaching $l_E$.
A node in~\cref{fig:summarized-arg} consists of a program location which the control flow is currently at and a formula $\blkf$ to encode all possible paths starting from the program location of the preceding node.
Note that $\blkf$ is indexed with the unrolling counter $i$ to distinguish between different iterations.

To discover the similarity between~\cref{eq:BMC} and~\cref{fig:summarized-arg}, we additionally label a node in~\cref{fig:summarized-arg} with the subformula in~\cref{eq:BMC} that $\blkf$ corresponds to.
From those labels, we observe that the formulas in the unrolled CFA nicely match the subformulas in~\cref{eq:BMC}.
We name the formula matching $I(s_0)$ \textit{prefix formula},
the formula matching $T(s_0,s_1)$ \textit{loop formula}, and
the formula matching $T(s_1,s_2) \land \ldots \land T(s_{k-1},s_k) \land \lnot P(s_k)$ \textit{suffix formula}.

We use the example CFA in~\cref{fig:even-cfa} to illustrate
how the LBE-based adoption of IMC
separates the analysis of the control flow and the semantical reasoning about program states.
The adoption avoids the usage of symbolic program counters in the formulas
and hence often leads to more helpful interpolants.
Recall that in~\cref{sect:symbolic-pc},
a copy of the transition relation means the execution of one program edge.
Therefore, the BMC query $I(s_0) \land T(s_0,s_1) \land \neg P(s_1)$
is equivalent to $(pc_0=l_3) \land (pc_0=l_3 \land x_1=0 \land pc_1=l_4) \land (pc_1=l_8)$,
which encodes only a single step from the initial program location $l_3$.
Since the error location $l_8$ is syntactically unreachable from $l_3$ via one edge,
the resulting interpolant does not need to concern the program variable $x$.
In practice, an interpolant for this query could be $pc_1 \neq l_8$.

By contrast,
if the required predicates $I(s),T(s,s'),P(s)$ are obtained with LBE,
the BMC query $I(s_0) \land T(s_0,s_1) \land \neg P(s_1)$
is equivalent to $x_0=0 \land \lnot(r_0=0)\land x_1=x_0+2 \land r_1=0 \land \lnot(x_1\%2=0)$,
where $r_0$ and $r_1$ denote the first and second returned values from the function \texttt{nondet()}, respectively.
It represents the semantics of all syntactically feasible paths from $l_3$ to $l_8$ that visit the loop-head location $l_4$ twice.
By Craig's interpolation theorem,
the interpolant has to talk about the program variable $x$.
In fact, as will be shown in~\cref{sect:approach-example},
IMC is able to prove the correctness of the example CFA
by deriving the loop invariant $x \% 2 = 0$ as an interpolant.

\subsection{Effect of Single-Loop Transformation}
\label{sect:effect-SLT}
In our approach,
programs with multiple loops are transformed to single-loop programs before IMC is applied.
The transformation introduces a fresh loop-head location,
which is the unique entry to the new single loop,
and a location variable to track which old loop should be entered next.
Auxiliary logic is added to the CFA to redirect the control flow
between the new loop head and the old ones based on the location variable.

Due to the existence of the location variable,
there might be trivially infeasible program paths
that enter a different loop from the one required by the location variable.
As discussed in~\cref{sect:symbolic-pc},
trivially unsatisfiable queries often result in weak interpolants
that prevent IMC from converging to a fixed point.
One solution to this issue is to use a dedicated CPA to track the location variable,
which works similarly to the Location CPA.
This CPA will not produce successors that enter a wrong loop,
and when used in a composite CPA,
it eliminates the aforementioned infeasible paths from the analysis.

In our experiments,
we evaluated the IMC adoption with and without the CPA tracking the location variable.
The latter treats the location variable as a normal variable and encodes it in the SMT formulas.
No significant difference was observed from the empirical results.
Unlike the issue of symbolic program counters discussed in~\cref{sect:symbolic-pc},
having the location variable (and the related infeasible program paths) in the BMC queries
does not slow down the convergence of IMC.
This is because the BMC queries obtained by LBE also include syntactically feasible paths.
To refute these syntactically feasible paths,
the interpolants cannot be trivial and have to concern the program variables.
Since the performance of the two alternatives are similar,
we stick to the one without the additional CPA for simplicity.

%% file: figures/single-loop-cfa.tex
\begin{tikzpicture}
\node (s) {};
\node (0) [below of = s, node distance=0.7cm,draw, circle,inner sep=0.03cm]{$l_0$};
\node (1) [below of = 0, node distance=1.5cm,draw, circle,inner sep=0.03cm]{$l_H$};
\node (2) [right of = 1, node distance=2cm,draw, circle,inner sep=0.03cm]{$l_B$};
\node (3) [below of = 1, node distance=1.5cm,draw, circle,inner sep=0.03cm]{$l_T$};
\node (4) [below of = 3, node distance=1.5cm,draw, circle,inner sep=0.03cm]{$l_E$};

\draw[->] (s) --(0);
\draw[->] (0) to node[left] {$\varphi_0$} (1);
\draw[->] (1) to node[above] {$C$} (2);
\draw[->] (2) edge[bend left] node[below] {$\varphi_L$} (1);
\draw[->] (1) to node[left] {$\neg C$} (3);
\draw[->] (3) to node[left] {$\varphi_E$} (4);
\draw[->] (2) to node[right] {$\varphi_E'$} (4);
\end{tikzpicture}

%% file: figures/single-loop-arg.tex
\begin{tikzpicture}
\node (0) {$(l_0,\true)$};
\node (1) [below of = 0, node distance=1cm]{$(l_H,\varphi_0|_{i=0}):I(s_0)$};
\node (2) [below of = 1, node distance=1cm]{$(l_H,(C\land\varphi_L)|_{i=1}):T(s_0,s_1)$};
\node (3) [below of = 2, node distance=1cm]{$(l_H,(C\land\varphi_L)|_{i=2}):T(s_1,s_2)$};
\node (4) [below of = 3, node distance=0.5cm]{$\vdots$};
\node (5) [below of = 4, node distance=0.5cm]{$(l_H,(C\land\varphi_L)|_{i=k}):T(s_{k-1},s_k)$};
\node (6) [below of = 5, node distance=1cm]{$(l_E,((\lnot C\land\varphi_E)\lor (C\land\varphi_E'))|_{i=k}):\lnot P(s_k)$};

\draw[->] (0) --(1);
\draw[->] (1) --(2);
\draw[->] (2) --(3);
\draw[->] (5) --(6);
\end{tikzpicture}

%% file: implementation.tex
\section{Implementation in \cpachecker}
\label{sect:implementation}
In this section,
we will describe an implementation to adopt IMC with large-block encoding.
We implemented the proposed adoption in the verification framework \cpachecker~\cite{CPACHECKER},
leveraging its flexibility provided by configurable program analysis~\cite{CPA}.
Before delving into implementation details,
we emphasize that the idea to extract a transition relation with LBE
is general and independent of the underlying framework.
We chose to implement the proposed adoption in~\cpachecker because it provides
(1) the necessary components for the adoption, which are highly configurable,
and (2) the implementations of various state-of-the-art software-verification algorithms,
which is convenient for the evaluation.

\subsection{Data Structures}
The \emph{Predicate~CPA} for ABE~\cite{ABE} serves as the core data structure
in our IMC adoption to store formulas that encode program semantics.
We add to an abstract state~$(\as,\labs{}{},\pf)$ of the Predicate CPA a \emph{block formula} $\blkf$,
which encodes all possible paths from the previous abstraction location
to the current abstraction location and is used to compute the abstraction formula.
In the implementation of~\cpachecker,
a block formula is already stored in the data structure for abstraction formulas.
We append it to an abstract state of the Predicate CPA in order to make the subsequent discussion more understandable.

With the help of ABE, we can achieve the effect of LBE via using the block-adjustment operator~$\blk^{l}$~\cite{ABE}.
The operator~$\blk^{l}$ will make the Predicate CPA convert an intermediate state to an abstraction state if the current program location is at the loop head or the error location.
Under this configuration, the unrolled ARG, if projected to abstraction states, will have a similar structure to~\cref{fig:summarized-arg}.
Therefore, we can easily obtain the required formulas by collecting and combining the block formulas from the corresponding abstraction states in the ARG.

It is worth noting that
here we take advantage of the flexibility of the Predicate~CPA:
By choosing an appropriate implementation for the block-adjustment operator,
we can configure the Predicate~CPA to be suitable for IMC
(together with the algorithms described in the following)
without further changes to its definition.
Other choices for its operators would allow it to implement different algorithms
like \impact, predicate abstraction, and \kinduction~\cite{AlgorithmComparison-JAR}.
Using the Predicate~CPA as common framework
not only highlights conceptual differences and similarities between the approaches
but also allows for comparing them experimentally
with the set of confounding variables kept to a minimum.

\subsection{Algorithmic Procedures}
We present an algorithm for the adoption of IMC to software verification in~\cref{imc-algo-main},
which is based on the \cpapa algorithm~\cite{AlgorithmComparison-JAR}.
The algorithm assumes single-loop programs as input.
We apply single-loop transformation~\cite{DragonBook,kIndForDMARaces}
to input programs with multiple loops as a preprocessing.
The algorithm takes as input an upper limit $k_{max}$ for a counter~$k$
that tracks the number of loop iterations on a program path%
\footnote{
    While the algorithm \cpapa unrolls the program $k$~times,
    the algorithm~\cref{imc-algo-main} uses the last unrolling only for encoding the predicate~$P(s)$
    and thus only $k-1$~copies of $T(s,s')$ appear in its BMC query.
    This is done for consistency with other algorithms
    expressed on top of the same unifying framework~\cite{AlgorithmComparison-JAR}.
}
and a composite CPA consisting of the Location CPA,
the Predicate CPA,
and the Loop-Bound CPA.

\begin{algorithm}[t]
    \caption{IMC: main procedure}
    \label{imc-algo-main}
    \begin{algorithmic}[1]
        \newcommand\lc{i}
        \REQUIRE
        an upper limit $k_{max}$ for the loop unrolling bound $k$,\\ \hspace{4.5mm}
        a composite CPA $\cpa$ with components: the Location CPA $\loccpa$,\\ \hspace{4.5mm}
        the Predicate CPA $\predcpa$, and the Loop-Bound CPA $\boundscpa$\\
        \ENSURE
        \FALSE{} if an error path to $l_E$ is found,\\ \hspace{6.5mm}
        \TRUE{} if a fixed point is obtained,\\ \hspace{6.5mm}
        \textbf{unknown} otherwise
        \STATE $k := 1$
        \STATE $\initial\astate := (\initial\pc,(\true,\initial\pc,\true,\true),\{l_H\mapsto-1\})$\hfill\COMMENT{Create initial abstract state at $\pci$}
        \STATE $\reached := \wait := \{\initial\astate\}$
        \WHILE{$k \leq k_{max}$}
        \STATE $(\reached, \wait) := \textsf{\cpapa}(\cpa, \reached, \wait, k)$\label{imc-unroll-cfa}
        \STATE $\blkf_p := \blkf \mid (l_H,(\cdot,\cdot,\cdot,\blkf),\{l_H\mapsto0\}) \in \reached$\label{imc-collect-formula-start}
        \STATE $\blkf_l := \true$
        \IF{$k>1$}
        \STATE $\blkf_l := \blkf \mid (l_H,(\cdot,\cdot,\cdot,\blkf),\{l_H\mapsto1\}) \in \reached$
        \ENDIF
        \STATE $\blkf_s := \bigwedge_{i=2}^{k-1} \blkf \mid (l_H,(\cdot,\cdot,\cdot,\blkf),\{l_H\mapsto\lc\}) \in \reached ~~\land$\\
        \hspace{7.2mm}$\bigvee \left\{ \blkf \mid (l_E,(\cdot,\cdot,\cdot,\blkf),\{l_H\mapsto(k-1)\}) \in \reached\right\}$\label{imc-collect-formula-end}
        \IF{\textsf{sat}($\blkf_p\wedge \blkf_l\wedge \blkf_s$)}\label{imc-bmc-step}
        \RETURN \FALSE{} \hfill\COMMENT{Found an error path via BMC query}
        \ENDIF
        \IF{$k>1$ \textbf{and} \textsf{reach\_fixed\_point($\blkf_p$,$\blkf_l$,$\blkf_s$)}}
        \RETURN \TRUE{} \hfill\COMMENT{Obtained a fixed point via interpolation}
        \ENDIF
        \STATE $k:=k+1$
        \ENDWHILE
        \RETURN \textbf{unknown}
    \end{algorithmic}
\end{algorithm}

\begin{algorithm}[t]
    \caption{IMC: \textsf{reach\_fixed\_point($\blkf_p$,$\blkf_l$,$\blkf_s$)}}
    \label{imc-algo-fixed}
    \begin{algorithmic}[1]
        \REQUIRE
        prefix formula $\blkf_p$, loop formula $\blkf_l$, and suffix formula $\blkf_s$
        \ENSURE \TRUE{} if a fixed point is reached, \FALSE{} otherwise
        \STATE \textsf{image} $:=$ \textsf{start} $:= \blkf_p$ \hfill \COMMENT{Set current reachable and starting states to initial states}
        \WHILE{$\lnot\mathsf{sat}$($\mathsf{start}\land\blkf_l\land\blkf_s$)}\label{imc-inner-while-loop}
        \STATE $\itp:= \mathsf{get\_interpolant}(\mathsf{start}\land\blkf_l,\blkf_s)$ \hfill \COMMENT{formula $A$: $\mathsf{start}\land\blkf_l$; formula $B$: $\blkf_s$}
        \STATE $\itp:= \mathsf{shift\_variable\_index}(\itp,\blkf_p)$
        \IF{$\lnot\mathsf{sat}$($\itp\land\lnot\mathsf{image}$)}
        \RETURN \TRUE{} \hfill \COMMENT{Interpolant implies image: fixed point}
        \ENDIF
        \STATE $\mathsf{image}:=\mathsf{image}\lor\itp$ \hfill \COMMENT{Find new states: enlarge image}
        \STATE $\mathsf{start}:=\itp$ \hfill \COMMENT{Start next iteration from new states}
        \ENDWHILE
        \RETURN \FALSE{} \hfill \COMMENT{Reach error: might be wrong alarm}
    \end{algorithmic}
\end{algorithm}

An abstract state of the composite CPA is $(\pc, (\as,\labs{}{},\pf,\blkf), \{l_H\mapsto i\})$,
where the first element is an abstract state of the Location CPA representing the current program location $l$,
the second element is an abstract state of the Predicate CPA as explained above,
and the third element is an abstract state of the Loop-Bound CPA recording that
the loop body starting from $l_H$ has been completely traversed $i$~times already.
We also use the ARG CPA~$\argcpa$ to store the predecessor-successor relationship between abstract states.
To increase readability, we simply use abstract states as elements in the ARG
and explicitly give the unrolling upper bound~$k$ as a parameter to the \cpapa algorithm
(instead of passing it via the precision of the initial abstract state as in the literature~\cite{AlgorithmComparison-JAR}).
We also omit the aborting function and assume that the \cpapa algorithm never aborts early
(i.e., we pass $\mathsf{abort}^\mathit{never} = \{\cdot\mapsto\false\}$).

After unrolling the CFA with the \cpapa algorithm (\cref{imc-unroll-cfa}), we have to collect prefix, loop, and suffix formulas to pose a BMC query and perform the fixed-point computation via interpolation.
The formula collection is described in \crefrange{imc-collect-formula-start}{imc-collect-formula-end}, where we write $\blkf | {(\pc, (\as,\labs{}{},\pf,\blkf), \{l_H\mapsto i\})}$ to denote the block formula $\blkf$ of the abstract state ${(\pc, (\as,\labs{}{},\pf,\blkf), \{l_H\mapsto i\})}$.
The prefix formula $\blkf_p$ is the block formula of the abstract state ${(l_H, (\cdot,\cdot,\cdot,\blkf), \{l_H\mapsto 0\})}$;
if the loop body has been completely traversed at least once, i.e., $k>1$,
the loop formula $\blkf_l$ is the block formula of the abstract state ${(l_H, (\cdot,\cdot,\cdot,\blkf), \{l_H\mapsto 1\})}$,
otherwise, it is set to $\true$;
the suffix formula $\blkf_s$ is the conjunction of the following two formulas:
the conjunction of block formulas of the abstract state
${(l_H, (\cdot,\cdot,\cdot,\blkf), \{l_H\mapsto i\})}$ for $i=2,\ldots,(k-1)$ and
the disjunction of the block formulas of the abstract states
${(l_E, (\cdot,\cdot,\cdot,\blkf), \{l_H\mapsto (k-1)\})}$.

Note that the above formula collection at abstract states whose locations equal $l_H$ is unambiguous, meaning that there is a unique abstract state satisfying the conditions imposed by the Location~CPA~$\loccpa$ and the Loop-Bound CPA~$\boundscpa$.
This is because we assume single-loop programs and use LBE to summarize all paths between two adjacent abstraction states.
After collecting these formulas, the BMC query is simply the conjunction of the prefix, loop, and suffix formulas.
If the BMC query is unsatisfiable, we try to compute a fixed point using~\cref{imc-algo-fixed},
which implements the procedure described in~\cref{sect:background-imc} to iteratively derive interpolants from unsatisfiable BMC queries and grow a fixed point as their union.

\Cref{imc-algo-fixed} first initializes both \textsf{image}, which stores an overapproximation of the reachable states,
and \textsf{start}, which stores the starting states of BMC queries, to be the prefix formula.
Using $\mathsf{start}\land\blkf_l$ as formula $A$ and $\blkf_s$ as formula $B$, we derive an interpolant $\itp$.
As discussed in~\cref{sect:background-imc},
the $i^{th}$~interpolant is an overapproximation of the reachable states after $i$~loop iterations.
We change the variables used in the interpolant to those in the prefix formula
and check whether the interpolant implies \textsf{image}.
If so, a fixed point has been reached, and we conclude the property is true;
otherwise, we enlarge \textsf{image} by adding the states contained in the interpolant to it
and pose another BMC query starting from the interpolant.
If any BMC query during the iteration is satisfiable,
we return back to~\cref{imc-algo-main} and increase the loop-unrolling counter $k$ to check whether the violation is a wrong alarm.

\subsection{Example}
\label{sect:approach-example}

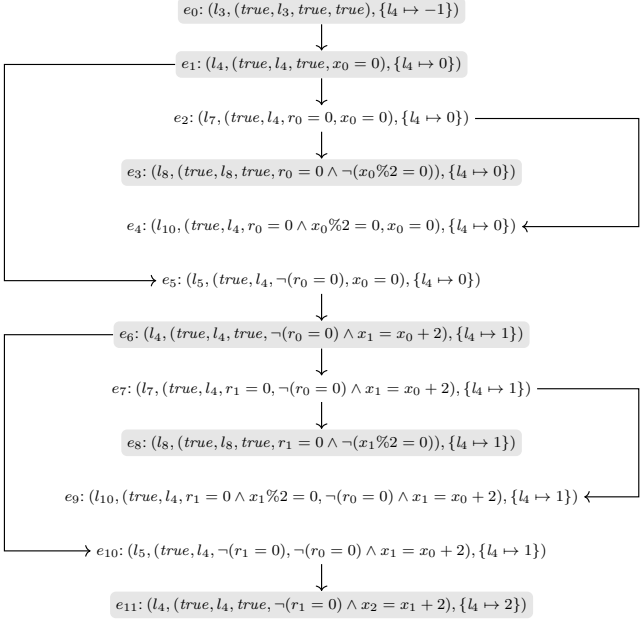
\begin{figure}[t]
    \centering
    \scalebox{0.9}{\input{figures/even-arg.tex}}
    \caption{ARG constructed by the \cpapa algorithm~\cite{AlgorithmComparison-JAR} for the CFA in \cref{fig:even-cfa} ($k=2$)}
    \label{fig:example-imc}
\end{figure}

We demonstrate step-by-step how to apply~\cref{imc-algo-main} and~\cref{imc-algo-fixed} to verify the CFA in~\cref{fig:even-cfa}.
The ARG constructed by the \cpapa algorithm when $k=2$ is shown in~\cref{fig:example-imc}.
In this figure, each abstract state is a tuple ${(\pc, (\as,\labs{}{},\pf,\blkf), \{\pc_4\mapsto i\})}$
of the abstract states of $\loccpa$, $\predcpa$, and $\boundscpa$.
Note that every abstract state in the ARG has an abstraction formula $\as$ (the first element in an abstract state of Predicate CPA) equal to $\true$ because IMC does not compute an abstraction formula.
Instead, it relies on interpolants for the abstraction of program states.
Abstract states whose predicate abstract state is an abstraction state
(where the path formula is always reset to $\true$) are highlighted in gray.
We use~$r$ to denote the returned value of the function~\texttt{nondet}.

The prefix formula~$\blkf_p$ is the block formula $x_0=0$ of the abstract state~$e_1$,
the loop formula~$\blkf_l$ is the block formula $\lnot(r_0=0)\land x_1=x_0+2$ of the abstract state~$e_6$,
and the suffix formula~$\blkf_s$ is the block formula $r_1=0\land\lnot(x_1\%2=0)$ of the abstract state~$e_8$
(note that the block formula of the abstract state~$e_3$, which also has location~$l_8$,
is not selected because $l_4 \mapsto 0$ does not match in \cref{imc-collect-formula-end} of \cref{imc-algo-main}).
As the BMC query ${x_0=0 \land \lnot(r_0=0)\land x_1=x_0+2} \land {r_1=0 \land \lnot(x_1\%2=0)}$ is unsatisfiable, we try to compute a fixed point using~\cref{imc-algo-fixed}.

Variables \textsf{image} and \textsf{start} are initialized to $x_0=0$.
Using $x_0=0 \land \lnot(r_0=0)\land x_1=x_0+2$ as formula $A$ and $r_1=0\land\lnot(x_1\%2=0)$ as formula $B$,
we can derive an interpolant $\itp$ from the unsatisfiable BMC query.
Assume that $\itp$ is $x_1\%2=0$, referring to the common variable $x_1$ of formulas $A$ and $B$.
After shifting the variable to the one used in $\blkf_p$, we obtain $x_0\%2=0$.
As the interpolant does not imply \textsf{image}, we enlarge the current image by disjoining it with the interpolant.
The computation is then repeated again, with \textsf{start} equal to $x_0\%2=0$ this time.
The BMC query in the second iteration becomes $x_0\%2=0 \land \lnot(r_0=0)\land x_1=x_0+2 \land r_1=0 \land\lnot(x_1\%2=0)$, which is still unsatisfiable.
Assume the interpolant is again $x_1\%2=0$.
Obviously, we have reached a fixed point, as the newly derived interpolant implies \textsf{image}.
Therefore, we conclude that the property holds.

\subsection{Correctness}
It is straightforward to see that~\cref{imc-algo-main} is \emph{precise}, i.e.,
does not produce wrong alarms, because if it returns \textbf{false},
then the BMC query for all paths from~$l_0$ to~$l_E$ at~\cref{imc-bmc-step} is satisfiable,
which implies that the CFA has a feasible path to~$l_E$.
More interesting is the soundness of~\cref{imc-algo-main},
i.e., whether it may produce wrong proofs,
which we discuss in the following.
Its soundness follows from that of large-block encoding~\cite{LBE} and the original IMC algorithm~\cite{McMillanCraig}.
We state the soundness of~\cref{imc-algo-main} when it is applied to a single-loop CFA in~\cref{thm:imc-soundness}.
For CFAs with multiple loops,
the soundness will also depend on that of the single-loop transformation~\cite{DragonBook,kIndForDMARaces}.

\begin{theorem}\label{thm:imc-soundness}
    Given a single-loop CFA~$A$ and its corresponding composite CPA~$\cpa$ as input,
    if \cref{imc-algo-main} returns \textbf{true} upon~$\cpa$,
    then $A$ does not have a feasible path to~$l_E$.
\end{theorem}
\begin{proof}
    We prove the statement by contradiction.
    Suppose~\cref{imc-algo-main} returns \textbf{true} when the value of the loop-unrolling counter $k$ equals~$\hat{k}$,
    but the single-loop CFA has a feasible path to $l_E$.
    We split into two cases based on the number $\hat{h}$ of the visits to $l_H$ on the error path.

    First, assume $\hat{h}\leq\hat{k}$.
    Thanks to the sound summarization of LBE,
    the formula of the error path must imply $\blkf_p\land\blkf_l\land\blkf_s$ when $k=\hat{h}$.
    Therefore, \cref{imc-algo-main} should have returned \textbf{false} at $k=\hat{h}$,
    because the BMC query at~\cref{imc-bmc-step} of~\cref{imc-algo-main} is satisfiable.
    This result contradicts the assumption that~\cref{imc-algo-main} returns \textbf{true}.

    Second, assume $\hat{h} > \hat{k}$.
    Such an error path indicates the existence of a state $\hat{s}$
    that is reachable from $l_0$ by traversing the loop $\hat{h}-\hat{k}$ times
    and will reach $l_E$ after further traversing the loop $\hat{k}-1$ times.
    We will show that~\cref{imc-algo-fixed} will return \textbf{false} after discovering $\hat{s}$ via interpolation.
    Note that~\cref{imc-algo-fixed} cannot return \textbf{true} before finding $\hat{s}$
    because the state must be contained in the computed fixed point.

    According to the property of the original IMC algorithm described in~\cref{sect:background-imc},
    the interpolant derived in the $i^{th}$ while-loop iteration of~\cref{imc-algo-fixed} is an overapproximation of the set of states reachable from $l_0$ by traversing the loop $i$ times.
    Therefore, $\hat{s}$ must belong to the interpolant $\itp$
    derived in the $(\hat{h}-\hat{k})^{th}$ while-loop iteration of~\cref{imc-algo-fixed},
    which will be used as new starting states in the next iteration.
    Moreover, because of the soundness of LBE,
    the formula from the $(\hat{h}-\hat{k}+1)^{th}$ $l_H$ to $l_E$ (involving $\hat{k}$ visits to $l_H$) on the error path
    must imply $\blkf_l\land\blkf_s$ when we enter~\cref{imc-algo-fixed} with $k=\hat{k}$.
    Thus, in the beginning of the next iteration,
    the satisfiability query at~\cref{imc-inner-while-loop} of~\cref{imc-algo-fixed} must be satisfiable,
    which makes~\cref{imc-algo-fixed} return \textbf{false}.
    This in turn prevents~\cref{imc-algo-main} from returning \textbf{true} when $k=\hat{k}$,
    contradicting our assumption.

    Having analyzed the above two possibilities,
    we conclude that such a feasible error path does not exist,
    and hence~\cref{imc-algo-main} is sound.
\end{proof}

\subsection{Backward Derivation of Interpolants}
Notice that in the example of~\cref{sect:approach-example}, the ``quality'' of interpolants heavily affects the convergence of the fixed-point computation.
For example, instead of $x_1\%2=0$, which is actually the loop invariant, suppose the interpolant derived by the solver is $x_1=2$.
Starting from this interpolant, we might be trapped in a sequence of interpolants $x_1=4$, $x_1=6$, $x_1=8, \ldots$ and never reach a fixed point.

While in general it is difficult to control the interpolation process of the solver, there is a trick to mitigate this problem.
First, we switch the labels of the two formulas, i.e., we label the original formula~$B$ as the new formula~$A$ and the original formula~$A$ as the new formula~$B$.
Second, we ask the solver to derive an interpolant for the new formulas and then negate it.
The negated interpolant is a valid interpolant for the original formulas~$A$ and~$B$.
In other words, instead of $\mathsf{get\_interpolant}(A, B)$, we use $\lnot \mathsf{get\_interpolant}(B, A)$.

Using this trick in IMC, we are actually deriving interpolants backwards from the safety property.
Therefore, we call it \textit{backward derivation} of interpolants.
With the backward derivation, we can in practice often avoid the bad interpolant $x_1=2$ and obtain the good one $x_1\%2=0$ for fast convergence of the example program in~\cref{sect:approach-example}.
Empirically, we found that the backward derivation performs slightly better than the forward derivation.
This phenomenon might be attributed to the fact that deriving the interpolants backward from the safety property side is likely to yield interpolants summarizing information more relevant to proving the property.
As a result, we use it as default in our implementation.

%% file: figures/even-arg.tex
\scalebox{.8}{
\begin{tikzpicture}[
    semithick,
    on grid,
    start chain=going below,
  ]

  \newcommand\intnode[8][]{\node[on chain,#1] (a#2) {\intstate{#2}{#3}{#4}{#5}{#6}{#7}{#8}};}

  \intnode[abstraction state]{0}{l_3}{\true}{l_3}{\true}{\true}{-1};
  \intnode[abstraction state]{1}{l_4}{\true}{l_4}{\true}{x_0=0}{0};
  
  \intnode{2}{l_7}{\true}{l_4}{r_0=0}{x_0=0}{0};
  \intnode[abstraction state]{3}{l_8}{\true}{l_8}{\true}{r_0=0 \land \lnot(x_0 \% 2 = 0)}{0};
  \intnode{4}{l_{10}}{\true}{l_4}{r_0=0 \land x_0 \% 2 = 0}{x_0=0}{0};
  
  \intnode{5}{l_5}{\true}{l_4}{\lnot(r_0=0)}{x_0=0}{0};
  \intnode[abstraction state]{6}{l_4}{\true}{l_4}{\true}{\lnot(r_0=0) \land x_1 = x_0 + 2}{1};
  
  \intnode{7}{l_7}{\true}{l_4}{r_1 = 0}{\lnot(r_0=0) \land x_1 = x_0 + 2}{1}
  \intnode[abstraction state]{8}{l_8}{\true}{l_8}{\true}{r_1 = 0 \land \lnot(x_1 \% 2 = 0)}{1}
  \intnode{9}{l_{10}}{\true}{l_4}{r_1=0 \land x_1 \% 2 = 0}{\lnot(r_0=0) \land x_1 = x_0 + 2}{1};
  
  \intnode{10}{l_5}{\true}{l_4}{\lnot(r_1=0)}{\lnot(r_0=0) \land x_1 = x_0 + 2}{1};
  \intnode[abstraction state]{11}{l_4}{\true}{l_4}{\true}{\lnot(r_1=0) \land x_2 = x_1 + 2}{2};

  \draw[arg edge]  (a0) --  (a1);
  \draw[arg edge]  (a1) --  (a2);
  \draw[arg edge]  (a2) --  (a3);
  \draw[arg edge]  (a2) -|  ($(a4-|a9.east)+(1,0)$) -- (a4);
  \draw[arg edge]  (a1) -|  ($(a5-|a9.west)-(1,0)$) -- (a5);
  \draw[arg edge]  (a5) --  (a6);
  \draw[arg edge]  (a6) --  (a7);
  \draw[arg edge]  (a7) --  (a8);
  \draw[arg edge]  (a7) -|  ($(a9-|a9.east)+(1,0)$) -- (a9);
  \draw[arg edge]  (a6) -|  ($(a10-|a9.west)-(1,0)$) -- (a10);
  \draw[arg edge]  (a10) --  (a11);
\end{tikzpicture}
}

%% file: evaluation.tex
\newcommand{\experimentRevision}{revision \href{https://svn.sosy-lab.org/software/cpachecker/branches/cfa-single-loop-transformation/?p=43042}{\num{43042}} of branch \texttt{cfa-single-loop-transformation}\xspace}
\newcommand\plotpath{evaluation/tex}
\newcommand{\ntasks}{\num{6024}}
\newcommand{\ntruetasks}{\num{4231}}
\newcommand{\nfalsetasks}{\num{1793}}
\newcommand{\nimcvsimpactproofs}{\num{328}}
\newcommand{\nimcvsimpactalarms}{\num{25}}
\newcommand{\nimcwrongalarms}{\num{3}}
\newcommand{\nimcvspdrtasks}{\num{1199}}
\newcommand{\nimcvsbmctasks}{\num{929}}
\newcommand{\nimcvskitasks}{\num{100}}
\newcommand{\nimcvspredtasks}{\num{323}}
\newcommand{\nimcvsimpacttasks}{\num{185}}
\newcommand{\nimcuniquetasks}{\num{7}}
\newcommand{\necatasks}{\num{1265}}
\newcommand{\nimcecatasks}{\num{565}}
\newcommand{\nkiecatasks}{\num{607}}
\newcommand{\npredecatasks}{\num{476}}
\newcommand{\nimpactecatasks}{\num{555}}
\newcommand{\ntrueecatasks}{\num{785}}
\newcommand{\nimcecaproofs}{\num{423}}
\newcommand{\nkiecaproofs}{\num{497}}
\newcommand{\npredecaproofs}{\num{348}}
\newcommand{\nimpactecaproofs}{\num{390}}
\newcommand{\nimcimpactbothsolved}{\num{382}}
\newcommand{\nimcfeweritpcalls}{\num{354}}

\section{Evaluation}
\label{sect:evaluation}
To evaluate the proposed adoption of IMC~\cite{McMillanCraig} and understand its characteristics,
we carried out two parts of experiments to answer the research questions below:
\begin{itemize}
  \item Part 1: IMC vs. other SMT-based algorithms
        \begin{itemize}
          \item \textbf{RQ1}: Can IMC solve more safety-verification tasks?
          \item \textbf{RQ2}: Can IMC solve safety-verification tasks faster?
          \item \textbf{RQ3}: Can IMC solve tasks unsolvable by existing approaches?
        \end{itemize}
  \item Part 2: IMC vs. \impact~\cite{IMPACT} (a closely related interpolation-based algorithm)
        \begin{itemize}
          \item \textbf{RQ4}: Why can IMC deliver more proofs than \impact?
        \end{itemize}
\end{itemize}
We evaluated the adoption of IMC based on large-block encoding
against several state-of-the-art SMT-based algorithms
on the largest publicly available benchmark suite of C safety-verification tasks~\cite{SVCOMP22}.
We excluded the naive adoption of IMC with symbolic program counters from the evaluation
because it was shown infeasible in the experiment described in~\cref{sect:symbolic-pc}.

\subsection{Evaluated Approaches}
We assessed IMC against five SMT-based verification algorithms,
including BMC~\cite{BMC},
\kinduction~\cite{k-Induction},
predicate abstraction~\cite{AbstractionsFromProofs},
\impact~\cite{IMPACT},
and PDR~\cite{IC3}.
All of the compared approaches are implemented in \cpachecker.
The implementations of BMC, \kinduction, predicate abstraction, and \impact are built on top of the \cpapa algorithm in a unified manner~\cite{AlgorithmComparison-JAR}.
The implementation of PDR in \cpachecker follows a software-verification adaptation named \ctigar~\cite{CTIGAR},
which was compared against other PDR-related approaches recently in the literature~\cite{PDR}.
We did not include other state-of-the-art verifiers in the evaluation
to keep confounding variables at a minimum
(same parser, same libraries, same SMT solver, etc.).
We chose \cpachecker because it is a flexible framework
that performed well in the competitions.
Empirical results of \cpachecker against other software verifiers
are available from the report~\cite{SVCOMP22} of the 2022 Competition on Software Verification (SV-COMP\,'22).

\subsection{Benchmark Set}
As the benchmark set,
we used the verification tasks~\cite{SVCOMP22-SVBENCHMARKS-artifact} from SV-COMP\,'22.
We used only verification tasks where the safety property is the reachability of a program location.
From those,
we further excluded verification tasks that are not supported by at least one of the compared approaches,
e.g., those from the categories
\textit{ReachSafety-Recursive} and
\textit{ConcurrencySafety-Main}.
The resulting benchmark set consists of a total of~\ntasks~verification tasks from
the subcategories
\textit{AWS-C-Common-ReachSafety},
\textit{DeviceDriversLinux64-ReachSafety},
\textit{DeviceDriversLinux64Large-ReachSafety}, and
\textit{uthash-ReachSafety}
of the category \textit{SoftwareSystems}
and from the following subcategories of the category \textit{ReachSafety}:
\textit{Arrays},
\textit{Bitvectors},
\textit{ControlFlow},
\textit{ECA},
\textit{Floats},
\textit{Heap},
\textit{Loops},
\textit{ProductLines},
\textit{Sequentialized},
\textit{XCSP}, and
\textit{Combinations}.
A total of~\nfalsetasks~tasks in the benchmark set contain a known specification violation,
while the other~\ntruetasks~tasks are assumed to be correct.

\subsection{Experimental Setup}
Our experiments were performed on machines with
one 3.4\,GHz CPU (Intel Xeon E3-1230~v5) with 8~processing units and 33\,GB of RAM each.
The operating system was Ubuntu~22.04 (64~bit),
using Linux~5.15 and OpenJDK~17.0.5.
Each verification task was limited to
two CPU cores,
a CPU time of \SI{15}{min},
and a memory usage of \SI{15}{GB}.
We used \benchexec\footnote{\url{\benchexecurl}}~\cite{Benchmarking-STTT} to achieve reliable benchmarking
and \experimentRevision of \cpachecker for evaluation.
We configured \cpachecker to use the SMT theories
of equality with uninterpreted functions, bit vectors, floats, and arrays.
All SMT queries were handled by \mathsat~\cite{MATHSAT5}.

\input{\plotpath/data-commands}
\begin{table*}[t]
  \centering
  \caption{Summary of the results for~\ntasks~reachability-safety verification tasks}
  \label{tab:results}
  \newcommand\stoh[1]{\fpeval{#1/3600}}
  \newcommand\precnum[1]{\tablenum[round-mode=off,table-format=4]{#1}}
  \newcommand\rndnum[1]{\tablenum[round-mode=figures,round-precision=2,table-format=4.1]{#1}}
  \begin{tabular}{lccc@{\hspace{3.5mm}}c@{\hspace{3.5mm}}c@{\hspace{3.5mm}}c}
    \toprule
    Algorithm          & {IMC}                                                                                       & {PDR} & {BMC} & {\kInduction} & {Predicate Abstraction} & {\impact} \\
    \midrule
    Correct results    & \precnum{\JarFullInterpolationModelCheckingImcCorrectCount}
                       & \precnum{\JarFullInterpolationModelCheckingPdrCorrectCount}
                       & \precnum{\JarFullInterpolationModelCheckingBmcCorrectCount}
                       & \precnum{\JarFullInterpolationModelCheckingKInductionCorrectCount}
                       & \precnum{\JarFullInterpolationModelCheckingPredicateAbstractionCorrectCount}
                       & \precnum{\JarFullInterpolationModelCheckingImpactCorrectCount}                                                                                                    \\
    \quad proofs       & \precnum{\JarFullInterpolationModelCheckingImcCorrectTrueCount}
                       & \precnum{\JarFullInterpolationModelCheckingPdrCorrectTrueCount}
                       & \precnum{\JarFullInterpolationModelCheckingBmcCorrectTrueCount}
                       & \precnum{\JarFullInterpolationModelCheckingKInductionCorrectTrueCount}
                       & \precnum{\JarFullInterpolationModelCheckingPredicateAbstractionCorrectTrueCount}
                       & \precnum{\JarFullInterpolationModelCheckingImpactCorrectTrueCount}                                                                                                \\
    \quad alarms       & \precnum{\JarFullInterpolationModelCheckingImcCorrectFalseCount}
                       & \precnum{\JarFullInterpolationModelCheckingPdrCorrectFalseCount}
                       & \precnum{\JarFullInterpolationModelCheckingBmcCorrectFalseCount}
                       & \precnum{\JarFullInterpolationModelCheckingKInductionCorrectFalseCount}
                       & \precnum{\JarFullInterpolationModelCheckingPredicateAbstractionCorrectFalseCount}
                       & \precnum{\JarFullInterpolationModelCheckingImpactCorrectFalseCount}                                                                                               \\
    Wrong proofs       & \precnum{\JarFullInterpolationModelCheckingImcWrongTrueCount}
                       & \precnum{\JarFullInterpolationModelCheckingPdrWrongTrueCount}
                       & \precnum{\JarFullInterpolationModelCheckingBmcWrongTrueCount}
                       & \precnum{\JarFullInterpolationModelCheckingKInductionWrongTrueCount}
                       & \precnum{\JarFullInterpolationModelCheckingPredicateAbstractionWrongTrueCount}
                       & \precnum{\JarFullInterpolationModelCheckingImpactWrongTrueCount}                                                                                                  \\
    Wrong alarms       & \precnum{\JarFullInterpolationModelCheckingImcWrongFalseCount}
                       & \precnum{\JarFullInterpolationModelCheckingPdrWrongFalseCount}
                       & \precnum{\JarFullInterpolationModelCheckingBmcWrongFalseCount}
                       & \precnum{\JarFullInterpolationModelCheckingKInductionWrongFalseCount}
                       & \precnum{\JarFullInterpolationModelCheckingPredicateAbstractionWrongFalseCount}
                       & \precnum{\JarFullInterpolationModelCheckingImpactWrongFalseCount}                                                                                                 \\
    Timeouts           & \precnum{\JarFullInterpolationModelCheckingImcErrorTimeoutCount}
                       & \precnum{\JarFullInterpolationModelCheckingPdrErrorTimeoutCount}
                       & \precnum{\JarFullInterpolationModelCheckingBmcErrorTimeoutCount}
                       & \precnum{\JarFullInterpolationModelCheckingKInductionErrorTimeoutCount}
                       & \precnum{\JarFullInterpolationModelCheckingPredicateAbstractionErrorTimeoutCount}
                       & \precnum{\JarFullInterpolationModelCheckingImpactErrorTimeoutCount}                                                                                               \\
    Out of memory      & \precnum{\JarFullInterpolationModelCheckingImcErrorOutOfMemoryCount}
                       & \precnum{\JarFullInterpolationModelCheckingPdrErrorOutOfMemoryCount}
                       & \precnum{\JarFullInterpolationModelCheckingBmcErrorOutOfMemoryCount}
                       & \precnum{\JarFullInterpolationModelCheckingKInductionErrorOutOfMemoryCount}
                       & \precnum{\JarFullInterpolationModelCheckingPredicateAbstractionErrorOutOfMemoryCount}
                       & \precnum{\JarFullInterpolationModelCheckingImpactErrorOutOfMemoryCount}                                                                                           \\
    Other inconclusive & \precnum{\JarFullInterpolationModelCheckingImcErrorOtherInconclusiveCount}
                       & \precnum{\JarFullInterpolationModelCheckingPdrErrorOtherInconclusiveCount}
                       & \precnum{\JarFullInterpolationModelCheckingBmcErrorOtherInconclusiveCount}
                       & \precnum{\JarFullInterpolationModelCheckingKInductionErrorOtherInconclusiveCount}
                       & \precnum{\JarFullInterpolationModelCheckingPredicateAbstractionErrorOtherInconclusiveCount}
                       & \precnum{\JarFullInterpolationModelCheckingImpactErrorOtherInconclusiveCount}                                                                                     \\
    \bottomrule
  \end{tabular}
\end{table*}

\subsection{Results}

\inlineheadingbf{RQ1: Effectiveness of IMC}
The experimental results of all compared approaches are summarized in~\cref{tab:results}.
Observe that IMC produced the most correct results for both proofs and alarms
among the interpolation-based approaches (IMC, PDR, predicate abstraction, and \impact)
and was second only to \kinduction in the evaluation.
In comparison to the most-related approach \impact,
IMC proved the safety of~\nimcvsimpactproofs~more programs
and found~\nimcvsimpactalarms~more bugs (an increase of 21\,\% and 3\,\%, respectively).
We will study the underlying mechanism that enables IMC to deliver more proofs than \impact in RQ4.
Meanwhile, BMC generated the most correct alarms as expected,
and \kinduction correctly solved the most tasks, with the most correct proofs and the second-most correct alarms.
Moreover, although IMC is a new addition to \cpachecker,
it did not produce any wrong proof in the evaluation,
identical to the other long-established approaches in the software-verification framework.
We consider the~\nimcwrongalarms~wrong alarms of IMC not caused by our implementation.
They are related to the program encoding of~\cpachecker,
and other approaches, such as predicate abstraction, also failed to solve these tasks correctly.

\inlineheadingbf{RQ2: Efficiency of IMC}
To study the run-time efficiency of IMC,
we present the quantile plots for the compared approaches in~\cref{fig:evaluation:quantile-cputime}
and the scatter plots for CPU time spent on correctly solved tasks in~\cref{fig:evaluation:scatter-cputime}.
The quantile plots for the correct proofs and alarms of the compared approaches are shown in~\cref{fig:evaluation:quantile-true} and~\cref{fig:evaluation:quantile-false}, respectively.
A data point $(x,y)$ in the plots indicates that there are $x$ tasks correctly solved by the respective algorithm within a CPU time of $y$~seconds each.
Note that IMC is not only effective in producing proofs and alarms but is also efficient.
From~\cref{fig:evaluation:quantile-cputime},
we see that it is the most efficient and effective interpolation-based approach in the evaluation.

\input{\plotpath/plot-defs}
\newsavebox{\quantileTrue}
\begin{lrbox}{\quantileTrue}
  \begin{minipage}[b]{0.5\textwidth}
    \scalebox{0.85}{\input{\plotpath/quantile-true}}
  \end{minipage}
\end{lrbox}

\newsavebox{\quantileFalse}
\begin{lrbox}{\quantileFalse}
  \begin{minipage}[b]{0.5\textwidth}
    \scalebox{0.85}{\input{\plotpath/quantile-false}}
  \end{minipage}
\end{lrbox}

\begin{figure*}[t]
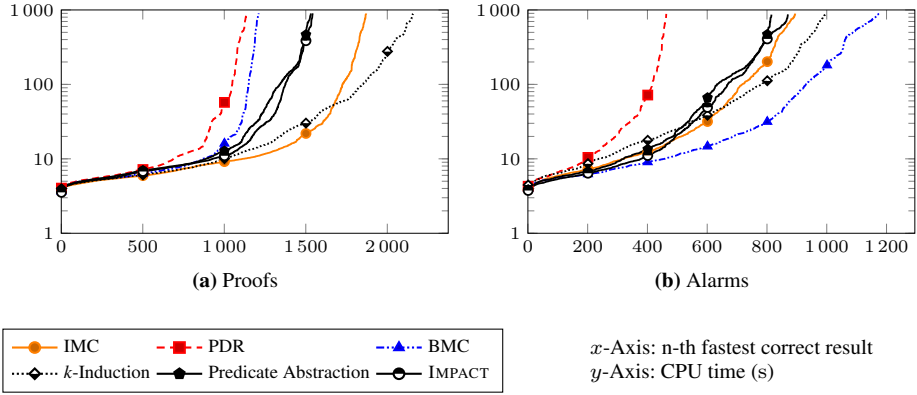

  \centering
  \subfloat[Proofs]{\usebox{\quantileTrue}\label{fig:evaluation:quantile-true}}
  \subfloat[Alarms]{\usebox{\quantileFalse}\label{fig:evaluation:quantile-false}}\\
  \bigskip
  \begin{minipage}[c]{.55\linewidth}
    \pgfplotslegendfromname{legend:quantile-true}
  \end{minipage}
  \hfill
  \begin{minipage}[c]{.37\linewidth}
    $x$-Axis: n-th fastest correct result\\
    $y$-Axis: CPU time (\second)
  \end{minipage}
  \caption{Quantile plots for all correct proofs and alarms}
  \label{fig:evaluation:quantile-cputime}
\end{figure*}

\newsavebox{\scatterPdr}
\begin{lrbox}{\scatterPdr}
  \begin{minipage}[b]{0.5\textwidth}
    \scalebox{0.8}{\input{\plotpath/scatter-pdr}}
  \end{minipage}
\end{lrbox}

\newsavebox{\scatterKidf}
\begin{lrbox}{\scatterKidf}
  \begin{minipage}[b]{0.5\textwidth}
    \scalebox{0.8}{\input{\plotpath/scatter-ki-df}}
  \end{minipage}
\end{lrbox}

\newsavebox{\scatterPredicate}
\begin{lrbox}{\scatterPredicate}
  \begin{minipage}[b]{0.5\textwidth}
    \scalebox{0.8}{\input{\plotpath/scatter-predicate}}
  \end{minipage}
\end{lrbox}

\newsavebox{\scatterImpact}
\begin{lrbox}{\scatterImpact}
  \begin{minipage}[b]{0.5\textwidth}
    \scalebox{0.8}{\input{\plotpath/scatter-predicate-impact}}
  \end{minipage}
\end{lrbox}

\begin{figure*}[t]
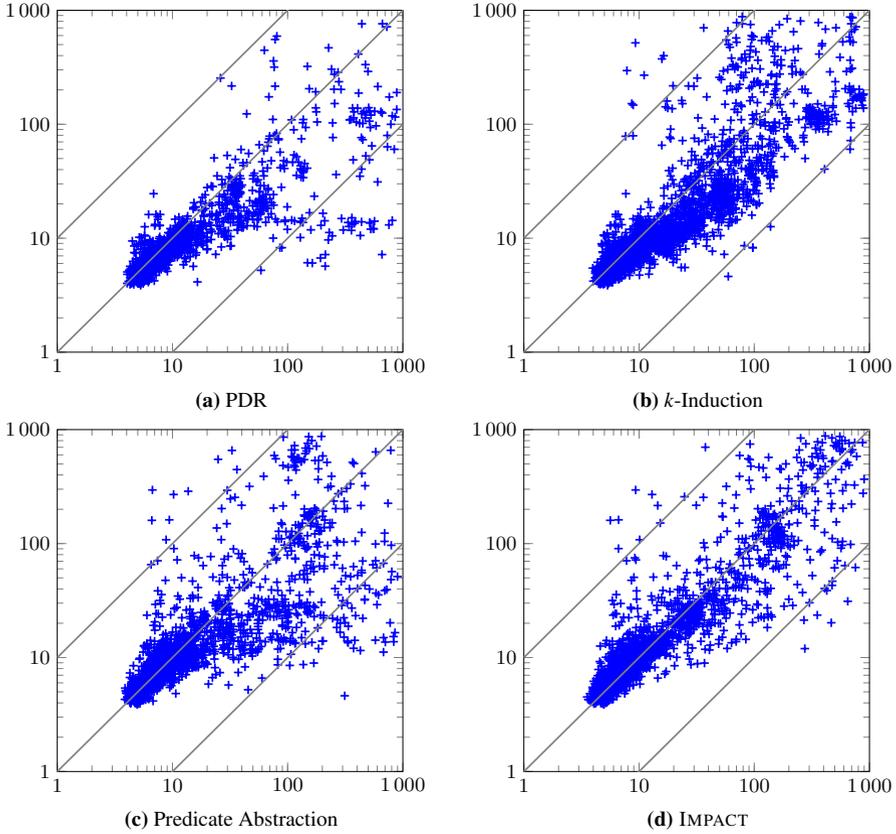

  \centering
  \subfloat[PDR]{\usebox{\scatterPdr}\label{fig:evaluation:scatter-pdr}}
  \subfloat[\kInduction]{\usebox{\scatterKidf}\label{fig:evaluation:scatter-ki}}\\
  \subfloat[Predicate Abstraction]{\usebox{\scatterPredicate}\label{fig:evaluation:scatter-predicate}}
  \subfloat[\impact]{\usebox{\scatterImpact}\label{fig:evaluation:scatter-impact}}
  \caption{Scatter plots of CPU time in seconds for all correct results with IMC in $y$-axis and compared approaches in $x$-axis}
  \label{fig:evaluation:scatter-cputime}
\end{figure*}

The scatter plots for the correctly solved tasks (including both proofs and alarms) of the compared approaches are shown in~\cref{fig:evaluation:scatter-cputime}.
We omitted the scatter plot for BMC as it is mainly inclined to bug hunting, while other approaches have more balanced behavior.
A data point $(x,y)$ in the plots indicates that there is a task correctly solved by both IMC and a compared approach,
while IMC took a CPU time of $y$~seconds and the other approach took a CPU time of $x$~seconds.
Observe that IMC is often more efficient than a compared approach.
For example, while it solved fewer tasks compared to \kinduction, its time efficiency is often better than \kinduction on the tasks which can be solved by both algorithms.
This phenomenon could be explained by the fact that, unlike \kinduction,
which relies on an external procedure to generate auxiliary invariants,
IMC generates interpolants from BMC queries and uses them to construct fixed points purely internally.
Moreover, when representing \textit{helpful} loop invariants,
i.e., those that help to prove the safety property of a program,
requires complex formulas,
the interval-based data-flow analysis~\cite{CPA-DF}
used by the default configuration~\cite{kInduction} of \kinduction in \cpachecker
is disadvantageous because the expressiveness of candidate invariants is limited.
By contrast, IMC is favorable in such cases
since it constructs a candidate fixed point (also a loop invariant) as a union of previously derived interpolants, which in principle can encode any combination of reachable states.

\inlineheadingbf{RQ3: Enhancing Software Verification with IMC}
To highlight IMC's contribution to software verification,
we report the numbers of tasks solvable by IMC but not by a compared approach
because it ran out of resources.
In our evaluation, IMC solved
\nimcvspdrtasks,
\nimcvsbmctasks,
\nimcvskitasks,
\nimcvspredtasks, and
\nimcvsimpacttasks~tasks for which
PDR,
BMC,
\kinduction,
predicate abstraction, and
\impact, respectively,
failed to solve within the time or memory limits.
Overall, it uniquely solved~\nimcuniquetasks~tasks for which all other approaches ran out of resources.

IMC performed best for the category \textit{ReachSafety-ECA}.
These event-condition-action (ECA)~\cite{RERS12} programs have a loop to receive external inputs,
generate outputs, and update internal variables based on the ECA rules,
implemented by a complex control flow inside the loop.
Conceptually, the working of these programs is similar to that of sequential Boolean-logic circuits.
IMC naturally performs well on them because it originated from hardware verification.
Out of a total~\necatasks~ECA programs,
IMC solved a second most~\nimcecatasks~tasks,
while predicate abstraction,
\impact,
and \kinduction solved
\npredecatasks,
\nimpactecatasks,
and~\nkiecatasks,
respectively.
We will use the ECA tasks without property violation (namely, safe ECA tasks)
to study the performance characteristics of IMC
and answer why it can deliver more correctness proofs than \impact,
a closely related interpolation-based approach.

\inlineheadingbf{RQ4: Performance Characteristics of IMC}
IMC is the best interpolation-based verification algorithm in our evaluation.
To profile its performance characteristics and understand why it can deliver more proofs,
we compared IMC and \impact on the \textit{ReachSafety-ECA} tasks without property violation.
Among the~\ntrueecatasks~safe ECA tasks,
IMC and \impact proved the correctness of~\nimcecaproofs~and~\nimpactecaproofs~of them, respectively.
The quantile plot in~\cref{fig:evaluation:quantile-imc-impact} shows that
IMC not only delivered more proofs than \impact but also spent less CPU time finding the proofs.
The scatter plot in~\cref{fig:evaluation:scatter-imc-impact-cpu} further demonstrates that
IMC usually obtained a proof faster than \impact when both methods succeeded.

\newsavebox{\quantileECACPU}
\begin{lrbox}{\quantileECACPU}
  \begin{minipage}[b]{0.5\textwidth}
    \scalebox{0.8}{\input{\plotpath/quantile-impact-ECA-true}}
  \end{minipage}
\end{lrbox}

\newsavebox{\scatterECACPU}
\begin{lrbox}{\scatterECACPU}
  \begin{minipage}[b]{0.5\textwidth}
    \scalebox{0.8}{\input{\plotpath/scatter-impact-ECA-cputime}}
  \end{minipage}
\end{lrbox}

\begin{figure*}[t]
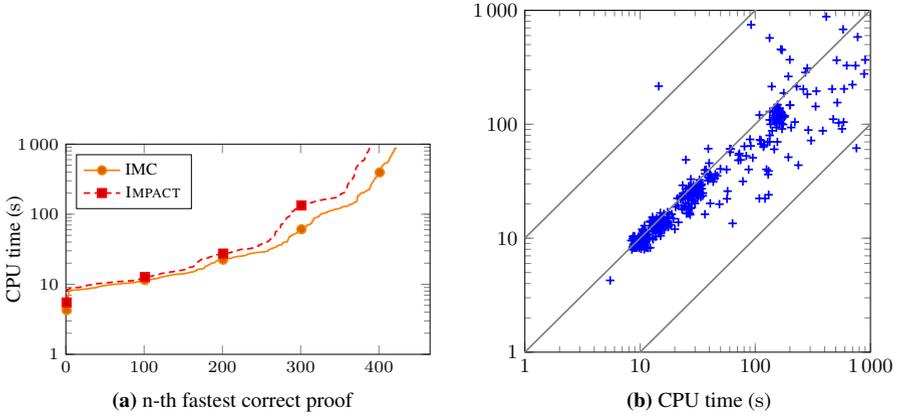

  \centering
  \subfloat[n-th fastest correct proof]{\usebox{\quantileECACPU}\label{fig:evaluation:quantile-imc-impact}}
  \subfloat[CPU time (\second)]{\usebox{\scatterECACPU}\label{fig:evaluation:scatter-imc-impact-cpu}}\\
  \caption{Comparing IMC and \impact on safe ECA tasks: (a) quantile plot for proofs and (b)~scatter plot for elapsed CPU time of proofs with IMC in $y$-axis and \impact in $x$-axis}
  \label{fig:evaluation:imc-impact-eca}
\end{figure*}

\newsavebox{\scatterECAITP}
\begin{lrbox}{\scatterECAITP}
  \begin{minipage}[b]{0.45\textwidth}
    \scalebox{0.7}{\input{\plotpath/scatter-impact-ECA-itptime}}
  \end{minipage}
\end{lrbox}

\newsavebox{\scatterECAITPCall}
\begin{lrbox}{\scatterECAITPCall}
  \begin{minipage}[b]{0.45\textwidth}
    \scalebox{0.5}{\input{\plotpath/scatter-impact-ECA-itpcall}}
  \end{minipage}
\end{lrbox}

\begin{figure}[t]
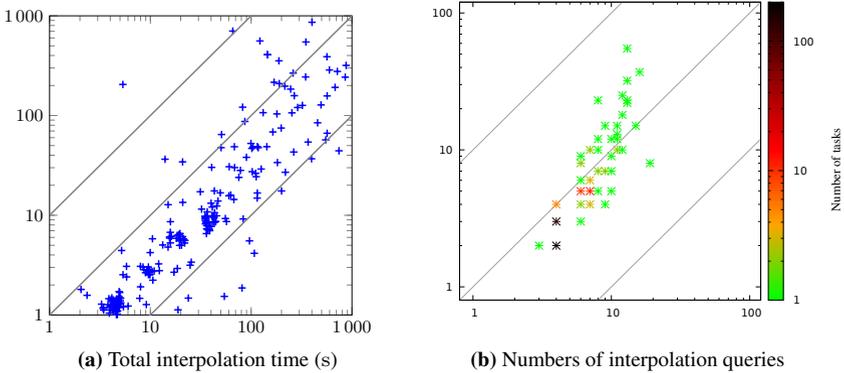

  \centering
  \subfloat[Total interpolation time (\second)]{\usebox{\scatterECAITP}\label{fig:evaluation:scatter-imc-impact-itp-time}}
  \subfloat[Numbers of interpolation queries]{\usebox{\scatterECAITPCall}\label{fig:evaluation:scatter-imc-impact-itp-call}}
  \caption{Scatter plots of (a) total interpolation time and (b) numbers of interpolation queries used to prove safe ECA tasks with IMC in $y$-axis and \impact in $x$-axis}
  \label{fig:evaluation:imc-impact-eca-itp}
\end{figure}

To understand the advantage of IMC over \impact for finding proofs,
we investigate the most essential and time-consuming step in their computation, namely, interpolation.
The total interpolation time and numbers of interpolation queries used by IMC and \impact
to produce proofs are compared in the scatter plots~\cref{fig:evaluation:scatter-imc-impact-itp-time}
and~\cref{fig:evaluation:scatter-imc-impact-itp-call}, respectively.
In~\cref{fig:evaluation:scatter-imc-impact-itp-call},
the color of a data point indicates the number of tasks falling into this coordinate.
From the scatter plots,
observe that IMC usually required fewer interpolation queries
and less interpolation time to prove a task.
Among the~\nimcimpactbothsolved~safe ECA tasks proved by both IMC and \impact,
IMC invoked fewer interpolation calls for~\nimcfeweritpcalls~of them.
This phenomenon indicates that the quality of interpolants derived by IMC were high,
which enabled it to generalize better than \impact on these tasks.

We attribute the quality of interpolants to two factors.
First, IMC is known to generalize better than approaches
based on interpolation sequences in hardware verification~\cite{CabodiDATE2011}.
Unlike algorithms based on interpolation sequences~\cite{IMPACT,VizelFMCAD09},
IMC derives interpolants from not only the initial states but also the previous interpolants.
Such eager abstraction decreases the numbers of interpolation queries required to reach a fixed point.
We observed the same effect for software verification,
as exhibited in~\cref{fig:evaluation:scatter-imc-impact-itp-call}.
Second, the proposed adoption of IMC analyzes the control-flow structures separately
and only encodes syntactically feasible program paths
without using symbolic program counters in the formulas.
Therefore, the underlying SMT solver can focus on the semantics of the program
and derive useful interpolants about the actual program variables.
The proposed adoption is crucial for unlocking the potential of IMC for software verification.

\inlineheadingbf{Answers to the Research Questions}
For the first part of our evaluation
where IMC was compared against five SMT-based verification approaches,
the proposed approach with large-block encoding is effective and efficient.
Adopted with the proposed method, IMC,
the first interpolation-based formal-verification approach ever invented,
is competitive against other state-of-the-art algorithms,
which have been investigated much more by the research community.
The conclusion is well supported by the experimental results:
Our IMC implementation not only solved the second most verification tasks (\cref{tab:results}) in the evaluation
but was also efficient compared to other SMT-based approaches (\cref{fig:evaluation:scatter-cputime}).
In our experiments, it was the most efficient and effective interpolation-based approach (\cref{fig:evaluation:quantile-cputime}).
It uniquely solved~\nimcuniquetasks~tasks for which all other approaches ran out of resources,
indicating its unique value to complement existing approaches.

In the second part of the evaluation,
IMC was compared to \impact on a subset of the SV-COMP\,'22 benchmark set
to study its strength to find proofs.
We observed that IMC spent less effort on interpolation than \impact (\cref{fig:evaluation:imc-impact-eca-itp}),
indicating that it derives high-quality interpolants and generalizes better.
The reason behind this phenomenon is that IMC eagerly computes interpolants
from not only the initial states but also the previous interpolants.
The same effect is also reported for hardware verification~\cite{CabodiDATE2011}.

\subsection{Threats to Validity\label{sec:threats}}
Here we discuss some threats that may affect the validity of our conclusions and how we limited them.
To ensure internal validity, all the compared algorithms are implemented in
the verification framework~\cpachecker~\cite{CPACHECKER}.
This practice minimizes the confounding variables (front ends and utilities) and rules out differences unrelated to the algorithms.
We also use \benchexec~\cite{Benchmarking-STTT} to ensure best possible measurement accuracy.
To reduce the external threat resulting from the selection bias of verification tasks,
we conduct the experiments using the largest publicly available benchmark set~\cite{SVCOMP22-SVBENCHMARKS-artifact}
of C safety-verification tasks.
Other external threats arise from the selection of the compared approaches and underlying framework.
It is clear from the literature~\cite{AlgorithmComparison-JAR} that the compared approaches in this paper indeed represent the state of the art of software verification;
the only missing main related state-of-the-art approach is trace abstraction~\cite{UAUTOMIZER2013},
for which the implementation in the framework is not yet mature enough.
Moreover, the chosen platform~\cpachecker is a well-maintained software project that performs well in the competitions, and the relative performance between~\cpachecker and other verifiers is available from SV-COMP\,'22~\cite{SVCOMP22}.


%% file: evaluation/tex/data-commands.tex
\providecommand\StoreBenchExecResult[7]{\expandafter\newcommand\csname#1#2#3#4#5#6\endcsname{#7}}%
\StoreBenchExecResult{JarFullInterpolationModelChecking}{Imc}{Total}{}{Count}{}{6024}%
\StoreBenchExecResult{JarFullInterpolationModelChecking}{Imc}{Total}{}{Cputime}{}{2164111.498525171}%
\StoreBenchExecResult{JarFullInterpolationModelChecking}{Imc}{Total}{}{Cputime}{Avg}{359.2482567272860225763612218}%
\StoreBenchExecResult{JarFullInterpolationModelChecking}{Imc}{Total}{}{Cputime}{Median}{45.158349052}%
\StoreBenchExecResult{JarFullInterpolationModelChecking}{Imc}{Total}{}{Cputime}{Min}{3.862362769}%
\StoreBenchExecResult{JarFullInterpolationModelChecking}{Imc}{Total}{}{Cputime}{Max}{962.827682563}%
\StoreBenchExecResult{JarFullInterpolationModelChecking}{Imc}{Total}{}{Cputime}{Stdev}{414.7410892215723515536387162}%
\StoreBenchExecResult{JarFullInterpolationModelChecking}{Imc}{Total}{}{Walltime}{}{2053613.2799366971037267}%
\StoreBenchExecResult{JarFullInterpolationModelChecking}{Imc}{Total}{}{Walltime}{Avg}{340.9052589536349773782702523}%
\StoreBenchExecResult{JarFullInterpolationModelChecking}{Imc}{Total}{}{Walltime}{Median}{35.7974436000222325}%
\StoreBenchExecResult{JarFullInterpolationModelChecking}{Imc}{Total}{}{Walltime}{Min}{2.083037902833894}%
\StoreBenchExecResult{JarFullInterpolationModelChecking}{Imc}{Total}{}{Walltime}{Max}{907.6059522270225}%
\StoreBenchExecResult{JarFullInterpolationModelChecking}{Imc}{Total}{}{Walltime}{Stdev}{401.4695051983782053573738234}%
\StoreBenchExecResult{JarFullInterpolationModelChecking}{Imc}{Correct}{}{Count}{}{2766}%
\StoreBenchExecResult{JarFullInterpolationModelChecking}{Imc}{Correct}{}{Cputime}{}{140646.773385412}%
\StoreBenchExecResult{JarFullInterpolationModelChecking}{Imc}{Correct}{}{Cputime}{Avg}{50.84843578648300795372378886}%
\StoreBenchExecResult{JarFullInterpolationModelChecking}{Imc}{Correct}{}{Cputime}{Median}{9.6684238495}%
\StoreBenchExecResult{JarFullInterpolationModelChecking}{Imc}{Correct}{}{Cputime}{Min}{3.862362769}%
\StoreBenchExecResult{JarFullInterpolationModelChecking}{Imc}{Correct}{}{Cputime}{Max}{897.711205451}%
\StoreBenchExecResult{JarFullInterpolationModelChecking}{Imc}{Correct}{}{Cputime}{Stdev}{126.4544511599228828775552545}%
\StoreBenchExecResult{JarFullInterpolationModelChecking}{Imc}{Correct}{}{Walltime}{}{123518.9649852856526513}%
\StoreBenchExecResult{JarFullInterpolationModelChecking}{Imc}{Correct}{}{Walltime}{Avg}{44.65616955361014195636297903}%
\StoreBenchExecResult{JarFullInterpolationModelChecking}{Imc}{Correct}{}{Walltime}{Median}{5.441057613003068}%
\StoreBenchExecResult{JarFullInterpolationModelChecking}{Imc}{Correct}{}{Walltime}{Min}{2.083037902833894}%
\StoreBenchExecResult{JarFullInterpolationModelChecking}{Imc}{Correct}{}{Walltime}{Max}{889.3461187209468}%
\StoreBenchExecResult{JarFullInterpolationModelChecking}{Imc}{Correct}{}{Walltime}{Stdev}{122.3568985928529628654203527}%
\StoreBenchExecResult{JarFullInterpolationModelChecking}{Imc}{Correct}{False}{Count}{}{895}%
\StoreBenchExecResult{JarFullInterpolationModelChecking}{Imc}{Correct}{False}{Cputime}{}{72701.260159966}%
\StoreBenchExecResult{JarFullInterpolationModelChecking}{Imc}{Correct}{False}{Cputime}{Avg}{81.23045827929162011173184358}%
\StoreBenchExecResult{JarFullInterpolationModelChecking}{Imc}{Correct}{False}{Cputime}{Median}{14.568396163}%
\StoreBenchExecResult{JarFullInterpolationModelChecking}{Imc}{Correct}{False}{Cputime}{Min}{3.862362769}%
\StoreBenchExecResult{JarFullInterpolationModelChecking}{Imc}{Correct}{False}{Cputime}{Max}{894.106196777}%
\StoreBenchExecResult{JarFullInterpolationModelChecking}{Imc}{Correct}{False}{Cputime}{Stdev}{161.5020378375417056738578009}%
\StoreBenchExecResult{JarFullInterpolationModelChecking}{Imc}{Correct}{False}{Walltime}{}{65718.3142509493044171}%
\StoreBenchExecResult{JarFullInterpolationModelChecking}{Imc}{Correct}{False}{Walltime}{Avg}{73.42828407927296582916201117}%
\StoreBenchExecResult{JarFullInterpolationModelChecking}{Imc}{Correct}{False}{Walltime}{Median}{8.829153769882396}%
\StoreBenchExecResult{JarFullInterpolationModelChecking}{Imc}{Correct}{False}{Walltime}{Min}{2.083037902833894}%
\StoreBenchExecResult{JarFullInterpolationModelChecking}{Imc}{Correct}{False}{Walltime}{Max}{839.310741835041}%
\StoreBenchExecResult{JarFullInterpolationModelChecking}{Imc}{Correct}{False}{Walltime}{Stdev}{155.3467982246166523223214598}%
\StoreBenchExecResult{JarFullInterpolationModelChecking}{Imc}{Correct}{True}{Count}{}{1871}%
\StoreBenchExecResult{JarFullInterpolationModelChecking}{Imc}{Correct}{True}{Cputime}{}{67945.513225446}%
\StoreBenchExecResult{JarFullInterpolationModelChecking}{Imc}{Correct}{True}{Cputime}{Avg}{36.31507922257936932121859968}%
\StoreBenchExecResult{JarFullInterpolationModelChecking}{Imc}{Correct}{True}{Cputime}{Median}{8.681276629}%
\StoreBenchExecResult{JarFullInterpolationModelChecking}{Imc}{Correct}{True}{Cputime}{Min}{3.933354622}%
\StoreBenchExecResult{JarFullInterpolationModelChecking}{Imc}{Correct}{True}{Cputime}{Max}{897.711205451}%
\StoreBenchExecResult{JarFullInterpolationModelChecking}{Imc}{Correct}{True}{Cputime}{Stdev}{102.5198718561547877811856671}%
\StoreBenchExecResult{JarFullInterpolationModelChecking}{Imc}{Correct}{True}{Walltime}{}{57800.6507343363482342}%
\StoreBenchExecResult{JarFullInterpolationModelChecking}{Imc}{Correct}{True}{Walltime}{Avg}{30.89291861803118558749331908}%
\StoreBenchExecResult{JarFullInterpolationModelChecking}{Imc}{Correct}{True}{Walltime}{Median}{4.7979150449391454}%
\StoreBenchExecResult{JarFullInterpolationModelChecking}{Imc}{Correct}{True}{Walltime}{Min}{2.101005893899128}%
\StoreBenchExecResult{JarFullInterpolationModelChecking}{Imc}{Correct}{True}{Walltime}{Max}{889.3461187209468}%
\StoreBenchExecResult{JarFullInterpolationModelChecking}{Imc}{Correct}{True}{Walltime}{Stdev}{100.0169328252263886025041361}%

\StoreBenchExecResult{JarFullInterpolationModelChecking}{Imc}{Error}{}{Count}{}{3255}%
\StoreBenchExecResult{JarFullInterpolationModelChecking}{Imc}{Error}{}{Cputime}{}{2023446.643199279}%
\StoreBenchExecResult{JarFullInterpolationModelChecking}{Imc}{Error}{}{Cputime}{Avg}{621.6425939168291858678955453}%
\StoreBenchExecResult{JarFullInterpolationModelChecking}{Imc}{Error}{}{Cputime}{Median}{900.573159294}%
\StoreBenchExecResult{JarFullInterpolationModelChecking}{Imc}{Error}{}{Cputime}{Min}{4.037974454}%
\StoreBenchExecResult{JarFullInterpolationModelChecking}{Imc}{Error}{}{Cputime}{Max}{962.827682563}%
\StoreBenchExecResult{JarFullInterpolationModelChecking}{Imc}{Error}{}{Cputime}{Stdev}{393.6515409436106166387836432}%
\StoreBenchExecResult{JarFullInterpolationModelChecking}{Imc}{Error}{}{Walltime}{}{1930084.2107928134527544}%
\StoreBenchExecResult{JarFullInterpolationModelChecking}{Imc}{Error}{}{Walltime}{Avg}{592.9598189839672665912135177}%
\StoreBenchExecResult{JarFullInterpolationModelChecking}{Imc}{Error}{}{Walltime}{Median}{865.6697617350146}%
\StoreBenchExecResult{JarFullInterpolationModelChecking}{Imc}{Error}{}{Walltime}{Min}{2.1457877659704536}%
\StoreBenchExecResult{JarFullInterpolationModelChecking}{Imc}{Error}{}{Walltime}{Max}{907.6059522270225}%
\StoreBenchExecResult{JarFullInterpolationModelChecking}{Imc}{Error}{}{Walltime}{Stdev}{383.8657528468688560233588115}%
\StoreBenchExecResult{JarFullInterpolationModelChecking}{Imc}{Error}{Error}{Count}{}{1036}%
\StoreBenchExecResult{JarFullInterpolationModelChecking}{Imc}{Error}{Error}{Cputime}{}{84866.015684697}%
\StoreBenchExecResult{JarFullInterpolationModelChecking}{Imc}{Error}{Error}{Cputime}{Avg}{81.91700355665733590733590734}%
\StoreBenchExecResult{JarFullInterpolationModelChecking}{Imc}{Error}{Error}{Cputime}{Median}{18.9952293365}%
\StoreBenchExecResult{JarFullInterpolationModelChecking}{Imc}{Error}{Error}{Cputime}{Min}{4.037974454}%
\StoreBenchExecResult{JarFullInterpolationModelChecking}{Imc}{Error}{Error}{Cputime}{Max}{869.243807084}%
\StoreBenchExecResult{JarFullInterpolationModelChecking}{Imc}{Error}{Error}{Cputime}{Stdev}{147.7175850456655244389747609}%
\StoreBenchExecResult{JarFullInterpolationModelChecking}{Imc}{Error}{Error}{Walltime}{}{74155.9133022993333374}%
\StoreBenchExecResult{JarFullInterpolationModelChecking}{Imc}{Error}{Error}{Walltime}{Avg}{71.57906689411132561525096525}%
\StoreBenchExecResult{JarFullInterpolationModelChecking}{Imc}{Error}{Error}{Walltime}{Median}{13.2267677170457315}%
\StoreBenchExecResult{JarFullInterpolationModelChecking}{Imc}{Error}{Error}{Walltime}{Min}{2.1457877659704536}%
\StoreBenchExecResult{JarFullInterpolationModelChecking}{Imc}{Error}{Error}{Walltime}{Max}{849.5166652218904}%
\StoreBenchExecResult{JarFullInterpolationModelChecking}{Imc}{Error}{Error}{Walltime}{Stdev}{137.9995789536295916553698746}%
\StoreBenchExecResult{JarFullInterpolationModelChecking}{Imc}{Error}{OutOfJavaMemory}{Count}{}{3}%
\StoreBenchExecResult{JarFullInterpolationModelChecking}{Imc}{Error}{OutOfJavaMemory}{Cputime}{}{188.626247028}%
\StoreBenchExecResult{JarFullInterpolationModelChecking}{Imc}{Error}{OutOfJavaMemory}{Cputime}{Avg}{62.875415676}%
\StoreBenchExecResult{JarFullInterpolationModelChecking}{Imc}{Error}{OutOfJavaMemory}{Cputime}{Median}{62.406496604}%
\StoreBenchExecResult{JarFullInterpolationModelChecking}{Imc}{Error}{OutOfJavaMemory}{Cputime}{Min}{61.922917769}%
\StoreBenchExecResult{JarFullInterpolationModelChecking}{Imc}{Error}{OutOfJavaMemory}{Cputime}{Max}{64.296832655}%
\StoreBenchExecResult{JarFullInterpolationModelChecking}{Imc}{Error}{OutOfJavaMemory}{Cputime}{Stdev}{1.024298717353196121827834184}%
\StoreBenchExecResult{JarFullInterpolationModelChecking}{Imc}{Error}{OutOfJavaMemory}{Walltime}{}{103.163706794613973}%
\StoreBenchExecResult{JarFullInterpolationModelChecking}{Imc}{Error}{OutOfJavaMemory}{Walltime}{Avg}{34.38790226487132433333333333}%
\StoreBenchExecResult{JarFullInterpolationModelChecking}{Imc}{Error}{OutOfJavaMemory}{Walltime}{Median}{34.135868118843064}%
\StoreBenchExecResult{JarFullInterpolationModelChecking}{Imc}{Error}{OutOfJavaMemory}{Walltime}{Min}{33.736304262885824}%
\StoreBenchExecResult{JarFullInterpolationModelChecking}{Imc}{Error}{OutOfJavaMemory}{Walltime}{Max}{35.291534412885085}%
\StoreBenchExecResult{JarFullInterpolationModelChecking}{Imc}{Error}{OutOfJavaMemory}{Walltime}{Stdev}{0.6594574097644349556215469049}%
\StoreBenchExecResult{JarFullInterpolationModelChecking}{Imc}{Error}{OutOfMemory}{Count}{}{160}%
\StoreBenchExecResult{JarFullInterpolationModelChecking}{Imc}{Error}{OutOfMemory}{Cputime}{}{106128.930224972}%
\StoreBenchExecResult{JarFullInterpolationModelChecking}{Imc}{Error}{OutOfMemory}{Cputime}{Avg}{663.305813906075}%
\StoreBenchExecResult{JarFullInterpolationModelChecking}{Imc}{Error}{OutOfMemory}{Cputime}{Median}{699.4391186745}%
\StoreBenchExecResult{JarFullInterpolationModelChecking}{Imc}{Error}{OutOfMemory}{Cputime}{Min}{105.414431958}%
\StoreBenchExecResult{JarFullInterpolationModelChecking}{Imc}{Error}{OutOfMemory}{Cputime}{Max}{897.380499878}%
\StoreBenchExecResult{JarFullInterpolationModelChecking}{Imc}{Error}{OutOfMemory}{Cputime}{Stdev}{170.1878309900762616123309270}%
\StoreBenchExecResult{JarFullInterpolationModelChecking}{Imc}{Error}{OutOfMemory}{Walltime}{}{97159.39285320462653}%
\StoreBenchExecResult{JarFullInterpolationModelChecking}{Imc}{Error}{OutOfMemory}{Walltime}{Avg}{607.2462053325289158125}%
\StoreBenchExecResult{JarFullInterpolationModelChecking}{Imc}{Error}{OutOfMemory}{Walltime}{Median}{640.4313238725299}%
\StoreBenchExecResult{JarFullInterpolationModelChecking}{Imc}{Error}{OutOfMemory}{Walltime}{Min}{88.26022180705331}%
\StoreBenchExecResult{JarFullInterpolationModelChecking}{Imc}{Error}{OutOfMemory}{Walltime}{Max}{853.1459841749165}%
\StoreBenchExecResult{JarFullInterpolationModelChecking}{Imc}{Error}{OutOfMemory}{Walltime}{Stdev}{166.2169754118184021011780019}%
\StoreBenchExecResult{JarFullInterpolationModelChecking}{Imc}{Error}{OutOfNativeMemory}{Count}{}{2}%
\StoreBenchExecResult{JarFullInterpolationModelChecking}{Imc}{Error}{OutOfNativeMemory}{Cputime}{}{49.802667447}%
\StoreBenchExecResult{JarFullInterpolationModelChecking}{Imc}{Error}{OutOfNativeMemory}{Cputime}{Avg}{24.9013337235}%
\StoreBenchExecResult{JarFullInterpolationModelChecking}{Imc}{Error}{OutOfNativeMemory}{Cputime}{Median}{24.9013337235}%
\StoreBenchExecResult{JarFullInterpolationModelChecking}{Imc}{Error}{OutOfNativeMemory}{Cputime}{Min}{21.903921345}%
\StoreBenchExecResult{JarFullInterpolationModelChecking}{Imc}{Error}{OutOfNativeMemory}{Cputime}{Max}{27.898746102}%
\StoreBenchExecResult{JarFullInterpolationModelChecking}{Imc}{Error}{OutOfNativeMemory}{Cputime}{Stdev}{2.9974123785000}%
\StoreBenchExecResult{JarFullInterpolationModelChecking}{Imc}{Error}{OutOfNativeMemory}{Walltime}{}{39.372262360993773}%
\StoreBenchExecResult{JarFullInterpolationModelChecking}{Imc}{Error}{OutOfNativeMemory}{Walltime}{Avg}{19.6861311804968865}%
\StoreBenchExecResult{JarFullInterpolationModelChecking}{Imc}{Error}{OutOfNativeMemory}{Walltime}{Median}{19.6861311804968865}%
\StoreBenchExecResult{JarFullInterpolationModelChecking}{Imc}{Error}{OutOfNativeMemory}{Walltime}{Min}{16.614605971146375}%
\StoreBenchExecResult{JarFullInterpolationModelChecking}{Imc}{Error}{OutOfNativeMemory}{Walltime}{Max}{22.757656389847398}%
\StoreBenchExecResult{JarFullInterpolationModelChecking}{Imc}{Error}{OutOfNativeMemory}{Walltime}{Stdev}{3.071525209350511500000000000}%
\StoreBenchExecResult{JarFullInterpolationModelChecking}{Imc}{Error}{SegmentationFault}{Count}{}{36}%
\StoreBenchExecResult{JarFullInterpolationModelChecking}{Imc}{Error}{SegmentationFault}{Cputime}{}{4559.365654214}%
\StoreBenchExecResult{JarFullInterpolationModelChecking}{Imc}{Error}{SegmentationFault}{Cputime}{Avg}{126.6490459503888888888888889}%
\StoreBenchExecResult{JarFullInterpolationModelChecking}{Imc}{Error}{SegmentationFault}{Cputime}{Median}{47.399956632}%
\StoreBenchExecResult{JarFullInterpolationModelChecking}{Imc}{Error}{SegmentationFault}{Cputime}{Min}{5.252537549}%
\StoreBenchExecResult{JarFullInterpolationModelChecking}{Imc}{Error}{SegmentationFault}{Cputime}{Max}{856.910329647}%
\StoreBenchExecResult{JarFullInterpolationModelChecking}{Imc}{Error}{SegmentationFault}{Cputime}{Stdev}{206.9228366341963332562178280}%
\StoreBenchExecResult{JarFullInterpolationModelChecking}{Imc}{Error}{SegmentationFault}{Walltime}{}{4408.1763407108374110}%
\StoreBenchExecResult{JarFullInterpolationModelChecking}{Imc}{Error}{SegmentationFault}{Walltime}{Avg}{122.4493427975232614166666667}%
\StoreBenchExecResult{JarFullInterpolationModelChecking}{Imc}{Error}{SegmentationFault}{Walltime}{Median}{43.2389636384323225}%
\StoreBenchExecResult{JarFullInterpolationModelChecking}{Imc}{Error}{SegmentationFault}{Walltime}{Min}{3.1299324929714203}%
\StoreBenchExecResult{JarFullInterpolationModelChecking}{Imc}{Error}{SegmentationFault}{Walltime}{Max}{852.0255157009233}%
\StoreBenchExecResult{JarFullInterpolationModelChecking}{Imc}{Error}{SegmentationFault}{Walltime}{Stdev}{205.6132445949046716906691706}%
\StoreBenchExecResult{JarFullInterpolationModelChecking}{Imc}{Error}{Timeout}{Count}{}{2018}%
\StoreBenchExecResult{JarFullInterpolationModelChecking}{Imc}{Error}{Timeout}{Cputime}{}{1827653.902720921}%
\StoreBenchExecResult{JarFullInterpolationModelChecking}{Imc}{Error}{Timeout}{Cputime}{Avg}{905.6758685435683845391476710}%
\StoreBenchExecResult{JarFullInterpolationModelChecking}{Imc}{Error}{Timeout}{Cputime}{Median}{901.3602202925}%
\StoreBenchExecResult{JarFullInterpolationModelChecking}{Imc}{Error}{Timeout}{Cputime}{Min}{900.07670757}%
\StoreBenchExecResult{JarFullInterpolationModelChecking}{Imc}{Error}{Timeout}{Cputime}{Max}{962.827682563}%
\StoreBenchExecResult{JarFullInterpolationModelChecking}{Imc}{Error}{Timeout}{Cputime}{Stdev}{11.74252927826565069840340701}%
\StoreBenchExecResult{JarFullInterpolationModelChecking}{Imc}{Error}{Timeout}{Walltime}{}{1754218.19232744304773}%
\StoreBenchExecResult{JarFullInterpolationModelChecking}{Imc}{Error}{Timeout}{Walltime}{Avg}{869.2855264258885271209117939}%
\StoreBenchExecResult{JarFullInterpolationModelChecking}{Imc}{Error}{Timeout}{Walltime}{Median}{892.02566822513475}%
\StoreBenchExecResult{JarFullInterpolationModelChecking}{Imc}{Error}{Timeout}{Walltime}{Min}{483.2500354589429}%
\StoreBenchExecResult{JarFullInterpolationModelChecking}{Imc}{Error}{Timeout}{Walltime}{Max}{907.6059522270225}%
\StoreBenchExecResult{JarFullInterpolationModelChecking}{Imc}{Error}{Timeout}{Walltime}{Stdev}{65.49315470927919870588901703}%
\StoreBenchExecResult{JarFullInterpolationModelChecking}{Imc}{Wrong}{}{Count}{}{3}%
\StoreBenchExecResult{JarFullInterpolationModelChecking}{Imc}{Wrong}{}{Cputime}{}{18.081940480}%
\StoreBenchExecResult{JarFullInterpolationModelChecking}{Imc}{Wrong}{}{Cputime}{Avg}{6.027313493333333333333333333}%
\StoreBenchExecResult{JarFullInterpolationModelChecking}{Imc}{Wrong}{}{Cputime}{Median}{5.610179935}%
\StoreBenchExecResult{JarFullInterpolationModelChecking}{Imc}{Wrong}{}{Cputime}{Min}{4.945310144}%
\StoreBenchExecResult{JarFullInterpolationModelChecking}{Imc}{Wrong}{}{Cputime}{Max}{7.526450401}%
\StoreBenchExecResult{JarFullInterpolationModelChecking}{Imc}{Wrong}{}{Cputime}{Stdev}{1.094249076061563421211317295}%
\StoreBenchExecResult{JarFullInterpolationModelChecking}{Imc}{Wrong}{}{Walltime}{}{10.104158597998321}%
\StoreBenchExecResult{JarFullInterpolationModelChecking}{Imc}{Wrong}{}{Walltime}{Avg}{3.368052865999440333333333333}%
\StoreBenchExecResult{JarFullInterpolationModelChecking}{Imc}{Wrong}{}{Walltime}{Median}{2.952091026119888}%
\StoreBenchExecResult{JarFullInterpolationModelChecking}{Imc}{Wrong}{}{Walltime}{Min}{2.659910501912236}%
\StoreBenchExecResult{JarFullInterpolationModelChecking}{Imc}{Wrong}{}{Walltime}{Max}{4.492157069966197}%
\StoreBenchExecResult{JarFullInterpolationModelChecking}{Imc}{Wrong}{}{Walltime}{Stdev}{0.8037620131661553523482915483}%
\StoreBenchExecResult{JarFullInterpolationModelChecking}{Imc}{Wrong}{False}{Count}{}{3}%
\StoreBenchExecResult{JarFullInterpolationModelChecking}{Imc}{Wrong}{False}{Cputime}{}{18.081940480}%
\StoreBenchExecResult{JarFullInterpolationModelChecking}{Imc}{Wrong}{False}{Cputime}{Avg}{6.027313493333333333333333333}%
\StoreBenchExecResult{JarFullInterpolationModelChecking}{Imc}{Wrong}{False}{Cputime}{Median}{5.610179935}%
\StoreBenchExecResult{JarFullInterpolationModelChecking}{Imc}{Wrong}{False}{Cputime}{Min}{4.945310144}%
\StoreBenchExecResult{JarFullInterpolationModelChecking}{Imc}{Wrong}{False}{Cputime}{Max}{7.526450401}%
\StoreBenchExecResult{JarFullInterpolationModelChecking}{Imc}{Wrong}{False}{Cputime}{Stdev}{1.094249076061563421211317295}%
\StoreBenchExecResult{JarFullInterpolationModelChecking}{Imc}{Wrong}{False}{Walltime}{}{10.104158597998321}%
\StoreBenchExecResult{JarFullInterpolationModelChecking}{Imc}{Wrong}{False}{Walltime}{Avg}{3.368052865999440333333333333}%
\StoreBenchExecResult{JarFullInterpolationModelChecking}{Imc}{Wrong}{False}{Walltime}{Median}{2.952091026119888}%
\StoreBenchExecResult{JarFullInterpolationModelChecking}{Imc}{Wrong}{False}{Walltime}{Min}{2.659910501912236}%
\StoreBenchExecResult{JarFullInterpolationModelChecking}{Imc}{Wrong}{False}{Walltime}{Max}{4.492157069966197}%
\StoreBenchExecResult{JarFullInterpolationModelChecking}{Imc}{Wrong}{False}{Walltime}{Stdev}{0.8037620131661553523482915483}%
\providecommand\StoreBenchExecResult[7]{\expandafter\newcommand\csname#1#2#3#4#5#6\endcsname{#7}}%
\StoreBenchExecResult{JarFullInterpolationModelChecking}{Pdr}{Total}{}{Count}{}{6024}%
\StoreBenchExecResult{JarFullInterpolationModelChecking}{Pdr}{Total}{}{Cputime}{}{3206173.105286605}%
\StoreBenchExecResult{JarFullInterpolationModelChecking}{Pdr}{Total}{}{Cputime}{Avg}{532.2332512095957835325365206}%
\StoreBenchExecResult{JarFullInterpolationModelChecking}{Pdr}{Total}{}{Cputime}{Median}{900.7356754305}%
\StoreBenchExecResult{JarFullInterpolationModelChecking}{Pdr}{Total}{}{Cputime}{Min}{4.043025449}%
\StoreBenchExecResult{JarFullInterpolationModelChecking}{Pdr}{Total}{}{Cputime}{Max}{962.086140631}%
\StoreBenchExecResult{JarFullInterpolationModelChecking}{Pdr}{Total}{}{Cputime}{Stdev}{427.7642787708063742679640873}%
\StoreBenchExecResult{JarFullInterpolationModelChecking}{Pdr}{Total}{}{Walltime}{}{3081394.9849556814871622}%
\StoreBenchExecResult{JarFullInterpolationModelChecking}{Pdr}{Total}{}{Walltime}{Avg}{511.5197518186722256245351926}%
\StoreBenchExecResult{JarFullInterpolationModelChecking}{Pdr}{Total}{}{Walltime}{Median}{857.5925252470188}%
\StoreBenchExecResult{JarFullInterpolationModelChecking}{Pdr}{Total}{}{Walltime}{Min}{2.122047150041908}%
\StoreBenchExecResult{JarFullInterpolationModelChecking}{Pdr}{Total}{}{Walltime}{Max}{906.6350679579191}%
\StoreBenchExecResult{JarFullInterpolationModelChecking}{Pdr}{Total}{}{Walltime}{Stdev}{420.0495688184942506360555863}%
\StoreBenchExecResult{JarFullInterpolationModelChecking}{Pdr}{Correct}{}{Count}{}{1599}%
\StoreBenchExecResult{JarFullInterpolationModelChecking}{Pdr}{Correct}{}{Cputime}{}{70873.610385422}%
\StoreBenchExecResult{JarFullInterpolationModelChecking}{Pdr}{Correct}{}{Cputime}{Avg}{44.32370880889430894308943089}%
\StoreBenchExecResult{JarFullInterpolationModelChecking}{Pdr}{Correct}{}{Cputime}{Median}{8.535056964}%
\StoreBenchExecResult{JarFullInterpolationModelChecking}{Pdr}{Correct}{}{Cputime}{Min}{4.043025449}%
\StoreBenchExecResult{JarFullInterpolationModelChecking}{Pdr}{Correct}{}{Cputime}{Max}{884.940913132}%
\StoreBenchExecResult{JarFullInterpolationModelChecking}{Pdr}{Correct}{}{Cputime}{Stdev}{115.7624231543679644328642147}%
\StoreBenchExecResult{JarFullInterpolationModelChecking}{Pdr}{Correct}{}{Walltime}{}{62131.6246359643990078}%
\StoreBenchExecResult{JarFullInterpolationModelChecking}{Pdr}{Correct}{}{Walltime}{Avg}{38.85655074169130644640400250}%
\StoreBenchExecResult{JarFullInterpolationModelChecking}{Pdr}{Correct}{}{Walltime}{Median}{4.549815858947113}%
\StoreBenchExecResult{JarFullInterpolationModelChecking}{Pdr}{Correct}{}{Walltime}{Min}{2.122047150041908}%
\StoreBenchExecResult{JarFullInterpolationModelChecking}{Pdr}{Correct}{}{Walltime}{Max}{882.0178654058836}%
\StoreBenchExecResult{JarFullInterpolationModelChecking}{Pdr}{Correct}{}{Walltime}{Stdev}{112.1593917932918285990520570}%
\StoreBenchExecResult{JarFullInterpolationModelChecking}{Pdr}{Correct}{False}{Count}{}{463}%
\StoreBenchExecResult{JarFullInterpolationModelChecking}{Pdr}{Correct}{False}{Cputime}{}{22462.744186960}%
\StoreBenchExecResult{JarFullInterpolationModelChecking}{Pdr}{Correct}{False}{Cputime}{Avg}{48.51564619213822894168466523}%
\StoreBenchExecResult{JarFullInterpolationModelChecking}{Pdr}{Correct}{False}{Cputime}{Median}{12.496315061}%
\StoreBenchExecResult{JarFullInterpolationModelChecking}{Pdr}{Correct}{False}{Cputime}{Min}{4.283366683}%
\StoreBenchExecResult{JarFullInterpolationModelChecking}{Pdr}{Correct}{False}{Cputime}{Max}{884.940913132}%
\StoreBenchExecResult{JarFullInterpolationModelChecking}{Pdr}{Correct}{False}{Cputime}{Stdev}{111.4884247530187342578222422}%
\StoreBenchExecResult{JarFullInterpolationModelChecking}{Pdr}{Correct}{False}{Walltime}{}{19962.1805010498499894}%
\StoreBenchExecResult{JarFullInterpolationModelChecking}{Pdr}{Correct}{False}{Walltime}{Avg}{43.11486069341220300086393089}%
\StoreBenchExecResult{JarFullInterpolationModelChecking}{Pdr}{Correct}{False}{Walltime}{Median}{6.701444719918072}%
\StoreBenchExecResult{JarFullInterpolationModelChecking}{Pdr}{Correct}{False}{Walltime}{Min}{2.2728986411821097}%
\StoreBenchExecResult{JarFullInterpolationModelChecking}{Pdr}{Correct}{False}{Walltime}{Max}{882.0178654058836}%
\StoreBenchExecResult{JarFullInterpolationModelChecking}{Pdr}{Correct}{False}{Walltime}{Stdev}{111.0252491115287290339671507}%
\StoreBenchExecResult{JarFullInterpolationModelChecking}{Pdr}{Correct}{True}{Count}{}{1136}%
\StoreBenchExecResult{JarFullInterpolationModelChecking}{Pdr}{Correct}{True}{Cputime}{}{48410.866198462}%
\StoreBenchExecResult{JarFullInterpolationModelChecking}{Pdr}{Correct}{True}{Cputime}{Avg}{42.61519911836443661971830986}%
\StoreBenchExecResult{JarFullInterpolationModelChecking}{Pdr}{Correct}{True}{Cputime}{Median}{7.7789547395}%
\StoreBenchExecResult{JarFullInterpolationModelChecking}{Pdr}{Correct}{True}{Cputime}{Min}{4.043025449}%
\StoreBenchExecResult{JarFullInterpolationModelChecking}{Pdr}{Correct}{True}{Cputime}{Max}{884.097980555}%
\StoreBenchExecResult{JarFullInterpolationModelChecking}{Pdr}{Correct}{True}{Cputime}{Stdev}{117.4168576922215148029906229}%
\StoreBenchExecResult{JarFullInterpolationModelChecking}{Pdr}{Correct}{True}{Walltime}{}{42169.4441349145490184}%
\StoreBenchExecResult{JarFullInterpolationModelChecking}{Pdr}{Correct}{True}{Walltime}{Avg}{37.12098955538252554436619718}%
\StoreBenchExecResult{JarFullInterpolationModelChecking}{Pdr}{Correct}{True}{Walltime}{Median}{4.10776288947090525}%
\StoreBenchExecResult{JarFullInterpolationModelChecking}{Pdr}{Correct}{True}{Walltime}{Min}{2.122047150041908}%
\StoreBenchExecResult{JarFullInterpolationModelChecking}{Pdr}{Correct}{True}{Walltime}{Max}{834.9564448681194}%
\StoreBenchExecResult{JarFullInterpolationModelChecking}{Pdr}{Correct}{True}{Walltime}{Stdev}{112.5721634241209622844023189}%

\StoreBenchExecResult{JarFullInterpolationModelChecking}{Pdr}{Error}{}{Count}{}{4424}%
\StoreBenchExecResult{JarFullInterpolationModelChecking}{Pdr}{Error}{}{Cputime}{}{3135291.751833912}%
\StoreBenchExecResult{JarFullInterpolationModelChecking}{Pdr}{Error}{}{Cputime}{Avg}{708.7006672318969258589511754}%
\StoreBenchExecResult{JarFullInterpolationModelChecking}{Pdr}{Error}{}{Cputime}{Median}{901.5183127485}%
\StoreBenchExecResult{JarFullInterpolationModelChecking}{Pdr}{Error}{}{Cputime}{Min}{4.503733422}%
\StoreBenchExecResult{JarFullInterpolationModelChecking}{Pdr}{Error}{}{Cputime}{Max}{962.086140631}%
\StoreBenchExecResult{JarFullInterpolationModelChecking}{Pdr}{Error}{}{Cputime}{Stdev}{356.4708853429061103950135433}%
\StoreBenchExecResult{JarFullInterpolationModelChecking}{Pdr}{Error}{}{Walltime}{}{3019258.5493473310483609}%
\StoreBenchExecResult{JarFullInterpolationModelChecking}{Pdr}{Error}{}{Walltime}{Avg}{682.4725473208252821792269439}%
\StoreBenchExecResult{JarFullInterpolationModelChecking}{Pdr}{Error}{}{Walltime}{Median}{887.10206043254585}%
\StoreBenchExecResult{JarFullInterpolationModelChecking}{Pdr}{Error}{}{Walltime}{Min}{2.42479420802556}%
\StoreBenchExecResult{JarFullInterpolationModelChecking}{Pdr}{Error}{}{Walltime}{Max}{906.6350679579191}%
\StoreBenchExecResult{JarFullInterpolationModelChecking}{Pdr}{Error}{}{Walltime}{Stdev}{354.5074439350812008669854130}%
\StoreBenchExecResult{JarFullInterpolationModelChecking}{Pdr}{Error}{Error}{Count}{}{1009}%
\StoreBenchExecResult{JarFullInterpolationModelChecking}{Pdr}{Error}{Error}{Cputime}{}{72879.759399349}%
\StoreBenchExecResult{JarFullInterpolationModelChecking}{Pdr}{Error}{Error}{Cputime}{Avg}{72.22969216982061446977205154}%
\StoreBenchExecResult{JarFullInterpolationModelChecking}{Pdr}{Error}{Error}{Cputime}{Median}{23.593858862}%
\StoreBenchExecResult{JarFullInterpolationModelChecking}{Pdr}{Error}{Error}{Cputime}{Min}{4.503733422}%
\StoreBenchExecResult{JarFullInterpolationModelChecking}{Pdr}{Error}{Error}{Cputime}{Max}{894.974472902}%
\StoreBenchExecResult{JarFullInterpolationModelChecking}{Pdr}{Error}{Error}{Cputime}{Stdev}{133.8885044022209812000217610}%
\StoreBenchExecResult{JarFullInterpolationModelChecking}{Pdr}{Error}{Error}{Walltime}{}{53504.3596741606015919}%
\StoreBenchExecResult{JarFullInterpolationModelChecking}{Pdr}{Error}{Error}{Walltime}{Avg}{53.02711563345946639435084242}%
\StoreBenchExecResult{JarFullInterpolationModelChecking}{Pdr}{Error}{Error}{Walltime}{Median}{12.258987782988697}%
\StoreBenchExecResult{JarFullInterpolationModelChecking}{Pdr}{Error}{Error}{Walltime}{Min}{2.42479420802556}%
\StoreBenchExecResult{JarFullInterpolationModelChecking}{Pdr}{Error}{Error}{Walltime}{Max}{822.7401875939686}%
\StoreBenchExecResult{JarFullInterpolationModelChecking}{Pdr}{Error}{Error}{Walltime}{Stdev}{109.3445575878451622453708315}%
\StoreBenchExecResult{JarFullInterpolationModelChecking}{Pdr}{Error}{Exception}{Count}{}{3}%
\StoreBenchExecResult{JarFullInterpolationModelChecking}{Pdr}{Error}{Exception}{Cputime}{}{25.379044620}%
\StoreBenchExecResult{JarFullInterpolationModelChecking}{Pdr}{Error}{Exception}{Cputime}{Avg}{8.459681540}%
\StoreBenchExecResult{JarFullInterpolationModelChecking}{Pdr}{Error}{Exception}{Cputime}{Median}{8.26059699}%
\StoreBenchExecResult{JarFullInterpolationModelChecking}{Pdr}{Error}{Exception}{Cputime}{Min}{7.719442319}%
\StoreBenchExecResult{JarFullInterpolationModelChecking}{Pdr}{Error}{Exception}{Cputime}{Max}{9.399005311}%
\StoreBenchExecResult{JarFullInterpolationModelChecking}{Pdr}{Error}{Exception}{Cputime}{Stdev}{0.6999804542797486289441500400}%
\StoreBenchExecResult{JarFullInterpolationModelChecking}{Pdr}{Error}{Exception}{Walltime}{}{16.278316826093943}%
\StoreBenchExecResult{JarFullInterpolationModelChecking}{Pdr}{Error}{Exception}{Walltime}{Avg}{5.426105608697981}%
\StoreBenchExecResult{JarFullInterpolationModelChecking}{Pdr}{Error}{Exception}{Walltime}{Median}{5.350970037048683}%
\StoreBenchExecResult{JarFullInterpolationModelChecking}{Pdr}{Error}{Exception}{Walltime}{Min}{4.83232405805029}%
\StoreBenchExecResult{JarFullInterpolationModelChecking}{Pdr}{Error}{Exception}{Walltime}{Max}{6.09502273099497}%
\StoreBenchExecResult{JarFullInterpolationModelChecking}{Pdr}{Error}{Exception}{Walltime}{Stdev}{0.5182251764496228345386754909}%
\StoreBenchExecResult{JarFullInterpolationModelChecking}{Pdr}{Error}{OutOfJavaMemory}{Count}{}{3}%
\StoreBenchExecResult{JarFullInterpolationModelChecking}{Pdr}{Error}{OutOfJavaMemory}{Cputime}{}{177.286971412}%
\StoreBenchExecResult{JarFullInterpolationModelChecking}{Pdr}{Error}{OutOfJavaMemory}{Cputime}{Avg}{59.09565713733333333333333333}%
\StoreBenchExecResult{JarFullInterpolationModelChecking}{Pdr}{Error}{OutOfJavaMemory}{Cputime}{Median}{60.040880206}%
\StoreBenchExecResult{JarFullInterpolationModelChecking}{Pdr}{Error}{OutOfJavaMemory}{Cputime}{Min}{47.582610185}%
\StoreBenchExecResult{JarFullInterpolationModelChecking}{Pdr}{Error}{OutOfJavaMemory}{Cputime}{Max}{69.663481021}%
\StoreBenchExecResult{JarFullInterpolationModelChecking}{Pdr}{Error}{OutOfJavaMemory}{Cputime}{Stdev}{9.039221913645132515261034113}%
\StoreBenchExecResult{JarFullInterpolationModelChecking}{Pdr}{Error}{OutOfJavaMemory}{Walltime}{}{96.812210662988944}%
\StoreBenchExecResult{JarFullInterpolationModelChecking}{Pdr}{Error}{OutOfJavaMemory}{Walltime}{Avg}{32.27073688766298133333333333}%
\StoreBenchExecResult{JarFullInterpolationModelChecking}{Pdr}{Error}{OutOfJavaMemory}{Walltime}{Median}{32.816900457954034}%
\StoreBenchExecResult{JarFullInterpolationModelChecking}{Pdr}{Error}{OutOfJavaMemory}{Walltime}{Min}{26.10979335103184}%
\StoreBenchExecResult{JarFullInterpolationModelChecking}{Pdr}{Error}{OutOfJavaMemory}{Walltime}{Max}{37.88551685400307}%
\StoreBenchExecResult{JarFullInterpolationModelChecking}{Pdr}{Error}{OutOfJavaMemory}{Walltime}{Stdev}{4.822906246150900536215535024}%
\StoreBenchExecResult{JarFullInterpolationModelChecking}{Pdr}{Error}{OutOfMemory}{Count}{}{23}%
\StoreBenchExecResult{JarFullInterpolationModelChecking}{Pdr}{Error}{OutOfMemory}{Cputime}{}{10170.879698242}%
\StoreBenchExecResult{JarFullInterpolationModelChecking}{Pdr}{Error}{OutOfMemory}{Cputime}{Avg}{442.2121607931304347826086957}%
\StoreBenchExecResult{JarFullInterpolationModelChecking}{Pdr}{Error}{OutOfMemory}{Cputime}{Median}{467.226430021}%
\StoreBenchExecResult{JarFullInterpolationModelChecking}{Pdr}{Error}{OutOfMemory}{Cputime}{Min}{137.724644141}%
\StoreBenchExecResult{JarFullInterpolationModelChecking}{Pdr}{Error}{OutOfMemory}{Cputime}{Max}{850.457096572}%
\StoreBenchExecResult{JarFullInterpolationModelChecking}{Pdr}{Error}{OutOfMemory}{Cputime}{Stdev}{195.9416498346822451346905002}%
\StoreBenchExecResult{JarFullInterpolationModelChecking}{Pdr}{Error}{OutOfMemory}{Walltime}{}{9019.63388577965086}%
\StoreBenchExecResult{JarFullInterpolationModelChecking}{Pdr}{Error}{OutOfMemory}{Walltime}{Avg}{392.1579950338978634782608696}%
\StoreBenchExecResult{JarFullInterpolationModelChecking}{Pdr}{Error}{OutOfMemory}{Walltime}{Median}{387.6860698510427}%
\StoreBenchExecResult{JarFullInterpolationModelChecking}{Pdr}{Error}{OutOfMemory}{Walltime}{Min}{116.33093119598925}%
\StoreBenchExecResult{JarFullInterpolationModelChecking}{Pdr}{Error}{OutOfMemory}{Walltime}{Max}{791.4044800831471}%
\StoreBenchExecResult{JarFullInterpolationModelChecking}{Pdr}{Error}{OutOfMemory}{Walltime}{Stdev}{189.8739242998544592498588033}%
\StoreBenchExecResult{JarFullInterpolationModelChecking}{Pdr}{Error}{SegmentationFault}{Count}{}{13}%
\StoreBenchExecResult{JarFullInterpolationModelChecking}{Pdr}{Error}{SegmentationFault}{Cputime}{}{5179.157200500}%
\StoreBenchExecResult{JarFullInterpolationModelChecking}{Pdr}{Error}{SegmentationFault}{Cputime}{Avg}{398.3967077307692307692307692}%
\StoreBenchExecResult{JarFullInterpolationModelChecking}{Pdr}{Error}{SegmentationFault}{Cputime}{Median}{478.529115503}%
\StoreBenchExecResult{JarFullInterpolationModelChecking}{Pdr}{Error}{SegmentationFault}{Cputime}{Min}{7.490094143}%
\StoreBenchExecResult{JarFullInterpolationModelChecking}{Pdr}{Error}{SegmentationFault}{Cputime}{Max}{734.516468994}%
\StoreBenchExecResult{JarFullInterpolationModelChecking}{Pdr}{Error}{SegmentationFault}{Cputime}{Stdev}{281.8583118050022635242968875}%
\StoreBenchExecResult{JarFullInterpolationModelChecking}{Pdr}{Error}{SegmentationFault}{Walltime}{}{5089.327404748648442}%
\StoreBenchExecResult{JarFullInterpolationModelChecking}{Pdr}{Error}{SegmentationFault}{Walltime}{Avg}{391.4867234422037263076923077}%
\StoreBenchExecResult{JarFullInterpolationModelChecking}{Pdr}{Error}{SegmentationFault}{Walltime}{Median}{469.6611400418915}%
\StoreBenchExecResult{JarFullInterpolationModelChecking}{Pdr}{Error}{SegmentationFault}{Walltime}{Min}{4.22160192206502}%
\StoreBenchExecResult{JarFullInterpolationModelChecking}{Pdr}{Error}{SegmentationFault}{Walltime}{Max}{726.7299602078274}%
\StoreBenchExecResult{JarFullInterpolationModelChecking}{Pdr}{Error}{SegmentationFault}{Walltime}{Stdev}{279.5978513068555194554827576}%
\StoreBenchExecResult{JarFullInterpolationModelChecking}{Pdr}{Error}{Timeout}{Count}{}{3373}%
\StoreBenchExecResult{JarFullInterpolationModelChecking}{Pdr}{Error}{Timeout}{Cputime}{}{3046859.289519789}%
\StoreBenchExecResult{JarFullInterpolationModelChecking}{Pdr}{Error}{Timeout}{Cputime}{Avg}{903.3084166972395493625852357}%
\StoreBenchExecResult{JarFullInterpolationModelChecking}{Pdr}{Error}{Timeout}{Cputime}{Median}{902.010515587}%
\StoreBenchExecResult{JarFullInterpolationModelChecking}{Pdr}{Error}{Timeout}{Cputime}{Min}{900.423365463}%
\StoreBenchExecResult{JarFullInterpolationModelChecking}{Pdr}{Error}{Timeout}{Cputime}{Max}{962.086140631}%
\StoreBenchExecResult{JarFullInterpolationModelChecking}{Pdr}{Error}{Timeout}{Cputime}{Stdev}{4.025120783863387710132241593}%
\StoreBenchExecResult{JarFullInterpolationModelChecking}{Pdr}{Error}{Timeout}{Walltime}{}{2951532.13785515306458}%
\StoreBenchExecResult{JarFullInterpolationModelChecking}{Pdr}{Error}{Timeout}{Walltime}{Avg}{875.0465869715840689534538986}%
\StoreBenchExecResult{JarFullInterpolationModelChecking}{Pdr}{Error}{Timeout}{Walltime}{Median}{890.1549008190632}%
\StoreBenchExecResult{JarFullInterpolationModelChecking}{Pdr}{Error}{Timeout}{Walltime}{Min}{469.9386572299991}%
\StoreBenchExecResult{JarFullInterpolationModelChecking}{Pdr}{Error}{Timeout}{Walltime}{Max}{906.6350679579191}%
\StoreBenchExecResult{JarFullInterpolationModelChecking}{Pdr}{Error}{Timeout}{Walltime}{Stdev}{58.49729739126034926604034327}%
\StoreBenchExecResult{JarFullInterpolationModelChecking}{Pdr}{Wrong}{}{Count}{}{1}%
\StoreBenchExecResult{JarFullInterpolationModelChecking}{Pdr}{Wrong}{}{Cputime}{}{7.743067271}%
\StoreBenchExecResult{JarFullInterpolationModelChecking}{Pdr}{Wrong}{}{Cputime}{Avg}{7.743067271}%
\StoreBenchExecResult{JarFullInterpolationModelChecking}{Pdr}{Wrong}{}{Cputime}{Median}{7.743067271}%
\StoreBenchExecResult{JarFullInterpolationModelChecking}{Pdr}{Wrong}{}{Cputime}{Min}{7.743067271}%
\StoreBenchExecResult{JarFullInterpolationModelChecking}{Pdr}{Wrong}{}{Cputime}{Max}{7.743067271}%
\StoreBenchExecResult{JarFullInterpolationModelChecking}{Pdr}{Wrong}{}{Cputime}{Stdev}{0E-14}%
\StoreBenchExecResult{JarFullInterpolationModelChecking}{Pdr}{Wrong}{}{Walltime}{}{4.8109723860397935}%
\StoreBenchExecResult{JarFullInterpolationModelChecking}{Pdr}{Wrong}{}{Walltime}{Avg}{4.8109723860397935}%
\StoreBenchExecResult{JarFullInterpolationModelChecking}{Pdr}{Wrong}{}{Walltime}{Median}{4.8109723860397935}%
\StoreBenchExecResult{JarFullInterpolationModelChecking}{Pdr}{Wrong}{}{Walltime}{Min}{4.8109723860397935}%
\StoreBenchExecResult{JarFullInterpolationModelChecking}{Pdr}{Wrong}{}{Walltime}{Max}{4.8109723860397935}%
\StoreBenchExecResult{JarFullInterpolationModelChecking}{Pdr}{Wrong}{}{Walltime}{Stdev}{0E-16}%
\StoreBenchExecResult{JarFullInterpolationModelChecking}{Pdr}{Wrong}{False}{Count}{}{1}%
\StoreBenchExecResult{JarFullInterpolationModelChecking}{Pdr}{Wrong}{False}{Cputime}{}{7.743067271}%
\StoreBenchExecResult{JarFullInterpolationModelChecking}{Pdr}{Wrong}{False}{Cputime}{Avg}{7.743067271}%
\StoreBenchExecResult{JarFullInterpolationModelChecking}{Pdr}{Wrong}{False}{Cputime}{Median}{7.743067271}%
\StoreBenchExecResult{JarFullInterpolationModelChecking}{Pdr}{Wrong}{False}{Cputime}{Min}{7.743067271}%
\StoreBenchExecResult{JarFullInterpolationModelChecking}{Pdr}{Wrong}{False}{Cputime}{Max}{7.743067271}%
\StoreBenchExecResult{JarFullInterpolationModelChecking}{Pdr}{Wrong}{False}{Cputime}{Stdev}{0E-14}%
\StoreBenchExecResult{JarFullInterpolationModelChecking}{Pdr}{Wrong}{False}{Walltime}{}{4.8109723860397935}%
\StoreBenchExecResult{JarFullInterpolationModelChecking}{Pdr}{Wrong}{False}{Walltime}{Avg}{4.8109723860397935}%
\StoreBenchExecResult{JarFullInterpolationModelChecking}{Pdr}{Wrong}{False}{Walltime}{Median}{4.8109723860397935}%
\StoreBenchExecResult{JarFullInterpolationModelChecking}{Pdr}{Wrong}{False}{Walltime}{Min}{4.8109723860397935}%
\StoreBenchExecResult{JarFullInterpolationModelChecking}{Pdr}{Wrong}{False}{Walltime}{Max}{4.8109723860397935}%
\StoreBenchExecResult{JarFullInterpolationModelChecking}{Pdr}{Wrong}{False}{Walltime}{Stdev}{0E-16}%
\providecommand\StoreBenchExecResult[7]{\expandafter\newcommand\csname#1#2#3#4#5#6\endcsname{#7}}%
\StoreBenchExecResult{JarFullInterpolationModelChecking}{Bmc}{Total}{}{Count}{}{6024}%
\StoreBenchExecResult{JarFullInterpolationModelChecking}{Bmc}{Total}{}{Cputime}{}{2445312.040219048}%
\StoreBenchExecResult{JarFullInterpolationModelChecking}{Bmc}{Total}{}{Cputime}{Avg}{405.9282935290584329349269588}%
\StoreBenchExecResult{JarFullInterpolationModelChecking}{Bmc}{Total}{}{Cputime}{Median}{104.7699774605}%
\StoreBenchExecResult{JarFullInterpolationModelChecking}{Bmc}{Total}{}{Cputime}{Min}{3.645498412}%
\StoreBenchExecResult{JarFullInterpolationModelChecking}{Bmc}{Total}{}{Cputime}{Max}{962.809485752}%
\StoreBenchExecResult{JarFullInterpolationModelChecking}{Bmc}{Total}{}{Cputime}{Stdev}{420.9838402515688107044025338}%
\StoreBenchExecResult{JarFullInterpolationModelChecking}{Bmc}{Total}{}{Walltime}{}{2336022.1786900197602357}%
\StoreBenchExecResult{JarFullInterpolationModelChecking}{Bmc}{Total}{}{Walltime}{Avg}{387.7858862367230677682104914}%
\StoreBenchExecResult{JarFullInterpolationModelChecking}{Bmc}{Total}{}{Walltime}{Median}{83.64615821558982}%
\StoreBenchExecResult{JarFullInterpolationModelChecking}{Bmc}{Total}{}{Walltime}{Min}{1.9342655108775944}%
\StoreBenchExecResult{JarFullInterpolationModelChecking}{Bmc}{Total}{}{Walltime}{Max}{907.5419994289987}%
\StoreBenchExecResult{JarFullInterpolationModelChecking}{Bmc}{Total}{}{Walltime}{Stdev}{410.2809581128537080330151830}%
\StoreBenchExecResult{JarFullInterpolationModelChecking}{Bmc}{Correct}{}{Count}{}{2388}%
\StoreBenchExecResult{JarFullInterpolationModelChecking}{Bmc}{Correct}{}{Cputime}{}{148664.310829722}%
\StoreBenchExecResult{JarFullInterpolationModelChecking}{Bmc}{Correct}{}{Cputime}{Avg}{62.25473652835929648241206030}%
\StoreBenchExecResult{JarFullInterpolationModelChecking}{Bmc}{Correct}{}{Cputime}{Median}{8.805498204}%
\StoreBenchExecResult{JarFullInterpolationModelChecking}{Bmc}{Correct}{}{Cputime}{Min}{3.645498412}%
\StoreBenchExecResult{JarFullInterpolationModelChecking}{Bmc}{Correct}{}{Cputime}{Max}{899.407843851}%
\StoreBenchExecResult{JarFullInterpolationModelChecking}{Bmc}{Correct}{}{Cputime}{Stdev}{154.1778230661479575961679905}%
\StoreBenchExecResult{JarFullInterpolationModelChecking}{Bmc}{Correct}{}{Walltime}{}{134463.7039162132425780}%
\StoreBenchExecResult{JarFullInterpolationModelChecking}{Bmc}{Correct}{}{Walltime}{Avg}{56.30808371700722051005025126}%
\StoreBenchExecResult{JarFullInterpolationModelChecking}{Bmc}{Correct}{}{Walltime}{Median}{4.812604096485302}%
\StoreBenchExecResult{JarFullInterpolationModelChecking}{Bmc}{Correct}{}{Walltime}{Min}{1.9342655108775944}%
\StoreBenchExecResult{JarFullInterpolationModelChecking}{Bmc}{Correct}{}{Walltime}{Max}{895.262673119083}%
\StoreBenchExecResult{JarFullInterpolationModelChecking}{Bmc}{Correct}{}{Walltime}{Stdev}{149.9405887061663642715337853}%
\StoreBenchExecResult{JarFullInterpolationModelChecking}{Bmc}{Correct}{False}{Count}{}{1177}%
\StoreBenchExecResult{JarFullInterpolationModelChecking}{Bmc}{Correct}{False}{Cputime}{}{113504.693763834}%
\StoreBenchExecResult{JarFullInterpolationModelChecking}{Bmc}{Correct}{False}{Cputime}{Avg}{96.43559368210195412064570943}%
\StoreBenchExecResult{JarFullInterpolationModelChecking}{Bmc}{Correct}{False}{Cputime}{Median}{14.368533363}%
\StoreBenchExecResult{JarFullInterpolationModelChecking}{Bmc}{Correct}{False}{Cputime}{Min}{4.016564143}%
\StoreBenchExecResult{JarFullInterpolationModelChecking}{Bmc}{Correct}{False}{Cputime}{Max}{899.075219406}%
\StoreBenchExecResult{JarFullInterpolationModelChecking}{Bmc}{Correct}{False}{Cputime}{Stdev}{189.3246812137500949968970817}%
\StoreBenchExecResult{JarFullInterpolationModelChecking}{Bmc}{Correct}{False}{Walltime}{}{103757.7570992764087484}%
\StoreBenchExecResult{JarFullInterpolationModelChecking}{Bmc}{Correct}{False}{Walltime}{Avg}{88.15442404356534303177570093}%
\StoreBenchExecResult{JarFullInterpolationModelChecking}{Bmc}{Correct}{False}{Walltime}{Median}{8.252878854051232}%
\StoreBenchExecResult{JarFullInterpolationModelChecking}{Bmc}{Correct}{False}{Walltime}{Min}{2.1174684520810843}%
\StoreBenchExecResult{JarFullInterpolationModelChecking}{Bmc}{Correct}{False}{Walltime}{Max}{880.9240680171642}%
\StoreBenchExecResult{JarFullInterpolationModelChecking}{Bmc}{Correct}{False}{Walltime}{Stdev}{183.8873217487808994945331962}%
\StoreBenchExecResult{JarFullInterpolationModelChecking}{Bmc}{Correct}{True}{Count}{}{1211}%
\StoreBenchExecResult{JarFullInterpolationModelChecking}{Bmc}{Correct}{True}{Cputime}{}{35159.617065888}%
\StoreBenchExecResult{JarFullInterpolationModelChecking}{Bmc}{Correct}{True}{Cputime}{Avg}{29.03354010395375722543352601}%
\StoreBenchExecResult{JarFullInterpolationModelChecking}{Bmc}{Correct}{True}{Cputime}{Median}{6.932303561}%
\StoreBenchExecResult{JarFullInterpolationModelChecking}{Bmc}{Correct}{True}{Cputime}{Min}{3.645498412}%
\StoreBenchExecResult{JarFullInterpolationModelChecking}{Bmc}{Correct}{True}{Cputime}{Max}{899.407843851}%
\StoreBenchExecResult{JarFullInterpolationModelChecking}{Bmc}{Correct}{True}{Cputime}{Stdev}{98.98258741131917480296084337}%
\StoreBenchExecResult{JarFullInterpolationModelChecking}{Bmc}{Correct}{True}{Walltime}{}{30705.9468169368338296}%
\StoreBenchExecResult{JarFullInterpolationModelChecking}{Bmc}{Correct}{True}{Walltime}{Avg}{25.35586029474552752237819983}%
\StoreBenchExecResult{JarFullInterpolationModelChecking}{Bmc}{Correct}{True}{Walltime}{Median}{3.6890768450684845}%
\StoreBenchExecResult{JarFullInterpolationModelChecking}{Bmc}{Correct}{True}{Walltime}{Min}{1.9342655108775944}%
\StoreBenchExecResult{JarFullInterpolationModelChecking}{Bmc}{Correct}{True}{Walltime}{Max}{895.262673119083}%
\StoreBenchExecResult{JarFullInterpolationModelChecking}{Bmc}{Correct}{True}{Walltime}{Stdev}{97.59213700526282139780279076}%

\StoreBenchExecResult{JarFullInterpolationModelChecking}{Bmc}{Error}{}{Count}{}{3635}%
\StoreBenchExecResult{JarFullInterpolationModelChecking}{Bmc}{Error}{}{Cputime}{}{2296639.298802926}%
\StoreBenchExecResult{JarFullInterpolationModelChecking}{Bmc}{Error}{}{Cputime}{Avg}{631.8127369471598349381017882}%
\StoreBenchExecResult{JarFullInterpolationModelChecking}{Bmc}{Error}{}{Cputime}{Median}{901.572600635}%
\StoreBenchExecResult{JarFullInterpolationModelChecking}{Bmc}{Error}{}{Cputime}{Min}{4.332348193}%
\StoreBenchExecResult{JarFullInterpolationModelChecking}{Bmc}{Error}{}{Cputime}{Max}{962.809485752}%
\StoreBenchExecResult{JarFullInterpolationModelChecking}{Bmc}{Error}{}{Cputime}{Stdev}{386.5601130283989224024022916}%
\StoreBenchExecResult{JarFullInterpolationModelChecking}{Bmc}{Error}{}{Walltime}{}{2201553.3419914846287043}%
\StoreBenchExecResult{JarFullInterpolationModelChecking}{Bmc}{Error}{}{Walltime}{Avg}{605.6542894061855924908665750}%
\StoreBenchExecResult{JarFullInterpolationModelChecking}{Bmc}{Error}{}{Walltime}{Median}{868.4944559328724}%
\StoreBenchExecResult{JarFullInterpolationModelChecking}{Bmc}{Error}{}{Walltime}{Min}{2.3266525180079043}%
\StoreBenchExecResult{JarFullInterpolationModelChecking}{Bmc}{Error}{}{Walltime}{Max}{907.5419994289987}%
\StoreBenchExecResult{JarFullInterpolationModelChecking}{Bmc}{Error}{}{Walltime}{Stdev}{380.1324819982369434660826987}%
\StoreBenchExecResult{JarFullInterpolationModelChecking}{Bmc}{Error}{Error}{Count}{}{1042}%
\StoreBenchExecResult{JarFullInterpolationModelChecking}{Bmc}{Error}{Error}{Cputime}{}{61268.054874053}%
\StoreBenchExecResult{JarFullInterpolationModelChecking}{Bmc}{Error}{Error}{Cputime}{Avg}{58.79851715360172744721689060}%
\StoreBenchExecResult{JarFullInterpolationModelChecking}{Bmc}{Error}{Error}{Cputime}{Median}{19.057687784}%
\StoreBenchExecResult{JarFullInterpolationModelChecking}{Bmc}{Error}{Error}{Cputime}{Min}{4.332348193}%
\StoreBenchExecResult{JarFullInterpolationModelChecking}{Bmc}{Error}{Error}{Cputime}{Max}{890.266962774}%
\StoreBenchExecResult{JarFullInterpolationModelChecking}{Bmc}{Error}{Error}{Cputime}{Stdev}{123.8811254857360524939166932}%
\StoreBenchExecResult{JarFullInterpolationModelChecking}{Bmc}{Error}{Error}{Walltime}{}{44646.3453216769266133}%
\StoreBenchExecResult{JarFullInterpolationModelChecking}{Bmc}{Error}{Error}{Walltime}{Avg}{42.84678053903735759433781190}%
\StoreBenchExecResult{JarFullInterpolationModelChecking}{Bmc}{Error}{Error}{Walltime}{Median}{9.939534702571109}%
\StoreBenchExecResult{JarFullInterpolationModelChecking}{Bmc}{Error}{Error}{Walltime}{Min}{2.3266525180079043}%
\StoreBenchExecResult{JarFullInterpolationModelChecking}{Bmc}{Error}{Error}{Walltime}{Max}{809.984625991201}%
\StoreBenchExecResult{JarFullInterpolationModelChecking}{Bmc}{Error}{Error}{Walltime}{Stdev}{102.6691224997618642312978306}%
\StoreBenchExecResult{JarFullInterpolationModelChecking}{Bmc}{Error}{OutOfJavaMemory}{Count}{}{3}%
\StoreBenchExecResult{JarFullInterpolationModelChecking}{Bmc}{Error}{OutOfJavaMemory}{Cputime}{}{189.998199742}%
\StoreBenchExecResult{JarFullInterpolationModelChecking}{Bmc}{Error}{OutOfJavaMemory}{Cputime}{Avg}{63.33273324733333333333333333}%
\StoreBenchExecResult{JarFullInterpolationModelChecking}{Bmc}{Error}{OutOfJavaMemory}{Cputime}{Median}{67.208959109}%
\StoreBenchExecResult{JarFullInterpolationModelChecking}{Bmc}{Error}{OutOfJavaMemory}{Cputime}{Min}{50.627145965}%
\StoreBenchExecResult{JarFullInterpolationModelChecking}{Bmc}{Error}{OutOfJavaMemory}{Cputime}{Max}{72.162094668}%
\StoreBenchExecResult{JarFullInterpolationModelChecking}{Bmc}{Error}{OutOfJavaMemory}{Cputime}{Stdev}{9.208957563452389837649001681}%
\StoreBenchExecResult{JarFullInterpolationModelChecking}{Bmc}{Error}{OutOfJavaMemory}{Walltime}{}{103.758004989009351}%
\StoreBenchExecResult{JarFullInterpolationModelChecking}{Bmc}{Error}{OutOfJavaMemory}{Walltime}{Avg}{34.586001663003117}%
\StoreBenchExecResult{JarFullInterpolationModelChecking}{Bmc}{Error}{OutOfJavaMemory}{Walltime}{Median}{36.313580164918676}%
\StoreBenchExecResult{JarFullInterpolationModelChecking}{Bmc}{Error}{OutOfJavaMemory}{Walltime}{Min}{27.801186074037105}%
\StoreBenchExecResult{JarFullInterpolationModelChecking}{Bmc}{Error}{OutOfJavaMemory}{Walltime}{Max}{39.64323875005357}%
\StoreBenchExecResult{JarFullInterpolationModelChecking}{Bmc}{Error}{OutOfJavaMemory}{Walltime}{Stdev}{4.986444859523184396941837957}%
\StoreBenchExecResult{JarFullInterpolationModelChecking}{Bmc}{Error}{OutOfMemory}{Count}{}{363}%
\StoreBenchExecResult{JarFullInterpolationModelChecking}{Bmc}{Error}{OutOfMemory}{Cputime}{}{215322.276003421}%
\StoreBenchExecResult{JarFullInterpolationModelChecking}{Bmc}{Error}{OutOfMemory}{Cputime}{Avg}{593.1743140590110192837465565}%
\StoreBenchExecResult{JarFullInterpolationModelChecking}{Bmc}{Error}{OutOfMemory}{Cputime}{Median}{659.600242587}%
\StoreBenchExecResult{JarFullInterpolationModelChecking}{Bmc}{Error}{OutOfMemory}{Cputime}{Min}{100.409339679}%
\StoreBenchExecResult{JarFullInterpolationModelChecking}{Bmc}{Error}{OutOfMemory}{Cputime}{Max}{898.855631433}%
\StoreBenchExecResult{JarFullInterpolationModelChecking}{Bmc}{Error}{OutOfMemory}{Cputime}{Stdev}{201.4677585538451881964397288}%
\StoreBenchExecResult{JarFullInterpolationModelChecking}{Bmc}{Error}{OutOfMemory}{Walltime}{}{206259.69942539324943}%
\StoreBenchExecResult{JarFullInterpolationModelChecking}{Bmc}{Error}{OutOfMemory}{Walltime}{Avg}{568.2085383619648744628099174}%
\StoreBenchExecResult{JarFullInterpolationModelChecking}{Bmc}{Error}{OutOfMemory}{Walltime}{Median}{629.0149876310024}%
\StoreBenchExecResult{JarFullInterpolationModelChecking}{Bmc}{Error}{OutOfMemory}{Walltime}{Min}{80.27841905807145}%
\StoreBenchExecResult{JarFullInterpolationModelChecking}{Bmc}{Error}{OutOfMemory}{Walltime}{Max}{885.7246328832116}%
\StoreBenchExecResult{JarFullInterpolationModelChecking}{Bmc}{Error}{OutOfMemory}{Walltime}{Stdev}{198.5294654571253600706388202}%
\StoreBenchExecResult{JarFullInterpolationModelChecking}{Bmc}{Error}{Timeout}{Count}{}{2227}%
\StoreBenchExecResult{JarFullInterpolationModelChecking}{Bmc}{Error}{Timeout}{Cputime}{}{2019858.969725710}%
\StoreBenchExecResult{JarFullInterpolationModelChecking}{Bmc}{Error}{Timeout}{Cputime}{Avg}{906.9865153685271665918275707}%
\StoreBenchExecResult{JarFullInterpolationModelChecking}{Bmc}{Error}{Timeout}{Cputime}{Median}{904.640956011}%
\StoreBenchExecResult{JarFullInterpolationModelChecking}{Bmc}{Error}{Timeout}{Cputime}{Min}{900.451973755}%
\StoreBenchExecResult{JarFullInterpolationModelChecking}{Bmc}{Error}{Timeout}{Cputime}{Max}{962.809485752}%
\StoreBenchExecResult{JarFullInterpolationModelChecking}{Bmc}{Error}{Timeout}{Cputime}{Stdev}{8.765865760031412238471785035}%
\StoreBenchExecResult{JarFullInterpolationModelChecking}{Bmc}{Error}{Timeout}{Walltime}{}{1950543.53923942544331}%
\StoreBenchExecResult{JarFullInterpolationModelChecking}{Bmc}{Error}{Timeout}{Walltime}{Avg}{875.8614904532669255994611585}%
\StoreBenchExecResult{JarFullInterpolationModelChecking}{Bmc}{Error}{Timeout}{Walltime}{Median}{888.9447286031209}%
\StoreBenchExecResult{JarFullInterpolationModelChecking}{Bmc}{Error}{Timeout}{Walltime}{Min}{468.7250957470387}%
\StoreBenchExecResult{JarFullInterpolationModelChecking}{Bmc}{Error}{Timeout}{Walltime}{Max}{907.5419994289987}%
\StoreBenchExecResult{JarFullInterpolationModelChecking}{Bmc}{Error}{Timeout}{Walltime}{Stdev}{51.15742735103114124328373743}%
\StoreBenchExecResult{JarFullInterpolationModelChecking}{Bmc}{Wrong}{}{Count}{}{1}%
\StoreBenchExecResult{JarFullInterpolationModelChecking}{Bmc}{Wrong}{}{Cputime}{}{8.4305864}%
\StoreBenchExecResult{JarFullInterpolationModelChecking}{Bmc}{Wrong}{}{Cputime}{Avg}{8.4305864}%
\StoreBenchExecResult{JarFullInterpolationModelChecking}{Bmc}{Wrong}{}{Cputime}{Median}{8.4305864}%
\StoreBenchExecResult{JarFullInterpolationModelChecking}{Bmc}{Wrong}{}{Cputime}{Min}{8.4305864}%
\StoreBenchExecResult{JarFullInterpolationModelChecking}{Bmc}{Wrong}{}{Cputime}{Max}{8.4305864}%
\StoreBenchExecResult{JarFullInterpolationModelChecking}{Bmc}{Wrong}{}{Cputime}{Stdev}{0E-14}%
\StoreBenchExecResult{JarFullInterpolationModelChecking}{Bmc}{Wrong}{}{Walltime}{}{5.1327823218889534}%
\StoreBenchExecResult{JarFullInterpolationModelChecking}{Bmc}{Wrong}{}{Walltime}{Avg}{5.1327823218889534}%
\StoreBenchExecResult{JarFullInterpolationModelChecking}{Bmc}{Wrong}{}{Walltime}{Median}{5.1327823218889534}%
\StoreBenchExecResult{JarFullInterpolationModelChecking}{Bmc}{Wrong}{}{Walltime}{Min}{5.1327823218889534}%
\StoreBenchExecResult{JarFullInterpolationModelChecking}{Bmc}{Wrong}{}{Walltime}{Max}{5.1327823218889534}%
\StoreBenchExecResult{JarFullInterpolationModelChecking}{Bmc}{Wrong}{}{Walltime}{Stdev}{0E-16}%
\StoreBenchExecResult{JarFullInterpolationModelChecking}{Bmc}{Wrong}{False}{Count}{}{1}%
\StoreBenchExecResult{JarFullInterpolationModelChecking}{Bmc}{Wrong}{False}{Cputime}{}{8.4305864}%
\StoreBenchExecResult{JarFullInterpolationModelChecking}{Bmc}{Wrong}{False}{Cputime}{Avg}{8.4305864}%
\StoreBenchExecResult{JarFullInterpolationModelChecking}{Bmc}{Wrong}{False}{Cputime}{Median}{8.4305864}%
\StoreBenchExecResult{JarFullInterpolationModelChecking}{Bmc}{Wrong}{False}{Cputime}{Min}{8.4305864}%
\StoreBenchExecResult{JarFullInterpolationModelChecking}{Bmc}{Wrong}{False}{Cputime}{Max}{8.4305864}%
\StoreBenchExecResult{JarFullInterpolationModelChecking}{Bmc}{Wrong}{False}{Cputime}{Stdev}{0E-14}%
\StoreBenchExecResult{JarFullInterpolationModelChecking}{Bmc}{Wrong}{False}{Walltime}{}{5.1327823218889534}%
\StoreBenchExecResult{JarFullInterpolationModelChecking}{Bmc}{Wrong}{False}{Walltime}{Avg}{5.1327823218889534}%
\StoreBenchExecResult{JarFullInterpolationModelChecking}{Bmc}{Wrong}{False}{Walltime}{Median}{5.1327823218889534}%
\StoreBenchExecResult{JarFullInterpolationModelChecking}{Bmc}{Wrong}{False}{Walltime}{Min}{5.1327823218889534}%
\StoreBenchExecResult{JarFullInterpolationModelChecking}{Bmc}{Wrong}{False}{Walltime}{Max}{5.1327823218889534}%
\StoreBenchExecResult{JarFullInterpolationModelChecking}{Bmc}{Wrong}{False}{Walltime}{Stdev}{0E-16}%
\providecommand\StoreBenchExecResult[7]{\expandafter\newcommand\csname#1#2#3#4#5#6\endcsname{#7}}%
\StoreBenchExecResult{JarFullInterpolationModelChecking}{KInduction}{Total}{}{Count}{}{6024}%
\StoreBenchExecResult{JarFullInterpolationModelChecking}{KInduction}{Total}{}{Cputime}{}{2077826.530267808}%
\StoreBenchExecResult{JarFullInterpolationModelChecking}{KInduction}{Total}{}{Cputime}{Avg}{344.9247228200212483399734396}%
\StoreBenchExecResult{JarFullInterpolationModelChecking}{KInduction}{Total}{}{Cputime}{Median}{69.5589021025}%
\StoreBenchExecResult{JarFullInterpolationModelChecking}{KInduction}{Total}{}{Cputime}{Min}{3.963766385}%
\StoreBenchExecResult{JarFullInterpolationModelChecking}{KInduction}{Total}{}{Cputime}{Max}{962.447629537}%
\StoreBenchExecResult{JarFullInterpolationModelChecking}{KInduction}{Total}{}{Cputime}{Stdev}{397.5239392799204470464477267}%
\StoreBenchExecResult{JarFullInterpolationModelChecking}{KInduction}{Total}{}{Walltime}{}{1298162.6877708691147763}%
\StoreBenchExecResult{JarFullInterpolationModelChecking}{KInduction}{Total}{}{Walltime}{Avg}{215.4984541452305967424136786}%
\StoreBenchExecResult{JarFullInterpolationModelChecking}{KInduction}{Total}{}{Walltime}{Median}{38.946428700000979}%
\StoreBenchExecResult{JarFullInterpolationModelChecking}{KInduction}{Total}{}{Walltime}{Min}{2.1290870630182326}%
\StoreBenchExecResult{JarFullInterpolationModelChecking}{KInduction}{Total}{}{Walltime}{Max}{897.7362117860466}%
\StoreBenchExecResult{JarFullInterpolationModelChecking}{KInduction}{Total}{}{Walltime}{Stdev}{272.5222548687343781778980489}%
\StoreBenchExecResult{JarFullInterpolationModelChecking}{KInduction}{Correct}{}{Count}{}{3157}%
\StoreBenchExecResult{JarFullInterpolationModelChecking}{KInduction}{Correct}{}{Cputime}{}{226454.179104442}%
\StoreBenchExecResult{JarFullInterpolationModelChecking}{KInduction}{Correct}{}{Cputime}{Avg}{71.73081378031105479885967691}%
\StoreBenchExecResult{JarFullInterpolationModelChecking}{KInduction}{Correct}{}{Cputime}{Median}{15.166300322}%
\StoreBenchExecResult{JarFullInterpolationModelChecking}{KInduction}{Correct}{}{Cputime}{Min}{3.963766385}%
\StoreBenchExecResult{JarFullInterpolationModelChecking}{KInduction}{Correct}{}{Cputime}{Max}{890.589144981}%
\StoreBenchExecResult{JarFullInterpolationModelChecking}{KInduction}{Correct}{}{Cputime}{Stdev}{143.9559050374877216786360738}%
\StoreBenchExecResult{JarFullInterpolationModelChecking}{KInduction}{Correct}{}{Walltime}{}{123897.5025611596647077}%
\StoreBenchExecResult{JarFullInterpolationModelChecking}{KInduction}{Correct}{}{Walltime}{Avg}{39.24532865415257038571428571}%
\StoreBenchExecResult{JarFullInterpolationModelChecking}{KInduction}{Correct}{}{Walltime}{Median}{7.844452999997884}%
\StoreBenchExecResult{JarFullInterpolationModelChecking}{KInduction}{Correct}{}{Walltime}{Min}{2.1290870630182326}%
\StoreBenchExecResult{JarFullInterpolationModelChecking}{KInduction}{Correct}{}{Walltime}{Max}{764.3777275129687}%
\StoreBenchExecResult{JarFullInterpolationModelChecking}{KInduction}{Correct}{}{Walltime}{Stdev}{82.22225830330537415685521572}%
\StoreBenchExecResult{JarFullInterpolationModelChecking}{KInduction}{Correct}{False}{Count}{}{999}%
\StoreBenchExecResult{JarFullInterpolationModelChecking}{KInduction}{Correct}{False}{Cputime}{}{90234.806858915}%
\StoreBenchExecResult{JarFullInterpolationModelChecking}{KInduction}{Correct}{False}{Cputime}{Avg}{90.32513199090590590590590591}%
\StoreBenchExecResult{JarFullInterpolationModelChecking}{KInduction}{Correct}{False}{Cputime}{Median}{25.036180687}%
\StoreBenchExecResult{JarFullInterpolationModelChecking}{KInduction}{Correct}{False}{Cputime}{Min}{4.451336743}%
\StoreBenchExecResult{JarFullInterpolationModelChecking}{KInduction}{Correct}{False}{Cputime}{Max}{890.589144981}%
\StoreBenchExecResult{JarFullInterpolationModelChecking}{KInduction}{Correct}{False}{Cputime}{Stdev}{161.1267027604103699784479937}%
\StoreBenchExecResult{JarFullInterpolationModelChecking}{KInduction}{Correct}{False}{Walltime}{}{52575.5720185886600863}%
\StoreBenchExecResult{JarFullInterpolationModelChecking}{KInduction}{Correct}{False}{Walltime}{Avg}{52.62820021880746755385385385}%
\StoreBenchExecResult{JarFullInterpolationModelChecking}{KInduction}{Correct}{False}{Walltime}{Median}{13.174007270019501}%
\StoreBenchExecResult{JarFullInterpolationModelChecking}{KInduction}{Correct}{False}{Walltime}{Min}{2.372674386948347}%
\StoreBenchExecResult{JarFullInterpolationModelChecking}{KInduction}{Correct}{False}{Walltime}{Max}{764.3777275129687}%
\StoreBenchExecResult{JarFullInterpolationModelChecking}{KInduction}{Correct}{False}{Walltime}{Stdev}{99.04065622502925490170484522}%
\StoreBenchExecResult{JarFullInterpolationModelChecking}{KInduction}{Correct}{True}{Count}{}{2158}%
\StoreBenchExecResult{JarFullInterpolationModelChecking}{KInduction}{Correct}{True}{Cputime}{}{136219.372245527}%
\StoreBenchExecResult{JarFullInterpolationModelChecking}{KInduction}{Correct}{True}{Cputime}{Avg}{63.12297138346941612604263207}%
\StoreBenchExecResult{JarFullInterpolationModelChecking}{KInduction}{Correct}{True}{Cputime}{Median}{11.4321693375}%
\StoreBenchExecResult{JarFullInterpolationModelChecking}{KInduction}{Correct}{True}{Cputime}{Min}{3.963766385}%
\StoreBenchExecResult{JarFullInterpolationModelChecking}{KInduction}{Correct}{True}{Cputime}{Max}{886.207023849}%
\StoreBenchExecResult{JarFullInterpolationModelChecking}{KInduction}{Correct}{True}{Cputime}{Stdev}{134.4027290550173917963268469}%
\StoreBenchExecResult{JarFullInterpolationModelChecking}{KInduction}{Correct}{True}{Walltime}{}{71321.9305425710046214}%
\StoreBenchExecResult{JarFullInterpolationModelChecking}{KInduction}{Correct}{True}{Walltime}{Avg}{33.05001415318396877729379055}%
\StoreBenchExecResult{JarFullInterpolationModelChecking}{KInduction}{Correct}{True}{Walltime}{Median}{5.9353181730257345}%
\StoreBenchExecResult{JarFullInterpolationModelChecking}{KInduction}{Correct}{True}{Walltime}{Min}{2.1290870630182326}%
\StoreBenchExecResult{JarFullInterpolationModelChecking}{KInduction}{Correct}{True}{Walltime}{Max}{760.3620864299592}%
\StoreBenchExecResult{JarFullInterpolationModelChecking}{KInduction}{Correct}{True}{Walltime}{Stdev}{72.30452819417113961028239080}%

\StoreBenchExecResult{JarFullInterpolationModelChecking}{KInduction}{Error}{}{Count}{}{2866}%
\StoreBenchExecResult{JarFullInterpolationModelChecking}{KInduction}{Error}{}{Cputime}{}{1851358.870931396}%
\StoreBenchExecResult{JarFullInterpolationModelChecking}{KInduction}{Error}{}{Cputime}{Avg}{645.9730882524061409630146546}%
\StoreBenchExecResult{JarFullInterpolationModelChecking}{KInduction}{Error}{}{Cputime}{Median}{901.0453835645}%
\StoreBenchExecResult{JarFullInterpolationModelChecking}{KInduction}{Error}{}{Cputime}{Min}{5.101242669}%
\StoreBenchExecResult{JarFullInterpolationModelChecking}{KInduction}{Error}{}{Cputime}{Max}{962.447629537}%
\StoreBenchExecResult{JarFullInterpolationModelChecking}{KInduction}{Error}{}{Cputime}{Stdev}{369.3805015046350976211349053}%
\StoreBenchExecResult{JarFullInterpolationModelChecking}{KInduction}{Error}{}{Walltime}{}{1174258.0799390145087866}%
\StoreBenchExecResult{JarFullInterpolationModelChecking}{KInduction}{Error}{}{Walltime}{Avg}{409.7201953729987818515701326}%
\StoreBenchExecResult{JarFullInterpolationModelChecking}{KInduction}{Error}{}{Walltime}{Median}{452.3888167204568}%
\StoreBenchExecResult{JarFullInterpolationModelChecking}{KInduction}{Error}{}{Walltime}{Min}{2.6608412589412183}%
\StoreBenchExecResult{JarFullInterpolationModelChecking}{KInduction}{Error}{}{Walltime}{Max}{897.7362117860466}%
\StoreBenchExecResult{JarFullInterpolationModelChecking}{KInduction}{Error}{}{Walltime}{Stdev}{276.9476318361700050860037780}%
\StoreBenchExecResult{JarFullInterpolationModelChecking}{KInduction}{Error}{Error}{Count}{}{798}%
\StoreBenchExecResult{JarFullInterpolationModelChecking}{KInduction}{Error}{Error}{Cputime}{}{89847.032225088}%
\StoreBenchExecResult{JarFullInterpolationModelChecking}{KInduction}{Error}{Error}{Cputime}{Avg}{112.5902659462255639097744361}%
\StoreBenchExecResult{JarFullInterpolationModelChecking}{KInduction}{Error}{Error}{Cputime}{Median}{31.6422073155}%
\StoreBenchExecResult{JarFullInterpolationModelChecking}{KInduction}{Error}{Error}{Cputime}{Min}{5.101242669}%
\StoreBenchExecResult{JarFullInterpolationModelChecking}{KInduction}{Error}{Error}{Cputime}{Max}{890.631986316}%
\StoreBenchExecResult{JarFullInterpolationModelChecking}{KInduction}{Error}{Error}{Cputime}{Stdev}{169.6632510480425643232403248}%
\StoreBenchExecResult{JarFullInterpolationModelChecking}{KInduction}{Error}{Error}{Walltime}{}{71272.9855274055154146}%
\StoreBenchExecResult{JarFullInterpolationModelChecking}{KInduction}{Error}{Error}{Walltime}{Avg}{89.31451820476881630902255639}%
\StoreBenchExecResult{JarFullInterpolationModelChecking}{KInduction}{Error}{Error}{Walltime}{Median}{17.0299108700128265}%
\StoreBenchExecResult{JarFullInterpolationModelChecking}{KInduction}{Error}{Error}{Walltime}{Min}{2.6608412589412183}%
\StoreBenchExecResult{JarFullInterpolationModelChecking}{KInduction}{Error}{Error}{Walltime}{Max}{833.8062691839878}%
\StoreBenchExecResult{JarFullInterpolationModelChecking}{KInduction}{Error}{Error}{Walltime}{Stdev}{149.2751814403335763504564621}%
\StoreBenchExecResult{JarFullInterpolationModelChecking}{KInduction}{Error}{Exception}{Count}{}{1}%
\StoreBenchExecResult{JarFullInterpolationModelChecking}{KInduction}{Error}{Exception}{Cputime}{}{392.650572878}%
\StoreBenchExecResult{JarFullInterpolationModelChecking}{KInduction}{Error}{Exception}{Cputime}{Avg}{392.650572878}%
\StoreBenchExecResult{JarFullInterpolationModelChecking}{KInduction}{Error}{Exception}{Cputime}{Median}{392.650572878}%
\StoreBenchExecResult{JarFullInterpolationModelChecking}{KInduction}{Error}{Exception}{Cputime}{Min}{392.650572878}%
\StoreBenchExecResult{JarFullInterpolationModelChecking}{KInduction}{Error}{Exception}{Cputime}{Max}{392.650572878}%
\StoreBenchExecResult{JarFullInterpolationModelChecking}{KInduction}{Error}{Exception}{Cputime}{Stdev}{0E-14}%
\StoreBenchExecResult{JarFullInterpolationModelChecking}{KInduction}{Error}{Exception}{Walltime}{}{200.30770005797967}%
\StoreBenchExecResult{JarFullInterpolationModelChecking}{KInduction}{Error}{Exception}{Walltime}{Avg}{200.30770005797967}%
\StoreBenchExecResult{JarFullInterpolationModelChecking}{KInduction}{Error}{Exception}{Walltime}{Median}{200.30770005797967}%
\StoreBenchExecResult{JarFullInterpolationModelChecking}{KInduction}{Error}{Exception}{Walltime}{Min}{200.30770005797967}%
\StoreBenchExecResult{JarFullInterpolationModelChecking}{KInduction}{Error}{Exception}{Walltime}{Max}{200.30770005797967}%
\StoreBenchExecResult{JarFullInterpolationModelChecking}{KInduction}{Error}{Exception}{Walltime}{Stdev}{0E-14}%
\StoreBenchExecResult{JarFullInterpolationModelChecking}{KInduction}{Error}{OutOfJavaMemory}{Count}{}{5}%
\StoreBenchExecResult{JarFullInterpolationModelChecking}{KInduction}{Error}{OutOfJavaMemory}{Cputime}{}{1179.043787470}%
\StoreBenchExecResult{JarFullInterpolationModelChecking}{KInduction}{Error}{OutOfJavaMemory}{Cputime}{Avg}{235.808757494}%
\StoreBenchExecResult{JarFullInterpolationModelChecking}{KInduction}{Error}{OutOfJavaMemory}{Cputime}{Median}{62.023651437}%
\StoreBenchExecResult{JarFullInterpolationModelChecking}{KInduction}{Error}{OutOfJavaMemory}{Cputime}{Min}{57.733685042}%
\StoreBenchExecResult{JarFullInterpolationModelChecking}{KInduction}{Error}{OutOfJavaMemory}{Cputime}{Max}{510.085962125}%
\StoreBenchExecResult{JarFullInterpolationModelChecking}{KInduction}{Error}{OutOfJavaMemory}{Cputime}{Stdev}{215.7900548387406047691516818}%
\StoreBenchExecResult{JarFullInterpolationModelChecking}{KInduction}{Error}{OutOfJavaMemory}{Walltime}{}{602.058210332412272}%
\StoreBenchExecResult{JarFullInterpolationModelChecking}{KInduction}{Error}{OutOfJavaMemory}{Walltime}{Avg}{120.4116420664824544}%
\StoreBenchExecResult{JarFullInterpolationModelChecking}{KInduction}{Error}{OutOfJavaMemory}{Walltime}{Median}{33.548733675153926}%
\StoreBenchExecResult{JarFullInterpolationModelChecking}{KInduction}{Error}{OutOfJavaMemory}{Walltime}{Min}{30.940811682958156}%
\StoreBenchExecResult{JarFullInterpolationModelChecking}{KInduction}{Error}{OutOfJavaMemory}{Walltime}{Max}{257.21719994721934}%
\StoreBenchExecResult{JarFullInterpolationModelChecking}{KInduction}{Error}{OutOfJavaMemory}{Walltime}{Stdev}{107.6852847359463216380592994}%
\StoreBenchExecResult{JarFullInterpolationModelChecking}{KInduction}{Error}{OutOfMemory}{Count}{}{221}%
\StoreBenchExecResult{JarFullInterpolationModelChecking}{KInduction}{Error}{OutOfMemory}{Cputime}{}{96612.418763095}%
\StoreBenchExecResult{JarFullInterpolationModelChecking}{KInduction}{Error}{OutOfMemory}{Cputime}{Avg}{437.1602658963574660633484163}%
\StoreBenchExecResult{JarFullInterpolationModelChecking}{KInduction}{Error}{OutOfMemory}{Cputime}{Median}{419.170709036}%
\StoreBenchExecResult{JarFullInterpolationModelChecking}{KInduction}{Error}{OutOfMemory}{Cputime}{Min}{189.775137961}%
\StoreBenchExecResult{JarFullInterpolationModelChecking}{KInduction}{Error}{OutOfMemory}{Cputime}{Max}{895.778736101}%
\StoreBenchExecResult{JarFullInterpolationModelChecking}{KInduction}{Error}{OutOfMemory}{Cputime}{Stdev}{191.6366856878670122082129595}%
\StoreBenchExecResult{JarFullInterpolationModelChecking}{KInduction}{Error}{OutOfMemory}{Walltime}{}{48873.48123826156377}%
\StoreBenchExecResult{JarFullInterpolationModelChecking}{KInduction}{Error}{OutOfMemory}{Walltime}{Avg}{221.1469739287853564253393665}%
\StoreBenchExecResult{JarFullInterpolationModelChecking}{KInduction}{Error}{OutOfMemory}{Walltime}{Median}{210.38167036301456}%
\StoreBenchExecResult{JarFullInterpolationModelChecking}{KInduction}{Error}{OutOfMemory}{Walltime}{Min}{96.27279451582581}%
\StoreBenchExecResult{JarFullInterpolationModelChecking}{KInduction}{Error}{OutOfMemory}{Walltime}{Max}{449.1347853639163}%
\StoreBenchExecResult{JarFullInterpolationModelChecking}{KInduction}{Error}{OutOfMemory}{Walltime}{Stdev}{97.31748621511197690742072159}%
\StoreBenchExecResult{JarFullInterpolationModelChecking}{KInduction}{Error}{Timeout}{Count}{}{1841}%
\StoreBenchExecResult{JarFullInterpolationModelChecking}{KInduction}{Error}{Timeout}{Cputime}{}{1663327.725582865}%
\StoreBenchExecResult{JarFullInterpolationModelChecking}{KInduction}{Error}{Timeout}{Cputime}{Avg}{903.4914316039462248777838131}%
\StoreBenchExecResult{JarFullInterpolationModelChecking}{KInduction}{Error}{Timeout}{Cputime}{Median}{902.045743979}%
\StoreBenchExecResult{JarFullInterpolationModelChecking}{KInduction}{Error}{Timeout}{Cputime}{Min}{900.387217177}%
\StoreBenchExecResult{JarFullInterpolationModelChecking}{KInduction}{Error}{Timeout}{Cputime}{Max}{962.447629537}%
\StoreBenchExecResult{JarFullInterpolationModelChecking}{KInduction}{Error}{Timeout}{Cputime}{Stdev}{5.950885468169106111725037653}%
\StoreBenchExecResult{JarFullInterpolationModelChecking}{KInduction}{Error}{Timeout}{Walltime}{}{1053309.24726295703766}%
\StoreBenchExecResult{JarFullInterpolationModelChecking}{KInduction}{Error}{Timeout}{Walltime}{Avg}{572.1397323535888308853883759}%
\StoreBenchExecResult{JarFullInterpolationModelChecking}{KInduction}{Error}{Timeout}{Walltime}{Median}{455.6022394550964}%
\StoreBenchExecResult{JarFullInterpolationModelChecking}{KInduction}{Error}{Timeout}{Walltime}{Min}{451.1961625160184}%
\StoreBenchExecResult{JarFullInterpolationModelChecking}{KInduction}{Error}{Timeout}{Walltime}{Max}{897.7362117860466}%
\StoreBenchExecResult{JarFullInterpolationModelChecking}{KInduction}{Error}{Timeout}{Walltime}{Stdev}{182.1470989838344336834544130}%
\StoreBenchExecResult{JarFullInterpolationModelChecking}{KInduction}{Wrong}{}{Count}{}{1}%
\StoreBenchExecResult{JarFullInterpolationModelChecking}{KInduction}{Wrong}{}{Cputime}{}{13.48023197}%
\StoreBenchExecResult{JarFullInterpolationModelChecking}{KInduction}{Wrong}{}{Cputime}{Avg}{13.48023197}%
\StoreBenchExecResult{JarFullInterpolationModelChecking}{KInduction}{Wrong}{}{Cputime}{Median}{13.48023197}%
\StoreBenchExecResult{JarFullInterpolationModelChecking}{KInduction}{Wrong}{}{Cputime}{Min}{13.48023197}%
\StoreBenchExecResult{JarFullInterpolationModelChecking}{KInduction}{Wrong}{}{Cputime}{Max}{13.48023197}%
\StoreBenchExecResult{JarFullInterpolationModelChecking}{KInduction}{Wrong}{}{Cputime}{Stdev}{0E-14}%
\StoreBenchExecResult{JarFullInterpolationModelChecking}{KInduction}{Wrong}{}{Walltime}{}{7.105270694941282}%
\StoreBenchExecResult{JarFullInterpolationModelChecking}{KInduction}{Wrong}{}{Walltime}{Avg}{7.105270694941282}%
\StoreBenchExecResult{JarFullInterpolationModelChecking}{KInduction}{Wrong}{}{Walltime}{Median}{7.105270694941282}%
\StoreBenchExecResult{JarFullInterpolationModelChecking}{KInduction}{Wrong}{}{Walltime}{Min}{7.105270694941282}%
\StoreBenchExecResult{JarFullInterpolationModelChecking}{KInduction}{Wrong}{}{Walltime}{Max}{7.105270694941282}%
\StoreBenchExecResult{JarFullInterpolationModelChecking}{KInduction}{Wrong}{}{Walltime}{Stdev}{0E-15}%
\StoreBenchExecResult{JarFullInterpolationModelChecking}{KInduction}{Wrong}{False}{Count}{}{1}%
\StoreBenchExecResult{JarFullInterpolationModelChecking}{KInduction}{Wrong}{False}{Cputime}{}{13.48023197}%
\StoreBenchExecResult{JarFullInterpolationModelChecking}{KInduction}{Wrong}{False}{Cputime}{Avg}{13.48023197}%
\StoreBenchExecResult{JarFullInterpolationModelChecking}{KInduction}{Wrong}{False}{Cputime}{Median}{13.48023197}%
\StoreBenchExecResult{JarFullInterpolationModelChecking}{KInduction}{Wrong}{False}{Cputime}{Min}{13.48023197}%
\StoreBenchExecResult{JarFullInterpolationModelChecking}{KInduction}{Wrong}{False}{Cputime}{Max}{13.48023197}%
\StoreBenchExecResult{JarFullInterpolationModelChecking}{KInduction}{Wrong}{False}{Cputime}{Stdev}{0E-14}%
\StoreBenchExecResult{JarFullInterpolationModelChecking}{KInduction}{Wrong}{False}{Walltime}{}{7.105270694941282}%
\StoreBenchExecResult{JarFullInterpolationModelChecking}{KInduction}{Wrong}{False}{Walltime}{Avg}{7.105270694941282}%
\StoreBenchExecResult{JarFullInterpolationModelChecking}{KInduction}{Wrong}{False}{Walltime}{Median}{7.105270694941282}%
\StoreBenchExecResult{JarFullInterpolationModelChecking}{KInduction}{Wrong}{False}{Walltime}{Min}{7.105270694941282}%
\StoreBenchExecResult{JarFullInterpolationModelChecking}{KInduction}{Wrong}{False}{Walltime}{Max}{7.105270694941282}%
\StoreBenchExecResult{JarFullInterpolationModelChecking}{KInduction}{Wrong}{False}{Walltime}{Stdev}{0E-15}%
\providecommand\StoreBenchExecResult[7]{\expandafter\newcommand\csname#1#2#3#4#5#6\endcsname{#7}}%
\StoreBenchExecResult{JarFullInterpolationModelChecking}{PredicateAbstraction}{Total}{}{Count}{}{6024}%
\StoreBenchExecResult{JarFullInterpolationModelChecking}{PredicateAbstraction}{Total}{}{Cputime}{}{1937227.217203822}%
\StoreBenchExecResult{JarFullInterpolationModelChecking}{PredicateAbstraction}{Total}{}{Cputime}{Avg}{321.5848634136490703851261620}%
\StoreBenchExecResult{JarFullInterpolationModelChecking}{PredicateAbstraction}{Total}{}{Cputime}{Median}{23.6163748155}%
\StoreBenchExecResult{JarFullInterpolationModelChecking}{PredicateAbstraction}{Total}{}{Cputime}{Min}{3.858933994}%
\StoreBenchExecResult{JarFullInterpolationModelChecking}{PredicateAbstraction}{Total}{}{Cputime}{Max}{962.707443739}%
\StoreBenchExecResult{JarFullInterpolationModelChecking}{PredicateAbstraction}{Total}{}{Cputime}{Stdev}{412.3630326546115295191509429}%
\StoreBenchExecResult{JarFullInterpolationModelChecking}{PredicateAbstraction}{Total}{}{Walltime}{}{1846024.5507616293630601}%
\StoreBenchExecResult{JarFullInterpolationModelChecking}{PredicateAbstraction}{Total}{}{Walltime}{Avg}{306.4449785460872116633632138}%
\StoreBenchExecResult{JarFullInterpolationModelChecking}{PredicateAbstraction}{Total}{}{Walltime}{Median}{13.540833414532244}%
\StoreBenchExecResult{JarFullInterpolationModelChecking}{PredicateAbstraction}{Total}{}{Walltime}{Min}{2.0658672822173685}%
\StoreBenchExecResult{JarFullInterpolationModelChecking}{PredicateAbstraction}{Total}{}{Walltime}{Max}{907.2127315518446}%
\StoreBenchExecResult{JarFullInterpolationModelChecking}{PredicateAbstraction}{Total}{}{Walltime}{Stdev}{401.7900146558463500482511659}%
\StoreBenchExecResult{JarFullInterpolationModelChecking}{PredicateAbstraction}{Correct}{}{Count}{}{2346}%
\StoreBenchExecResult{JarFullInterpolationModelChecking}{PredicateAbstraction}{Correct}{}{Cputime}{}{124647.716309609}%
\StoreBenchExecResult{JarFullInterpolationModelChecking}{PredicateAbstraction}{Correct}{}{Cputime}{Avg}{53.13201888730136402387041773}%
\StoreBenchExecResult{JarFullInterpolationModelChecking}{PredicateAbstraction}{Correct}{}{Cputime}{Median}{9.581011784}%
\StoreBenchExecResult{JarFullInterpolationModelChecking}{PredicateAbstraction}{Correct}{}{Cputime}{Min}{3.858933994}%
\StoreBenchExecResult{JarFullInterpolationModelChecking}{PredicateAbstraction}{Correct}{}{Cputime}{Max}{897.662391862}%
\StoreBenchExecResult{JarFullInterpolationModelChecking}{PredicateAbstraction}{Correct}{}{Cputime}{Stdev}{113.8968846279495281603398551}%
\StoreBenchExecResult{JarFullInterpolationModelChecking}{PredicateAbstraction}{Correct}{}{Walltime}{}{106761.1445880443789685}%
\StoreBenchExecResult{JarFullInterpolationModelChecking}{PredicateAbstraction}{Correct}{}{Walltime}{Avg}{45.50773426600357159782608696}%
\StoreBenchExecResult{JarFullInterpolationModelChecking}{PredicateAbstraction}{Correct}{}{Walltime}{Median}{5.1995412154356015}%
\StoreBenchExecResult{JarFullInterpolationModelChecking}{PredicateAbstraction}{Correct}{}{Walltime}{Min}{2.0658672822173685}%
\StoreBenchExecResult{JarFullInterpolationModelChecking}{PredicateAbstraction}{Correct}{}{Walltime}{Max}{887.8194596508984}%
\StoreBenchExecResult{JarFullInterpolationModelChecking}{PredicateAbstraction}{Correct}{}{Walltime}{Stdev}{110.0561491221951493502099315}%
\StoreBenchExecResult{JarFullInterpolationModelChecking}{PredicateAbstraction}{Correct}{False}{Count}{}{815}%
\StoreBenchExecResult{JarFullInterpolationModelChecking}{PredicateAbstraction}{Correct}{False}{Cputime}{}{53085.129913634}%
\StoreBenchExecResult{JarFullInterpolationModelChecking}{PredicateAbstraction}{Correct}{False}{Cputime}{Avg}{65.13512872838527607361963190}%
\StoreBenchExecResult{JarFullInterpolationModelChecking}{PredicateAbstraction}{Correct}{False}{Cputime}{Median}{13.812731988}%
\StoreBenchExecResult{JarFullInterpolationModelChecking}{PredicateAbstraction}{Correct}{False}{Cputime}{Min}{3.858933994}%
\StoreBenchExecResult{JarFullInterpolationModelChecking}{PredicateAbstraction}{Correct}{False}{Cputime}{Max}{865.447987348}%
\StoreBenchExecResult{JarFullInterpolationModelChecking}{PredicateAbstraction}{Correct}{False}{Cputime}{Stdev}{114.8655532236740604870293191}%
\StoreBenchExecResult{JarFullInterpolationModelChecking}{PredicateAbstraction}{Correct}{False}{Walltime}{}{46305.4746082776694896}%
\StoreBenchExecResult{JarFullInterpolationModelChecking}{PredicateAbstraction}{Correct}{False}{Walltime}{Avg}{56.81653326169039201177914110}%
\StoreBenchExecResult{JarFullInterpolationModelChecking}{PredicateAbstraction}{Correct}{False}{Walltime}{Median}{7.591952315997332}%
\StoreBenchExecResult{JarFullInterpolationModelChecking}{PredicateAbstraction}{Correct}{False}{Walltime}{Min}{2.0658672822173685}%
\StoreBenchExecResult{JarFullInterpolationModelChecking}{PredicateAbstraction}{Correct}{False}{Walltime}{Max}{847.108346252935}%
\StoreBenchExecResult{JarFullInterpolationModelChecking}{PredicateAbstraction}{Correct}{False}{Walltime}{Stdev}{111.6463642549890001999417471}%
\StoreBenchExecResult{JarFullInterpolationModelChecking}{PredicateAbstraction}{Correct}{True}{Count}{}{1531}%
\StoreBenchExecResult{JarFullInterpolationModelChecking}{PredicateAbstraction}{Correct}{True}{Cputime}{}{71562.586395975}%
\StoreBenchExecResult{JarFullInterpolationModelChecking}{PredicateAbstraction}{Correct}{True}{Cputime}{Avg}{46.74238170867080339647289353}%
\StoreBenchExecResult{JarFullInterpolationModelChecking}{PredicateAbstraction}{Correct}{True}{Cputime}{Median}{8.524202585}%
\StoreBenchExecResult{JarFullInterpolationModelChecking}{PredicateAbstraction}{Correct}{True}{Cputime}{Min}{3.87526044}%
\StoreBenchExecResult{JarFullInterpolationModelChecking}{PredicateAbstraction}{Correct}{True}{Cputime}{Max}{897.662391862}%
\StoreBenchExecResult{JarFullInterpolationModelChecking}{PredicateAbstraction}{Correct}{True}{Cputime}{Stdev}{112.8583859805265812855655868}%
\StoreBenchExecResult{JarFullInterpolationModelChecking}{PredicateAbstraction}{Correct}{True}{Walltime}{}{60455.6699797667094789}%
\StoreBenchExecResult{JarFullInterpolationModelChecking}{PredicateAbstraction}{Correct}{True}{Walltime}{Avg}{39.48770083590248822919660353}%
\StoreBenchExecResult{JarFullInterpolationModelChecking}{PredicateAbstraction}{Correct}{True}{Walltime}{Median}{4.576883363071829}%
\StoreBenchExecResult{JarFullInterpolationModelChecking}{PredicateAbstraction}{Correct}{True}{Walltime}{Min}{2.0672398840542883}%
\StoreBenchExecResult{JarFullInterpolationModelChecking}{PredicateAbstraction}{Correct}{True}{Walltime}{Max}{887.8194596508984}%
\StoreBenchExecResult{JarFullInterpolationModelChecking}{PredicateAbstraction}{Correct}{True}{Walltime}{Stdev}{108.7214777869160962740831793}%
\StoreBenchExecResult{JarFullInterpolationModelChecking}{PredicateAbstraction}{Error}{}{Count}{}{3675}%
\StoreBenchExecResult{JarFullInterpolationModelChecking}{PredicateAbstraction}{Error}{}{Cputime}{}{1812516.757597624}%
\StoreBenchExecResult{JarFullInterpolationModelChecking}{PredicateAbstraction}{Error}{}{Cputime}{Avg}{493.2018388020745578231292517}%
\StoreBenchExecResult{JarFullInterpolationModelChecking}{PredicateAbstraction}{Error}{}{Cputime}{Median}{900.5492323}%
\StoreBenchExecResult{JarFullInterpolationModelChecking}{PredicateAbstraction}{Error}{}{Cputime}{Min}{3.946147536}%
\StoreBenchExecResult{JarFullInterpolationModelChecking}{PredicateAbstraction}{Error}{}{Cputime}{Max}{962.707443739}%
\StoreBenchExecResult{JarFullInterpolationModelChecking}{PredicateAbstraction}{Error}{}{Cputime}{Stdev}{441.4966094859310546238674617}%
\StoreBenchExecResult{JarFullInterpolationModelChecking}{PredicateAbstraction}{Error}{}{Walltime}{}{1739218.9638068859944486}%
\StoreBenchExecResult{JarFullInterpolationModelChecking}{PredicateAbstraction}{Error}{}{Walltime}{Avg}{473.2568608998329236594829932}%
\StoreBenchExecResult{JarFullInterpolationModelChecking}{PredicateAbstraction}{Error}{}{Walltime}{Median}{807.636398348026}%
\StoreBenchExecResult{JarFullInterpolationModelChecking}{PredicateAbstraction}{Error}{}{Walltime}{Min}{2.089004497975111}%
\StoreBenchExecResult{JarFullInterpolationModelChecking}{PredicateAbstraction}{Error}{}{Walltime}{Max}{907.2127315518446}%
\StoreBenchExecResult{JarFullInterpolationModelChecking}{PredicateAbstraction}{Error}{}{Walltime}{Stdev}{430.7304222591135109351801617}%
\StoreBenchExecResult{JarFullInterpolationModelChecking}{PredicateAbstraction}{Error}{Error}{Count}{}{1698}%
\StoreBenchExecResult{JarFullInterpolationModelChecking}{PredicateAbstraction}{Error}{Error}{Cputime}{}{36212.127744584}%
\StoreBenchExecResult{JarFullInterpolationModelChecking}{PredicateAbstraction}{Error}{Error}{Cputime}{Avg}{21.32634142790577149587750294}%
\StoreBenchExecResult{JarFullInterpolationModelChecking}{PredicateAbstraction}{Error}{Error}{Cputime}{Median}{11.4305151035}%
\StoreBenchExecResult{JarFullInterpolationModelChecking}{PredicateAbstraction}{Error}{Error}{Cputime}{Min}{3.946147536}%
\StoreBenchExecResult{JarFullInterpolationModelChecking}{PredicateAbstraction}{Error}{Error}{Cputime}{Max}{778.893130331}%
\StoreBenchExecResult{JarFullInterpolationModelChecking}{PredicateAbstraction}{Error}{Error}{Cputime}{Stdev}{45.27702677310657464997133582}%
\StoreBenchExecResult{JarFullInterpolationModelChecking}{PredicateAbstraction}{Error}{Error}{Walltime}{}{24657.1899758535436996}%
\StoreBenchExecResult{JarFullInterpolationModelChecking}{PredicateAbstraction}{Error}{Error}{Walltime}{Avg}{14.52131329555567944617196702}%
\StoreBenchExecResult{JarFullInterpolationModelChecking}{PredicateAbstraction}{Error}{Error}{Walltime}{Median}{6.319988669361919}%
\StoreBenchExecResult{JarFullInterpolationModelChecking}{PredicateAbstraction}{Error}{Error}{Walltime}{Min}{2.089004497975111}%
\StoreBenchExecResult{JarFullInterpolationModelChecking}{PredicateAbstraction}{Error}{Error}{Walltime}{Max}{769.5787020390853}%
\StoreBenchExecResult{JarFullInterpolationModelChecking}{PredicateAbstraction}{Error}{Error}{Walltime}{Stdev}{42.85272531625917472808514249}%
\StoreBenchExecResult{JarFullInterpolationModelChecking}{PredicateAbstraction}{Error}{Exception}{Count}{}{1}%
\StoreBenchExecResult{JarFullInterpolationModelChecking}{PredicateAbstraction}{Error}{Exception}{Cputime}{}{111.496480572}%
\StoreBenchExecResult{JarFullInterpolationModelChecking}{PredicateAbstraction}{Error}{Exception}{Cputime}{Avg}{111.496480572}%
\StoreBenchExecResult{JarFullInterpolationModelChecking}{PredicateAbstraction}{Error}{Exception}{Cputime}{Median}{111.496480572}%
\StoreBenchExecResult{JarFullInterpolationModelChecking}{PredicateAbstraction}{Error}{Exception}{Cputime}{Min}{111.496480572}%
\StoreBenchExecResult{JarFullInterpolationModelChecking}{PredicateAbstraction}{Error}{Exception}{Cputime}{Max}{111.496480572}%
\StoreBenchExecResult{JarFullInterpolationModelChecking}{PredicateAbstraction}{Error}{Exception}{Cputime}{Stdev}{0E-14}%
\StoreBenchExecResult{JarFullInterpolationModelChecking}{PredicateAbstraction}{Error}{Exception}{Walltime}{}{79.78363170009106}%
\StoreBenchExecResult{JarFullInterpolationModelChecking}{PredicateAbstraction}{Error}{Exception}{Walltime}{Avg}{79.78363170009106}%
\StoreBenchExecResult{JarFullInterpolationModelChecking}{PredicateAbstraction}{Error}{Exception}{Walltime}{Median}{79.78363170009106}%
\StoreBenchExecResult{JarFullInterpolationModelChecking}{PredicateAbstraction}{Error}{Exception}{Walltime}{Min}{79.78363170009106}%
\StoreBenchExecResult{JarFullInterpolationModelChecking}{PredicateAbstraction}{Error}{Exception}{Walltime}{Max}{79.78363170009106}%
\StoreBenchExecResult{JarFullInterpolationModelChecking}{PredicateAbstraction}{Error}{Exception}{Walltime}{Stdev}{0E-14}%
\StoreBenchExecResult{JarFullInterpolationModelChecking}{PredicateAbstraction}{Error}{OutOfJavaMemory}{Count}{}{4}%
\StoreBenchExecResult{JarFullInterpolationModelChecking}{PredicateAbstraction}{Error}{OutOfJavaMemory}{Cputime}{}{682.389509271}%
\StoreBenchExecResult{JarFullInterpolationModelChecking}{PredicateAbstraction}{Error}{OutOfJavaMemory}{Cputime}{Avg}{170.59737731775}%
\StoreBenchExecResult{JarFullInterpolationModelChecking}{PredicateAbstraction}{Error}{OutOfJavaMemory}{Cputime}{Median}{69.0363401185}%
\StoreBenchExecResult{JarFullInterpolationModelChecking}{PredicateAbstraction}{Error}{OutOfJavaMemory}{Cputime}{Min}{55.870619452}%
\StoreBenchExecResult{JarFullInterpolationModelChecking}{PredicateAbstraction}{Error}{OutOfJavaMemory}{Cputime}{Max}{488.446209582}%
\StoreBenchExecResult{JarFullInterpolationModelChecking}{PredicateAbstraction}{Error}{OutOfJavaMemory}{Cputime}{Stdev}{183.6176600683676963107308023}%
\StoreBenchExecResult{JarFullInterpolationModelChecking}{PredicateAbstraction}{Error}{OutOfJavaMemory}{Walltime}{}{364.626009460538633}%
\StoreBenchExecResult{JarFullInterpolationModelChecking}{PredicateAbstraction}{Error}{OutOfJavaMemory}{Walltime}{Avg}{91.15650236513465825}%
\StoreBenchExecResult{JarFullInterpolationModelChecking}{PredicateAbstraction}{Error}{OutOfJavaMemory}{Walltime}{Median}{37.783904478885235}%
\StoreBenchExecResult{JarFullInterpolationModelChecking}{PredicateAbstraction}{Error}{OutOfJavaMemory}{Walltime}{Min}{30.952984901843593}%
\StoreBenchExecResult{JarFullInterpolationModelChecking}{PredicateAbstraction}{Error}{OutOfJavaMemory}{Walltime}{Max}{258.10521560092457}%
\StoreBenchExecResult{JarFullInterpolationModelChecking}{PredicateAbstraction}{Error}{OutOfJavaMemory}{Walltime}{Stdev}{96.44585216494741603838259192}%
\StoreBenchExecResult{JarFullInterpolationModelChecking}{PredicateAbstraction}{Error}{OutOfMemory}{Count}{}{13}%
\StoreBenchExecResult{JarFullInterpolationModelChecking}{PredicateAbstraction}{Error}{OutOfMemory}{Cputime}{}{5966.100052786}%
\StoreBenchExecResult{JarFullInterpolationModelChecking}{PredicateAbstraction}{Error}{OutOfMemory}{Cputime}{Avg}{458.9307732912307692307692308}%
\StoreBenchExecResult{JarFullInterpolationModelChecking}{PredicateAbstraction}{Error}{OutOfMemory}{Cputime}{Median}{625.628779098}%
\StoreBenchExecResult{JarFullInterpolationModelChecking}{PredicateAbstraction}{Error}{OutOfMemory}{Cputime}{Min}{88.157027984}%
\StoreBenchExecResult{JarFullInterpolationModelChecking}{PredicateAbstraction}{Error}{OutOfMemory}{Cputime}{Max}{843.080218519}%
\StoreBenchExecResult{JarFullInterpolationModelChecking}{PredicateAbstraction}{Error}{OutOfMemory}{Cputime}{Stdev}{304.2718832231923363827462242}%
\StoreBenchExecResult{JarFullInterpolationModelChecking}{PredicateAbstraction}{Error}{OutOfMemory}{Walltime}{}{4983.13259506970633}%
\StoreBenchExecResult{JarFullInterpolationModelChecking}{PredicateAbstraction}{Error}{OutOfMemory}{Walltime}{Avg}{383.3178919284389484615384615}%
\StoreBenchExecResult{JarFullInterpolationModelChecking}{PredicateAbstraction}{Error}{OutOfMemory}{Walltime}{Median}{407.51882332889363}%
\StoreBenchExecResult{JarFullInterpolationModelChecking}{PredicateAbstraction}{Error}{OutOfMemory}{Walltime}{Min}{76.30541277583688}%
\StoreBenchExecResult{JarFullInterpolationModelChecking}{PredicateAbstraction}{Error}{OutOfMemory}{Walltime}{Max}{744.1892503770068}%
\StoreBenchExecResult{JarFullInterpolationModelChecking}{PredicateAbstraction}{Error}{OutOfMemory}{Walltime}{Stdev}{267.2300782110412447878719262}%
\StoreBenchExecResult{JarFullInterpolationModelChecking}{PredicateAbstraction}{Error}{OutOfNativeMemory}{Count}{}{2}%
\StoreBenchExecResult{JarFullInterpolationModelChecking}{PredicateAbstraction}{Error}{OutOfNativeMemory}{Cputime}{}{24.328657463}%
\StoreBenchExecResult{JarFullInterpolationModelChecking}{PredicateAbstraction}{Error}{OutOfNativeMemory}{Cputime}{Avg}{12.1643287315}%
\StoreBenchExecResult{JarFullInterpolationModelChecking}{PredicateAbstraction}{Error}{OutOfNativeMemory}{Cputime}{Median}{12.1643287315}%
\StoreBenchExecResult{JarFullInterpolationModelChecking}{PredicateAbstraction}{Error}{OutOfNativeMemory}{Cputime}{Min}{11.753098228}%
\StoreBenchExecResult{JarFullInterpolationModelChecking}{PredicateAbstraction}{Error}{OutOfNativeMemory}{Cputime}{Max}{12.575559235}%
\StoreBenchExecResult{JarFullInterpolationModelChecking}{PredicateAbstraction}{Error}{OutOfNativeMemory}{Cputime}{Stdev}{0.41123050350000}%
\StoreBenchExecResult{JarFullInterpolationModelChecking}{PredicateAbstraction}{Error}{OutOfNativeMemory}{Walltime}{}{14.969014051835984}%
\StoreBenchExecResult{JarFullInterpolationModelChecking}{PredicateAbstraction}{Error}{OutOfNativeMemory}{Walltime}{Avg}{7.484507025917992}%
\StoreBenchExecResult{JarFullInterpolationModelChecking}{PredicateAbstraction}{Error}{OutOfNativeMemory}{Walltime}{Median}{7.484507025917992}%
\StoreBenchExecResult{JarFullInterpolationModelChecking}{PredicateAbstraction}{Error}{OutOfNativeMemory}{Walltime}{Min}{7.215110598830506}%
\StoreBenchExecResult{JarFullInterpolationModelChecking}{PredicateAbstraction}{Error}{OutOfNativeMemory}{Walltime}{Max}{7.753903453005478}%
\StoreBenchExecResult{JarFullInterpolationModelChecking}{PredicateAbstraction}{Error}{OutOfNativeMemory}{Walltime}{Stdev}{0.2693964270874860000000000000}%
\StoreBenchExecResult{JarFullInterpolationModelChecking}{PredicateAbstraction}{Error}{SegmentationFault}{Count}{}{2}%
\StoreBenchExecResult{JarFullInterpolationModelChecking}{PredicateAbstraction}{Error}{SegmentationFault}{Cputime}{}{20.320702093}%
\StoreBenchExecResult{JarFullInterpolationModelChecking}{PredicateAbstraction}{Error}{SegmentationFault}{Cputime}{Avg}{10.1603510465}%
\StoreBenchExecResult{JarFullInterpolationModelChecking}{PredicateAbstraction}{Error}{SegmentationFault}{Cputime}{Median}{10.1603510465}%
\StoreBenchExecResult{JarFullInterpolationModelChecking}{PredicateAbstraction}{Error}{SegmentationFault}{Cputime}{Min}{7.232691805}%
\StoreBenchExecResult{JarFullInterpolationModelChecking}{PredicateAbstraction}{Error}{SegmentationFault}{Cputime}{Max}{13.088010288}%
\StoreBenchExecResult{JarFullInterpolationModelChecking}{PredicateAbstraction}{Error}{SegmentationFault}{Cputime}{Stdev}{2.9276592415000}%
\StoreBenchExecResult{JarFullInterpolationModelChecking}{PredicateAbstraction}{Error}{SegmentationFault}{Walltime}{}{11.595300039742142}%
\StoreBenchExecResult{JarFullInterpolationModelChecking}{PredicateAbstraction}{Error}{SegmentationFault}{Walltime}{Avg}{5.797650019871071}%
\StoreBenchExecResult{JarFullInterpolationModelChecking}{PredicateAbstraction}{Error}{SegmentationFault}{Walltime}{Median}{5.797650019871071}%
\StoreBenchExecResult{JarFullInterpolationModelChecking}{PredicateAbstraction}{Error}{SegmentationFault}{Walltime}{Min}{4.011805110843852}%
\StoreBenchExecResult{JarFullInterpolationModelChecking}{PredicateAbstraction}{Error}{SegmentationFault}{Walltime}{Max}{7.58349492889829}%
\StoreBenchExecResult{JarFullInterpolationModelChecking}{PredicateAbstraction}{Error}{SegmentationFault}{Walltime}{Stdev}{1.785844909027219000000000000}%
\StoreBenchExecResult{JarFullInterpolationModelChecking}{PredicateAbstraction}{Error}{Timeout}{Count}{}{1955}%
\StoreBenchExecResult{JarFullInterpolationModelChecking}{PredicateAbstraction}{Error}{Timeout}{Cputime}{}{1769499.994450855}%
\StoreBenchExecResult{JarFullInterpolationModelChecking}{PredicateAbstraction}{Error}{Timeout}{Cputime}{Avg}{905.1150866756291560102301790}%
\StoreBenchExecResult{JarFullInterpolationModelChecking}{PredicateAbstraction}{Error}{Timeout}{Cputime}{Median}{901.301634473}%
\StoreBenchExecResult{JarFullInterpolationModelChecking}{PredicateAbstraction}{Error}{Timeout}{Cputime}{Min}{900.012567316}%
\StoreBenchExecResult{JarFullInterpolationModelChecking}{PredicateAbstraction}{Error}{Timeout}{Cputime}{Max}{962.707443739}%
\StoreBenchExecResult{JarFullInterpolationModelChecking}{PredicateAbstraction}{Error}{Timeout}{Cputime}{Stdev}{10.23318339379310373367573565}%
\StoreBenchExecResult{JarFullInterpolationModelChecking}{PredicateAbstraction}{Error}{Timeout}{Walltime}{}{1709107.6672807105366}%
\StoreBenchExecResult{JarFullInterpolationModelChecking}{PredicateAbstraction}{Error}{Timeout}{Walltime}{Avg}{874.2238707318212463427109974}%
\StoreBenchExecResult{JarFullInterpolationModelChecking}{PredicateAbstraction}{Error}{Timeout}{Walltime}{Median}{889.9597784150392}%
\StoreBenchExecResult{JarFullInterpolationModelChecking}{PredicateAbstraction}{Error}{Timeout}{Walltime}{Min}{537.5722174451221}%
\StoreBenchExecResult{JarFullInterpolationModelChecking}{PredicateAbstraction}{Error}{Timeout}{Walltime}{Max}{907.2127315518446}%
\StoreBenchExecResult{JarFullInterpolationModelChecking}{PredicateAbstraction}{Error}{Timeout}{Walltime}{Stdev}{47.33773985094801168889278763}%
\StoreBenchExecResult{JarFullInterpolationModelChecking}{PredicateAbstraction}{Wrong}{}{Count}{}{3}%
\StoreBenchExecResult{JarFullInterpolationModelChecking}{PredicateAbstraction}{Wrong}{}{Cputime}{}{62.743296589}%
\StoreBenchExecResult{JarFullInterpolationModelChecking}{PredicateAbstraction}{Wrong}{}{Cputime}{Avg}{20.91443219633333333333333333}%
\StoreBenchExecResult{JarFullInterpolationModelChecking}{PredicateAbstraction}{Wrong}{}{Cputime}{Median}{8.527048792}%
\StoreBenchExecResult{JarFullInterpolationModelChecking}{PredicateAbstraction}{Wrong}{}{Cputime}{Min}{5.480026741}%
\StoreBenchExecResult{JarFullInterpolationModelChecking}{PredicateAbstraction}{Wrong}{}{Cputime}{Max}{48.736221056}%
\StoreBenchExecResult{JarFullInterpolationModelChecking}{PredicateAbstraction}{Wrong}{}{Cputime}{Stdev}{19.71226415811104877690487025}%
\StoreBenchExecResult{JarFullInterpolationModelChecking}{PredicateAbstraction}{Wrong}{}{Walltime}{}{44.442366698989643}%
\StoreBenchExecResult{JarFullInterpolationModelChecking}{PredicateAbstraction}{Wrong}{}{Walltime}{Avg}{14.81412223299654766666666667}%
\StoreBenchExecResult{JarFullInterpolationModelChecking}{PredicateAbstraction}{Wrong}{}{Walltime}{Median}{4.791924223070964}%
\StoreBenchExecResult{JarFullInterpolationModelChecking}{PredicateAbstraction}{Wrong}{}{Walltime}{Min}{2.900021356996149}%
\StoreBenchExecResult{JarFullInterpolationModelChecking}{PredicateAbstraction}{Wrong}{}{Walltime}{Max}{36.75042111892253}%
\StoreBenchExecResult{JarFullInterpolationModelChecking}{PredicateAbstraction}{Wrong}{}{Walltime}{Stdev}{15.53052329496495416580009669}%
\StoreBenchExecResult{JarFullInterpolationModelChecking}{PredicateAbstraction}{Wrong}{False}{Count}{}{2}%
\StoreBenchExecResult{JarFullInterpolationModelChecking}{PredicateAbstraction}{Wrong}{False}{Cputime}{}{14.007075533}%
\StoreBenchExecResult{JarFullInterpolationModelChecking}{PredicateAbstraction}{Wrong}{False}{Cputime}{Avg}{7.0035377665}%
\StoreBenchExecResult{JarFullInterpolationModelChecking}{PredicateAbstraction}{Wrong}{False}{Cputime}{Median}{7.0035377665}%
\StoreBenchExecResult{JarFullInterpolationModelChecking}{PredicateAbstraction}{Wrong}{False}{Cputime}{Min}{5.480026741}%
\StoreBenchExecResult{JarFullInterpolationModelChecking}{PredicateAbstraction}{Wrong}{False}{Cputime}{Max}{8.527048792}%
\StoreBenchExecResult{JarFullInterpolationModelChecking}{PredicateAbstraction}{Wrong}{False}{Cputime}{Stdev}{1.52351102550000}%
\StoreBenchExecResult{JarFullInterpolationModelChecking}{PredicateAbstraction}{Wrong}{False}{Walltime}{}{7.691945580067113}%
\StoreBenchExecResult{JarFullInterpolationModelChecking}{PredicateAbstraction}{Wrong}{False}{Walltime}{Avg}{3.8459727900335565}%
\StoreBenchExecResult{JarFullInterpolationModelChecking}{PredicateAbstraction}{Wrong}{False}{Walltime}{Median}{3.8459727900335565}%
\StoreBenchExecResult{JarFullInterpolationModelChecking}{PredicateAbstraction}{Wrong}{False}{Walltime}{Min}{2.900021356996149}%
\StoreBenchExecResult{JarFullInterpolationModelChecking}{PredicateAbstraction}{Wrong}{False}{Walltime}{Max}{4.791924223070964}%
\StoreBenchExecResult{JarFullInterpolationModelChecking}{PredicateAbstraction}{Wrong}{False}{Walltime}{Stdev}{0.9459514330374075000000000000}%
\StoreBenchExecResult{JarFullInterpolationModelChecking}{PredicateAbstraction}{Wrong}{True}{Count}{}{1}%
\StoreBenchExecResult{JarFullInterpolationModelChecking}{PredicateAbstraction}{Wrong}{True}{Cputime}{}{48.736221056}%
\StoreBenchExecResult{JarFullInterpolationModelChecking}{PredicateAbstraction}{Wrong}{True}{Cputime}{Avg}{48.736221056}%
\StoreBenchExecResult{JarFullInterpolationModelChecking}{PredicateAbstraction}{Wrong}{True}{Cputime}{Median}{48.736221056}%
\StoreBenchExecResult{JarFullInterpolationModelChecking}{PredicateAbstraction}{Wrong}{True}{Cputime}{Min}{48.736221056}%
\StoreBenchExecResult{JarFullInterpolationModelChecking}{PredicateAbstraction}{Wrong}{True}{Cputime}{Max}{48.736221056}%
\StoreBenchExecResult{JarFullInterpolationModelChecking}{PredicateAbstraction}{Wrong}{True}{Cputime}{Stdev}{0E-14}%
\StoreBenchExecResult{JarFullInterpolationModelChecking}{PredicateAbstraction}{Wrong}{True}{Walltime}{}{36.75042111892253}%
\StoreBenchExecResult{JarFullInterpolationModelChecking}{PredicateAbstraction}{Wrong}{True}{Walltime}{Avg}{36.75042111892253}%
\StoreBenchExecResult{JarFullInterpolationModelChecking}{PredicateAbstraction}{Wrong}{True}{Walltime}{Median}{36.75042111892253}%
\StoreBenchExecResult{JarFullInterpolationModelChecking}{PredicateAbstraction}{Wrong}{True}{Walltime}{Min}{36.75042111892253}%
\StoreBenchExecResult{JarFullInterpolationModelChecking}{PredicateAbstraction}{Wrong}{True}{Walltime}{Max}{36.75042111892253}%
\StoreBenchExecResult{JarFullInterpolationModelChecking}{PredicateAbstraction}{Wrong}{True}{Walltime}{Stdev}{0E-14}%
\providecommand\StoreBenchExecResult[7]{\expandafter\newcommand\csname#1#2#3#4#5#6\endcsname{#7}}%
\StoreBenchExecResult{JarFullInterpolationModelChecking}{Impact}{Total}{}{Count}{}{6024}%
\StoreBenchExecResult{JarFullInterpolationModelChecking}{Impact}{Total}{}{Cputime}{}{1788194.997935309}%
\StoreBenchExecResult{JarFullInterpolationModelChecking}{Impact}{Total}{}{Cputime}{Avg}{296.8451191791681606905710491}%
\StoreBenchExecResult{JarFullInterpolationModelChecking}{Impact}{Total}{}{Cputime}{Median}{19.1878772365}%
\StoreBenchExecResult{JarFullInterpolationModelChecking}{Impact}{Total}{}{Cputime}{Min}{3.542792912}%
\StoreBenchExecResult{JarFullInterpolationModelChecking}{Impact}{Total}{}{Cputime}{Max}{962.724753477}%
\StoreBenchExecResult{JarFullInterpolationModelChecking}{Impact}{Total}{}{Cputime}{Stdev}{400.2950117154691729191229698}%
\StoreBenchExecResult{JarFullInterpolationModelChecking}{Impact}{Total}{}{Walltime}{}{1712121.8609541309990782}%
\StoreBenchExecResult{JarFullInterpolationModelChecking}{Impact}{Total}{}{Walltime}{Avg}{284.2167763868079347739375830}%
\StoreBenchExecResult{JarFullInterpolationModelChecking}{Impact}{Total}{}{Walltime}{Median}{11.273857292602770}%
\StoreBenchExecResult{JarFullInterpolationModelChecking}{Impact}{Total}{}{Walltime}{Min}{1.8617133740335703}%
\StoreBenchExecResult{JarFullInterpolationModelChecking}{Impact}{Total}{}{Walltime}{Max}{908.7584513910115}%
\StoreBenchExecResult{JarFullInterpolationModelChecking}{Impact}{Total}{}{Walltime}{Stdev}{391.8403539596655636001566450}%
\StoreBenchExecResult{JarFullInterpolationModelChecking}{Impact}{Correct}{}{Count}{}{2413}%
\StoreBenchExecResult{JarFullInterpolationModelChecking}{Impact}{Correct}{}{Cputime}{}{144291.836490444}%
\StoreBenchExecResult{JarFullInterpolationModelChecking}{Impact}{Correct}{}{Cputime}{Avg}{59.79769435990219643597181931}%
\StoreBenchExecResult{JarFullInterpolationModelChecking}{Impact}{Correct}{}{Cputime}{Median}{9.184734811}%
\StoreBenchExecResult{JarFullInterpolationModelChecking}{Impact}{Correct}{}{Cputime}{Min}{3.542792912}%
\StoreBenchExecResult{JarFullInterpolationModelChecking}{Impact}{Correct}{}{Cputime}{Max}{899.22383568}%
\StoreBenchExecResult{JarFullInterpolationModelChecking}{Impact}{Correct}{}{Cputime}{Stdev}{132.5338735738400656829078029}%
\StoreBenchExecResult{JarFullInterpolationModelChecking}{Impact}{Correct}{}{Walltime}{}{126536.1317699567879398}%
\StoreBenchExecResult{JarFullInterpolationModelChecking}{Impact}{Correct}{}{Walltime}{Avg}{52.43934180271727639444674679}%
\StoreBenchExecResult{JarFullInterpolationModelChecking}{Impact}{Correct}{}{Walltime}{Median}{5.021363538922742}%
\StoreBenchExecResult{JarFullInterpolationModelChecking}{Impact}{Correct}{}{Walltime}{Min}{1.8617133740335703}%
\StoreBenchExecResult{JarFullInterpolationModelChecking}{Impact}{Correct}{}{Walltime}{Max}{885.4205059420783}%
\StoreBenchExecResult{JarFullInterpolationModelChecking}{Impact}{Correct}{}{Walltime}{Stdev}{126.3386564004298718297004698}%
\StoreBenchExecResult{JarFullInterpolationModelChecking}{Impact}{Correct}{False}{Count}{}{870}%
\StoreBenchExecResult{JarFullInterpolationModelChecking}{Impact}{Correct}{False}{Cputime}{}{77729.684372864}%
\StoreBenchExecResult{JarFullInterpolationModelChecking}{Impact}{Correct}{False}{Cputime}{Avg}{89.34446479639540229885057471}%
\StoreBenchExecResult{JarFullInterpolationModelChecking}{Impact}{Correct}{False}{Cputime}{Median}{12.934071658}%
\StoreBenchExecResult{JarFullInterpolationModelChecking}{Impact}{Correct}{False}{Cputime}{Min}{3.753468188}%
\StoreBenchExecResult{JarFullInterpolationModelChecking}{Impact}{Correct}{False}{Cputime}{Max}{872.099601231}%
\StoreBenchExecResult{JarFullInterpolationModelChecking}{Impact}{Correct}{False}{Cputime}{Stdev}{161.8012272632416741695650928}%
\StoreBenchExecResult{JarFullInterpolationModelChecking}{Impact}{Correct}{False}{Walltime}{}{69772.9733507966157022}%
\StoreBenchExecResult{JarFullInterpolationModelChecking}{Impact}{Correct}{False}{Walltime}{Avg}{80.19881994344438586459770115}%
\StoreBenchExecResult{JarFullInterpolationModelChecking}{Impact}{Correct}{False}{Walltime}{Median}{7.808266203966923}%
\StoreBenchExecResult{JarFullInterpolationModelChecking}{Impact}{Correct}{False}{Walltime}{Min}{2.0215585799887776}%
\StoreBenchExecResult{JarFullInterpolationModelChecking}{Impact}{Correct}{False}{Walltime}{Max}{823.5899079069495}%
\StoreBenchExecResult{JarFullInterpolationModelChecking}{Impact}{Correct}{False}{Walltime}{Stdev}{152.2592604939923229709349861}%
\StoreBenchExecResult{JarFullInterpolationModelChecking}{Impact}{Correct}{True}{Count}{}{1543}%
\StoreBenchExecResult{JarFullInterpolationModelChecking}{Impact}{Correct}{True}{Cputime}{}{66562.152117580}%
\StoreBenchExecResult{JarFullInterpolationModelChecking}{Impact}{Correct}{True}{Cputime}{Avg}{43.13814135941672067401166559}%
\StoreBenchExecResult{JarFullInterpolationModelChecking}{Impact}{Correct}{True}{Cputime}{Median}{8.418050224}%
\StoreBenchExecResult{JarFullInterpolationModelChecking}{Impact}{Correct}{True}{Cputime}{Min}{3.542792912}%
\StoreBenchExecResult{JarFullInterpolationModelChecking}{Impact}{Correct}{True}{Cputime}{Max}{899.22383568}%
\StoreBenchExecResult{JarFullInterpolationModelChecking}{Impact}{Correct}{True}{Cputime}{Stdev}{109.2626692082761562444052177}%
\StoreBenchExecResult{JarFullInterpolationModelChecking}{Impact}{Correct}{True}{Walltime}{}{56763.1584191601722376}%
\StoreBenchExecResult{JarFullInterpolationModelChecking}{Impact}{Correct}{True}{Walltime}{Avg}{36.78752975966310579235255995}%
\StoreBenchExecResult{JarFullInterpolationModelChecking}{Impact}{Correct}{True}{Walltime}{Median}{4.460522146895528}%
\StoreBenchExecResult{JarFullInterpolationModelChecking}{Impact}{Correct}{True}{Walltime}{Min}{1.8617133740335703}%
\StoreBenchExecResult{JarFullInterpolationModelChecking}{Impact}{Correct}{True}{Walltime}{Max}{885.4205059420783}%
\StoreBenchExecResult{JarFullInterpolationModelChecking}{Impact}{Correct}{True}{Walltime}{Stdev}{105.8786366596537097525392561}%

\StoreBenchExecResult{JarFullInterpolationModelChecking}{Impact}{Error}{}{Count}{}{3609}%
\StoreBenchExecResult{JarFullInterpolationModelChecking}{Impact}{Error}{}{Cputime}{}{1643891.751253716}%
\StoreBenchExecResult{JarFullInterpolationModelChecking}{Impact}{Error}{}{Cputime}{Avg}{455.4978529381313383208645054}%
\StoreBenchExecResult{JarFullInterpolationModelChecking}{Impact}{Error}{}{Cputime}{Median}{242.152750242}%
\StoreBenchExecResult{JarFullInterpolationModelChecking}{Impact}{Error}{}{Cputime}{Min}{3.55656136}%
\StoreBenchExecResult{JarFullInterpolationModelChecking}{Impact}{Error}{}{Cputime}{Max}{962.724753477}%
\StoreBenchExecResult{JarFullInterpolationModelChecking}{Impact}{Error}{}{Cputime}{Stdev}{439.2355759373467958105864038}%
\StoreBenchExecResult{JarFullInterpolationModelChecking}{Impact}{Error}{}{Walltime}{}{1585579.3901756613044081}%
\StoreBenchExecResult{JarFullInterpolationModelChecking}{Impact}{Error}{}{Walltime}{Avg}{439.3403685718097269072042117}%
\StoreBenchExecResult{JarFullInterpolationModelChecking}{Impact}{Error}{}{Walltime}{Median}{223.65466798399575}%
\StoreBenchExecResult{JarFullInterpolationModelChecking}{Impact}{Error}{}{Walltime}{Min}{1.9188382199499756}%
\StoreBenchExecResult{JarFullInterpolationModelChecking}{Impact}{Error}{}{Walltime}{Max}{908.7584513910115}%
\StoreBenchExecResult{JarFullInterpolationModelChecking}{Impact}{Error}{}{Walltime}{Stdev}{430.7945153735644768212051446}%
\StoreBenchExecResult{JarFullInterpolationModelChecking}{Impact}{Error}{Error}{Count}{}{1732}%
\StoreBenchExecResult{JarFullInterpolationModelChecking}{Impact}{Error}{Error}{Cputime}{}{38511.107637577}%
\StoreBenchExecResult{JarFullInterpolationModelChecking}{Impact}{Error}{Error}{Cputime}{Avg}{22.23505059906293302540415704}%
\StoreBenchExecResult{JarFullInterpolationModelChecking}{Impact}{Error}{Error}{Cputime}{Median}{10.353599354}%
\StoreBenchExecResult{JarFullInterpolationModelChecking}{Impact}{Error}{Error}{Cputime}{Min}{3.55656136}%
\StoreBenchExecResult{JarFullInterpolationModelChecking}{Impact}{Error}{Error}{Cputime}{Max}{878.952169717}%
\StoreBenchExecResult{JarFullInterpolationModelChecking}{Impact}{Error}{Error}{Cputime}{Stdev}{60.19695511096932965727362193}%
\StoreBenchExecResult{JarFullInterpolationModelChecking}{Impact}{Error}{Error}{Walltime}{}{27814.9280648236163498}%
\StoreBenchExecResult{JarFullInterpolationModelChecking}{Impact}{Error}{Error}{Walltime}{Avg}{16.05942728915913184168591224}%
\StoreBenchExecResult{JarFullInterpolationModelChecking}{Impact}{Error}{Error}{Walltime}{Median}{5.8285026280209425}%
\StoreBenchExecResult{JarFullInterpolationModelChecking}{Impact}{Error}{Error}{Walltime}{Min}{1.9188382199499756}%
\StoreBenchExecResult{JarFullInterpolationModelChecking}{Impact}{Error}{Error}{Walltime}{Max}{872.1848439609166}%
\StoreBenchExecResult{JarFullInterpolationModelChecking}{Impact}{Error}{Error}{Walltime}{Stdev}{58.47341984484134136562182479}%
\StoreBenchExecResult{JarFullInterpolationModelChecking}{Impact}{Error}{Exception}{Count}{}{35}%
\StoreBenchExecResult{JarFullInterpolationModelChecking}{Impact}{Error}{Exception}{Cputime}{}{1675.143102395}%
\StoreBenchExecResult{JarFullInterpolationModelChecking}{Impact}{Error}{Exception}{Cputime}{Avg}{47.861231497}%
\StoreBenchExecResult{JarFullInterpolationModelChecking}{Impact}{Error}{Exception}{Cputime}{Median}{24.73801727}%
\StoreBenchExecResult{JarFullInterpolationModelChecking}{Impact}{Error}{Exception}{Cputime}{Min}{7.034022277}%
\StoreBenchExecResult{JarFullInterpolationModelChecking}{Impact}{Error}{Exception}{Cputime}{Max}{239.770063879}%
\StoreBenchExecResult{JarFullInterpolationModelChecking}{Impact}{Error}{Exception}{Cputime}{Stdev}{51.74996119074568344412498305}%
\StoreBenchExecResult{JarFullInterpolationModelChecking}{Impact}{Error}{Exception}{Walltime}{}{1192.4541721493005603}%
\StoreBenchExecResult{JarFullInterpolationModelChecking}{Impact}{Error}{Exception}{Walltime}{Avg}{34.07011920426573029428571429}%
\StoreBenchExecResult{JarFullInterpolationModelChecking}{Impact}{Error}{Exception}{Walltime}{Median}{16.03093416406773}%
\StoreBenchExecResult{JarFullInterpolationModelChecking}{Impact}{Error}{Exception}{Walltime}{Min}{3.764273858163506}%
\StoreBenchExecResult{JarFullInterpolationModelChecking}{Impact}{Error}{Exception}{Walltime}{Max}{209.6009678079281}%
\StoreBenchExecResult{JarFullInterpolationModelChecking}{Impact}{Error}{Exception}{Walltime}{Stdev}{44.08877997681584659896580333}%
\StoreBenchExecResult{JarFullInterpolationModelChecking}{Impact}{Error}{OutOfJavaMemory}{Count}{}{3}%
\StoreBenchExecResult{JarFullInterpolationModelChecking}{Impact}{Error}{OutOfJavaMemory}{Cputime}{}{206.003813104}%
\StoreBenchExecResult{JarFullInterpolationModelChecking}{Impact}{Error}{OutOfJavaMemory}{Cputime}{Avg}{68.66793770133333333333333333}%
\StoreBenchExecResult{JarFullInterpolationModelChecking}{Impact}{Error}{OutOfJavaMemory}{Cputime}{Median}{77.205952927}%
\StoreBenchExecResult{JarFullInterpolationModelChecking}{Impact}{Error}{OutOfJavaMemory}{Cputime}{Min}{49.600928382}%
\StoreBenchExecResult{JarFullInterpolationModelChecking}{Impact}{Error}{OutOfJavaMemory}{Cputime}{Max}{79.196931795}%
\StoreBenchExecResult{JarFullInterpolationModelChecking}{Impact}{Error}{OutOfJavaMemory}{Cputime}{Stdev}{13.50689040208401132876147122}%
\StoreBenchExecResult{JarFullInterpolationModelChecking}{Impact}{Error}{OutOfJavaMemory}{Walltime}{}{112.423057845095173}%
\StoreBenchExecResult{JarFullInterpolationModelChecking}{Impact}{Error}{OutOfJavaMemory}{Walltime}{Avg}{37.47435261503172433333333333}%
\StoreBenchExecResult{JarFullInterpolationModelChecking}{Impact}{Error}{OutOfJavaMemory}{Walltime}{Median}{42.452769117895514}%
\StoreBenchExecResult{JarFullInterpolationModelChecking}{Impact}{Error}{OutOfJavaMemory}{Walltime}{Min}{26.938731593079865}%
\StoreBenchExecResult{JarFullInterpolationModelChecking}{Impact}{Error}{OutOfJavaMemory}{Walltime}{Max}{43.031557134119794}%
\StoreBenchExecResult{JarFullInterpolationModelChecking}{Impact}{Error}{OutOfJavaMemory}{Walltime}{Stdev}{7.453555376712363076733280037}%
\StoreBenchExecResult{JarFullInterpolationModelChecking}{Impact}{Error}{OutOfMemory}{Count}{}{107}%
\StoreBenchExecResult{JarFullInterpolationModelChecking}{Impact}{Error}{OutOfMemory}{Cputime}{}{39397.417949952}%
\StoreBenchExecResult{JarFullInterpolationModelChecking}{Impact}{Error}{OutOfMemory}{Cputime}{Avg}{368.2001677565607476635514019}%
\StoreBenchExecResult{JarFullInterpolationModelChecking}{Impact}{Error}{OutOfMemory}{Cputime}{Median}{259.945000014}%
\StoreBenchExecResult{JarFullInterpolationModelChecking}{Impact}{Error}{OutOfMemory}{Cputime}{Min}{102.750871624}%
\StoreBenchExecResult{JarFullInterpolationModelChecking}{Impact}{Error}{OutOfMemory}{Cputime}{Max}{897.291633463}%
\StoreBenchExecResult{JarFullInterpolationModelChecking}{Impact}{Error}{OutOfMemory}{Cputime}{Stdev}{222.6593447452256463149780227}%
\StoreBenchExecResult{JarFullInterpolationModelChecking}{Impact}{Error}{OutOfMemory}{Walltime}{}{36504.81188043695866}%
\StoreBenchExecResult{JarFullInterpolationModelChecking}{Impact}{Error}{OutOfMemory}{Walltime}{Avg}{341.1664661723080248598130841}%
\StoreBenchExecResult{JarFullInterpolationModelChecking}{Impact}{Error}{OutOfMemory}{Walltime}{Median}{238.685779272113}%
\StoreBenchExecResult{JarFullInterpolationModelChecking}{Impact}{Error}{OutOfMemory}{Walltime}{Min}{97.02376463986002}%
\StoreBenchExecResult{JarFullInterpolationModelChecking}{Impact}{Error}{OutOfMemory}{Walltime}{Max}{835.5387623871211}%
\StoreBenchExecResult{JarFullInterpolationModelChecking}{Impact}{Error}{OutOfMemory}{Walltime}{Stdev}{208.1057365796277088374364866}%
\StoreBenchExecResult{JarFullInterpolationModelChecking}{Impact}{Error}{OutOfNativeMemory}{Count}{}{2}%
\StoreBenchExecResult{JarFullInterpolationModelChecking}{Impact}{Error}{OutOfNativeMemory}{Cputime}{}{18.514631794}%
\StoreBenchExecResult{JarFullInterpolationModelChecking}{Impact}{Error}{OutOfNativeMemory}{Cputime}{Avg}{9.257315897}%
\StoreBenchExecResult{JarFullInterpolationModelChecking}{Impact}{Error}{OutOfNativeMemory}{Cputime}{Median}{9.257315897}%
\StoreBenchExecResult{JarFullInterpolationModelChecking}{Impact}{Error}{OutOfNativeMemory}{Cputime}{Min}{7.498870141}%
\StoreBenchExecResult{JarFullInterpolationModelChecking}{Impact}{Error}{OutOfNativeMemory}{Cputime}{Max}{11.015761653}%
\StoreBenchExecResult{JarFullInterpolationModelChecking}{Impact}{Error}{OutOfNativeMemory}{Cputime}{Stdev}{1.75844575600000}%
\StoreBenchExecResult{JarFullInterpolationModelChecking}{Impact}{Error}{OutOfNativeMemory}{Walltime}{}{10.184136079158634}%
\StoreBenchExecResult{JarFullInterpolationModelChecking}{Impact}{Error}{OutOfNativeMemory}{Walltime}{Avg}{5.092068039579317}%
\StoreBenchExecResult{JarFullInterpolationModelChecking}{Impact}{Error}{OutOfNativeMemory}{Walltime}{Median}{5.092068039579317}%
\StoreBenchExecResult{JarFullInterpolationModelChecking}{Impact}{Error}{OutOfNativeMemory}{Walltime}{Min}{3.959143767133355}%
\StoreBenchExecResult{JarFullInterpolationModelChecking}{Impact}{Error}{OutOfNativeMemory}{Walltime}{Max}{6.224992312025279}%
\StoreBenchExecResult{JarFullInterpolationModelChecking}{Impact}{Error}{OutOfNativeMemory}{Walltime}{Stdev}{1.132924272445962000000000000}%
\StoreBenchExecResult{JarFullInterpolationModelChecking}{Impact}{Error}{SegmentationFault}{Count}{}{5}%
\StoreBenchExecResult{JarFullInterpolationModelChecking}{Impact}{Error}{SegmentationFault}{Cputime}{}{276.650328765}%
\StoreBenchExecResult{JarFullInterpolationModelChecking}{Impact}{Error}{SegmentationFault}{Cputime}{Avg}{55.330065753}%
\StoreBenchExecResult{JarFullInterpolationModelChecking}{Impact}{Error}{SegmentationFault}{Cputime}{Median}{64.964182796}%
\StoreBenchExecResult{JarFullInterpolationModelChecking}{Impact}{Error}{SegmentationFault}{Cputime}{Min}{9.961667536}%
\StoreBenchExecResult{JarFullInterpolationModelChecking}{Impact}{Error}{SegmentationFault}{Cputime}{Max}{70.039327795}%
\StoreBenchExecResult{JarFullInterpolationModelChecking}{Impact}{Error}{SegmentationFault}{Cputime}{Stdev}{22.83653908804429399028903407}%
\StoreBenchExecResult{JarFullInterpolationModelChecking}{Impact}{Error}{SegmentationFault}{Walltime}{}{258.622938314918431}%
\StoreBenchExecResult{JarFullInterpolationModelChecking}{Impact}{Error}{SegmentationFault}{Walltime}{Avg}{51.7245876629836862}%
\StoreBenchExecResult{JarFullInterpolationModelChecking}{Impact}{Error}{SegmentationFault}{Walltime}{Median}{61.56855312897824}%
\StoreBenchExecResult{JarFullInterpolationModelChecking}{Impact}{Error}{SegmentationFault}{Walltime}{Min}{6.493545017903671}%
\StoreBenchExecResult{JarFullInterpolationModelChecking}{Impact}{Error}{SegmentationFault}{Walltime}{Max}{66.10055293701589}%
\StoreBenchExecResult{JarFullInterpolationModelChecking}{Impact}{Error}{SegmentationFault}{Walltime}{Stdev}{22.73571368976007524997584695}%
\StoreBenchExecResult{JarFullInterpolationModelChecking}{Impact}{Error}{Timeout}{Count}{}{1725}%
\StoreBenchExecResult{JarFullInterpolationModelChecking}{Impact}{Error}{Timeout}{Cputime}{}{1563806.913790129}%
\StoreBenchExecResult{JarFullInterpolationModelChecking}{Impact}{Error}{Timeout}{Cputime}{Avg}{906.5547326319588405797101449}%
\StoreBenchExecResult{JarFullInterpolationModelChecking}{Impact}{Error}{Timeout}{Cputime}{Median}{902.153832426}%
\StoreBenchExecResult{JarFullInterpolationModelChecking}{Impact}{Error}{Timeout}{Cputime}{Min}{900.45260663}%
\StoreBenchExecResult{JarFullInterpolationModelChecking}{Impact}{Error}{Timeout}{Cputime}{Max}{962.724753477}%
\StoreBenchExecResult{JarFullInterpolationModelChecking}{Impact}{Error}{Timeout}{Cputime}{Stdev}{11.77761613693790685966374134}%
\StoreBenchExecResult{JarFullInterpolationModelChecking}{Impact}{Error}{Timeout}{Walltime}{}{1519685.9659260122566}%
\StoreBenchExecResult{JarFullInterpolationModelChecking}{Impact}{Error}{Timeout}{Walltime}{Avg}{880.9773715513114531014492754}%
\StoreBenchExecResult{JarFullInterpolationModelChecking}{Impact}{Error}{Timeout}{Walltime}{Median}{894.6145567419007}%
\StoreBenchExecResult{JarFullInterpolationModelChecking}{Impact}{Error}{Timeout}{Walltime}{Min}{550.5088256280869}%
\StoreBenchExecResult{JarFullInterpolationModelChecking}{Impact}{Error}{Timeout}{Walltime}{Max}{908.7584513910115}%
\StoreBenchExecResult{JarFullInterpolationModelChecking}{Impact}{Error}{Timeout}{Walltime}{Stdev}{48.07337063405983104601807792}%
\StoreBenchExecResult{JarFullInterpolationModelChecking}{Impact}{Wrong}{}{Count}{}{2}%
\StoreBenchExecResult{JarFullInterpolationModelChecking}{Impact}{Wrong}{}{Cputime}{}{11.410191149}%
\StoreBenchExecResult{JarFullInterpolationModelChecking}{Impact}{Wrong}{}{Cputime}{Avg}{5.7050955745}%
\StoreBenchExecResult{JarFullInterpolationModelChecking}{Impact}{Wrong}{}{Cputime}{Median}{5.7050955745}%
\StoreBenchExecResult{JarFullInterpolationModelChecking}{Impact}{Wrong}{}{Cputime}{Min}{4.966879287}%
\StoreBenchExecResult{JarFullInterpolationModelChecking}{Impact}{Wrong}{}{Cputime}{Max}{6.443311862}%
\StoreBenchExecResult{JarFullInterpolationModelChecking}{Impact}{Wrong}{}{Cputime}{Stdev}{0.73821628750000}%
\StoreBenchExecResult{JarFullInterpolationModelChecking}{Impact}{Wrong}{}{Walltime}{}{6.3390085129067303}%
\StoreBenchExecResult{JarFullInterpolationModelChecking}{Impact}{Wrong}{}{Walltime}{Avg}{3.16950425645336515}%
\StoreBenchExecResult{JarFullInterpolationModelChecking}{Impact}{Wrong}{}{Walltime}{Median}{3.16950425645336515}%
\StoreBenchExecResult{JarFullInterpolationModelChecking}{Impact}{Wrong}{}{Walltime}{Min}{2.617243790999055}%
\StoreBenchExecResult{JarFullInterpolationModelChecking}{Impact}{Wrong}{}{Walltime}{Max}{3.7217647219076753}%
\StoreBenchExecResult{JarFullInterpolationModelChecking}{Impact}{Wrong}{}{Walltime}{Stdev}{0.5522604654543101500000000000}%
\StoreBenchExecResult{JarFullInterpolationModelChecking}{Impact}{Wrong}{False}{Count}{}{2}%
\StoreBenchExecResult{JarFullInterpolationModelChecking}{Impact}{Wrong}{False}{Cputime}{}{11.410191149}%
\StoreBenchExecResult{JarFullInterpolationModelChecking}{Impact}{Wrong}{False}{Cputime}{Avg}{5.7050955745}%
\StoreBenchExecResult{JarFullInterpolationModelChecking}{Impact}{Wrong}{False}{Cputime}{Median}{5.7050955745}%
\StoreBenchExecResult{JarFullInterpolationModelChecking}{Impact}{Wrong}{False}{Cputime}{Min}{4.966879287}%
\StoreBenchExecResult{JarFullInterpolationModelChecking}{Impact}{Wrong}{False}{Cputime}{Max}{6.443311862}%
\StoreBenchExecResult{JarFullInterpolationModelChecking}{Impact}{Wrong}{False}{Cputime}{Stdev}{0.73821628750000}%
\StoreBenchExecResult{JarFullInterpolationModelChecking}{Impact}{Wrong}{False}{Walltime}{}{6.3390085129067303}%
\StoreBenchExecResult{JarFullInterpolationModelChecking}{Impact}{Wrong}{False}{Walltime}{Avg}{3.16950425645336515}%
\StoreBenchExecResult{JarFullInterpolationModelChecking}{Impact}{Wrong}{False}{Walltime}{Median}{3.16950425645336515}%
\StoreBenchExecResult{JarFullInterpolationModelChecking}{Impact}{Wrong}{False}{Walltime}{Min}{2.617243790999055}%
\StoreBenchExecResult{JarFullInterpolationModelChecking}{Impact}{Wrong}{False}{Walltime}{Max}{3.7217647219076753}%
\StoreBenchExecResult{JarFullInterpolationModelChecking}{Impact}{Wrong}{False}{Walltime}{Stdev}{0.5522604654543101500000000000}%
\ifdefined\JarFullInterpolationModelCheckingImcTotalCount\else\edef\JarFullInterpolationModelCheckingImcTotalCount{0}\fi
\ifdefined\JarFullInterpolationModelCheckingImcCorrectCount\else\edef\JarFullInterpolationModelCheckingImcCorrectCount{0}\fi
\ifdefined\JarFullInterpolationModelCheckingImcCorrectTrueCount\else\edef\JarFullInterpolationModelCheckingImcCorrectTrueCount{0}\fi
\ifdefined\JarFullInterpolationModelCheckingImcCorrectFalseCount\else\edef\JarFullInterpolationModelCheckingImcCorrectFalseCount{0}\fi
\ifdefined\JarFullInterpolationModelCheckingImcWrongTrueCount\else\edef\JarFullInterpolationModelCheckingImcWrongTrueCount{0}\fi
\ifdefined\JarFullInterpolationModelCheckingImcWrongFalseCount\else\edef\JarFullInterpolationModelCheckingImcWrongFalseCount{0}\fi
\ifdefined\JarFullInterpolationModelCheckingImcErrorTimeoutCount\else\edef\JarFullInterpolationModelCheckingImcErrorTimeoutCount{0}\fi
\ifdefined\JarFullInterpolationModelCheckingImcErrorOutOfMemoryCount\else\edef\JarFullInterpolationModelCheckingImcErrorOutOfMemoryCount{0}\fi
\ifdefined\JarFullInterpolationModelCheckingImcCorrectCputime\else\edef\JarFullInterpolationModelCheckingImcCorrectCputime{0}\fi
\ifdefined\JarFullInterpolationModelCheckingImcCorrectCputimeAvg\else\edef\JarFullInterpolationModelCheckingImcCorrectCputimeAvg{None}\fi
\ifdefined\JarFullInterpolationModelCheckingImcCorrectWalltime\else\edef\JarFullInterpolationModelCheckingImcCorrectWalltime{0}\fi
\ifdefined\JarFullInterpolationModelCheckingImcCorrectWalltimeAvg\else\edef\JarFullInterpolationModelCheckingImcCorrectWalltimeAvg{None}\fi
\ifdefined\JarFullInterpolationModelCheckingPdrTotalCount\else\edef\JarFullInterpolationModelCheckingPdrTotalCount{0}\fi
\ifdefined\JarFullInterpolationModelCheckingPdrCorrectCount\else\edef\JarFullInterpolationModelCheckingPdrCorrectCount{0}\fi
\ifdefined\JarFullInterpolationModelCheckingPdrCorrectTrueCount\else\edef\JarFullInterpolationModelCheckingPdrCorrectTrueCount{0}\fi
\ifdefined\JarFullInterpolationModelCheckingPdrCorrectFalseCount\else\edef\JarFullInterpolationModelCheckingPdrCorrectFalseCount{0}\fi
\ifdefined\JarFullInterpolationModelCheckingPdrWrongTrueCount\else\edef\JarFullInterpolationModelCheckingPdrWrongTrueCount{0}\fi
\ifdefined\JarFullInterpolationModelCheckingPdrWrongFalseCount\else\edef\JarFullInterpolationModelCheckingPdrWrongFalseCount{0}\fi
\ifdefined\JarFullInterpolationModelCheckingPdrErrorTimeoutCount\else\edef\JarFullInterpolationModelCheckingPdrErrorTimeoutCount{0}\fi
\ifdefined\JarFullInterpolationModelCheckingPdrErrorOutOfMemoryCount\else\edef\JarFullInterpolationModelCheckingPdrErrorOutOfMemoryCount{0}\fi
\ifdefined\JarFullInterpolationModelCheckingPdrCorrectCputime\else\edef\JarFullInterpolationModelCheckingPdrCorrectCputime{0}\fi
\ifdefined\JarFullInterpolationModelCheckingPdrCorrectCputimeAvg\else\edef\JarFullInterpolationModelCheckingPdrCorrectCputimeAvg{None}\fi
\ifdefined\JarFullInterpolationModelCheckingPdrCorrectWalltime\else\edef\JarFullInterpolationModelCheckingPdrCorrectWalltime{0}\fi
\ifdefined\JarFullInterpolationModelCheckingPdrCorrectWalltimeAvg\else\edef\JarFullInterpolationModelCheckingPdrCorrectWalltimeAvg{None}\fi
\ifdefined\JarFullInterpolationModelCheckingBmcTotalCount\else\edef\JarFullInterpolationModelCheckingBmcTotalCount{0}\fi
\ifdefined\JarFullInterpolationModelCheckingBmcCorrectCount\else\edef\JarFullInterpolationModelCheckingBmcCorrectCount{0}\fi
\ifdefined\JarFullInterpolationModelCheckingBmcCorrectTrueCount\else\edef\JarFullInterpolationModelCheckingBmcCorrectTrueCount{0}\fi
\ifdefined\JarFullInterpolationModelCheckingBmcCorrectFalseCount\else\edef\JarFullInterpolationModelCheckingBmcCorrectFalseCount{0}\fi
\ifdefined\JarFullInterpolationModelCheckingBmcWrongTrueCount\else\edef\JarFullInterpolationModelCheckingBmcWrongTrueCount{0}\fi
\ifdefined\JarFullInterpolationModelCheckingBmcWrongFalseCount\else\edef\JarFullInterpolationModelCheckingBmcWrongFalseCount{0}\fi
\ifdefined\JarFullInterpolationModelCheckingBmcErrorTimeoutCount\else\edef\JarFullInterpolationModelCheckingBmcErrorTimeoutCount{0}\fi
\ifdefined\JarFullInterpolationModelCheckingBmcErrorOutOfMemoryCount\else\edef\JarFullInterpolationModelCheckingBmcErrorOutOfMemoryCount{0}\fi
\ifdefined\JarFullInterpolationModelCheckingBmcCorrectCputime\else\edef\JarFullInterpolationModelCheckingBmcCorrectCputime{0}\fi
\ifdefined\JarFullInterpolationModelCheckingBmcCorrectCputimeAvg\else\edef\JarFullInterpolationModelCheckingBmcCorrectCputimeAvg{None}\fi
\ifdefined\JarFullInterpolationModelCheckingBmcCorrectWalltime\else\edef\JarFullInterpolationModelCheckingBmcCorrectWalltime{0}\fi
\ifdefined\JarFullInterpolationModelCheckingBmcCorrectWalltimeAvg\else\edef\JarFullInterpolationModelCheckingBmcCorrectWalltimeAvg{None}\fi
\ifdefined\JarFullInterpolationModelCheckingKInductionTotalCount\else\edef\JarFullInterpolationModelCheckingKInductionTotalCount{0}\fi
\ifdefined\JarFullInterpolationModelCheckingKInductionCorrectCount\else\edef\JarFullInterpolationModelCheckingKInductionCorrectCount{0}\fi
\ifdefined\JarFullInterpolationModelCheckingKInductionCorrectTrueCount\else\edef\JarFullInterpolationModelCheckingKInductionCorrectTrueCount{0}\fi
\ifdefined\JarFullInterpolationModelCheckingKInductionCorrectFalseCount\else\edef\JarFullInterpolationModelCheckingKInductionCorrectFalseCount{0}\fi
\ifdefined\JarFullInterpolationModelCheckingKInductionWrongTrueCount\else\edef\JarFullInterpolationModelCheckingKInductionWrongTrueCount{0}\fi
\ifdefined\JarFullInterpolationModelCheckingKInductionWrongFalseCount\else\edef\JarFullInterpolationModelCheckingKInductionWrongFalseCount{0}\fi
\ifdefined\JarFullInterpolationModelCheckingKInductionErrorTimeoutCount\else\edef\JarFullInterpolationModelCheckingKInductionErrorTimeoutCount{0}\fi
\ifdefined\JarFullInterpolationModelCheckingKInductionErrorOutOfMemoryCount\else\edef\JarFullInterpolationModelCheckingKInductionErrorOutOfMemoryCount{0}\fi
\ifdefined\JarFullInterpolationModelCheckingKInductionCorrectCputime\else\edef\JarFullInterpolationModelCheckingKInductionCorrectCputime{0}\fi
\ifdefined\JarFullInterpolationModelCheckingKInductionCorrectCputimeAvg\else\edef\JarFullInterpolationModelCheckingKInductionCorrectCputimeAvg{None}\fi
\ifdefined\JarFullInterpolationModelCheckingKInductionCorrectWalltime\else\edef\JarFullInterpolationModelCheckingKInductionCorrectWalltime{0}\fi
\ifdefined\JarFullInterpolationModelCheckingKInductionCorrectWalltimeAvg\else\edef\JarFullInterpolationModelCheckingKInductionCorrectWalltimeAvg{None}\fi
\ifdefined\JarFullInterpolationModelCheckingPredicateAbstractionTotalCount\else\edef\JarFullInterpolationModelCheckingPredicateAbstractionTotalCount{0}\fi
\ifdefined\JarFullInterpolationModelCheckingPredicateAbstractionCorrectCount\else\edef\JarFullInterpolationModelCheckingPredicateAbstractionCorrectCount{0}\fi
\ifdefined\JarFullInterpolationModelCheckingPredicateAbstractionCorrectTrueCount\else\edef\JarFullInterpolationModelCheckingPredicateAbstractionCorrectTrueCount{0}\fi
\ifdefined\JarFullInterpolationModelCheckingPredicateAbstractionCorrectFalseCount\else\edef\JarFullInterpolationModelCheckingPredicateAbstractionCorrectFalseCount{0}\fi
\ifdefined\JarFullInterpolationModelCheckingPredicateAbstractionWrongTrueCount\else\edef\JarFullInterpolationModelCheckingPredicateAbstractionWrongTrueCount{0}\fi
\ifdefined\JarFullInterpolationModelCheckingPredicateAbstractionWrongFalseCount\else\edef\JarFullInterpolationModelCheckingPredicateAbstractionWrongFalseCount{0}\fi
\ifdefined\JarFullInterpolationModelCheckingPredicateAbstractionErrorTimeoutCount\else\edef\JarFullInterpolationModelCheckingPredicateAbstractionErrorTimeoutCount{0}\fi
\ifdefined\JarFullInterpolationModelCheckingPredicateAbstractionErrorOutOfMemoryCount\else\edef\JarFullInterpolationModelCheckingPredicateAbstractionErrorOutOfMemoryCount{0}\fi
\ifdefined\JarFullInterpolationModelCheckingPredicateAbstractionCorrectCputime\else\edef\JarFullInterpolationModelCheckingPredicateAbstractionCorrectCputime{0}\fi
\ifdefined\JarFullInterpolationModelCheckingPredicateAbstractionCorrectCputimeAvg\else\edef\JarFullInterpolationModelCheckingPredicateAbstractionCorrectCputimeAvg{None}\fi
\ifdefined\JarFullInterpolationModelCheckingPredicateAbstractionCorrectWalltime\else\edef\JarFullInterpolationModelCheckingPredicateAbstractionCorrectWalltime{0}\fi
\ifdefined\JarFullInterpolationModelCheckingPredicateAbstractionCorrectWalltimeAvg\else\edef\JarFullInterpolationModelCheckingPredicateAbstractionCorrectWalltimeAvg{None}\fi
\ifdefined\JarFullInterpolationModelCheckingImpactTotalCount\else\edef\JarFullInterpolationModelCheckingImpactTotalCount{0}\fi
\ifdefined\JarFullInterpolationModelCheckingImpactCorrectCount\else\edef\JarFullInterpolationModelCheckingImpactCorrectCount{0}\fi
\ifdefined\JarFullInterpolationModelCheckingImpactCorrectTrueCount\else\edef\JarFullInterpolationModelCheckingImpactCorrectTrueCount{0}\fi
\ifdefined\JarFullInterpolationModelCheckingImpactCorrectFalseCount\else\edef\JarFullInterpolationModelCheckingImpactCorrectFalseCount{0}\fi
\ifdefined\JarFullInterpolationModelCheckingImpactWrongTrueCount\else\edef\JarFullInterpolationModelCheckingImpactWrongTrueCount{0}\fi
\ifdefined\JarFullInterpolationModelCheckingImpactWrongFalseCount\else\edef\JarFullInterpolationModelCheckingImpactWrongFalseCount{0}\fi
\ifdefined\JarFullInterpolationModelCheckingImpactErrorTimeoutCount\else\edef\JarFullInterpolationModelCheckingImpactErrorTimeoutCount{0}\fi
\ifdefined\JarFullInterpolationModelCheckingImpactErrorOutOfMemoryCount\else\edef\JarFullInterpolationModelCheckingImpactErrorOutOfMemoryCount{0}\fi
\ifdefined\JarFullInterpolationModelCheckingImpactCorrectCputime\else\edef\JarFullInterpolationModelCheckingImpactCorrectCputime{0}\fi
\ifdefined\JarFullInterpolationModelCheckingImpactCorrectCputimeAvg\else\edef\JarFullInterpolationModelCheckingImpactCorrectCputimeAvg{None}\fi
\ifdefined\JarFullInterpolationModelCheckingImpactCorrectWalltime\else\edef\JarFullInterpolationModelCheckingImpactCorrectWalltime{0}\fi
\ifdefined\JarFullInterpolationModelCheckingImpactCorrectWalltimeAvg\else\edef\JarFullInterpolationModelCheckingImpactCorrectWalltimeAvg{None}\fi
\edef\JarFullInterpolationModelCheckingImcErrorOtherInconclusiveCount{\the\numexpr \JarFullInterpolationModelCheckingImcTotalCount - \JarFullInterpolationModelCheckingImcCorrectCount - \JarFullInterpolationModelCheckingImcWrongTrueCount - \JarFullInterpolationModelCheckingImcWrongFalseCount - \JarFullInterpolationModelCheckingImcErrorTimeoutCount - \JarFullInterpolationModelCheckingImcErrorOutOfMemoryCount \relax}
\edef\JarFullInterpolationModelCheckingPdrErrorOtherInconclusiveCount{\the\numexpr \JarFullInterpolationModelCheckingPdrTotalCount - \JarFullInterpolationModelCheckingPdrCorrectCount - \JarFullInterpolationModelCheckingPdrWrongTrueCount - \JarFullInterpolationModelCheckingPdrWrongFalseCount - \JarFullInterpolationModelCheckingPdrErrorTimeoutCount - \JarFullInterpolationModelCheckingPdrErrorOutOfMemoryCount \relax}
\edef\JarFullInterpolationModelCheckingBmcErrorOtherInconclusiveCount{\the\numexpr \JarFullInterpolationModelCheckingBmcTotalCount - \JarFullInterpolationModelCheckingBmcCorrectCount - \JarFullInterpolationModelCheckingBmcWrongTrueCount - \JarFullInterpolationModelCheckingBmcWrongFalseCount - \JarFullInterpolationModelCheckingBmcErrorTimeoutCount - \JarFullInterpolationModelCheckingBmcErrorOutOfMemoryCount \relax}
\edef\JarFullInterpolationModelCheckingKInductionErrorOtherInconclusiveCount{\the\numexpr \JarFullInterpolationModelCheckingKInductionTotalCount - \JarFullInterpolationModelCheckingKInductionCorrectCount - \JarFullInterpolationModelCheckingKInductionWrongTrueCount - \JarFullInterpolationModelCheckingKInductionWrongFalseCount - \JarFullInterpolationModelCheckingKInductionErrorTimeoutCount - \JarFullInterpolationModelCheckingKInductionErrorOutOfMemoryCount \relax}
\edef\JarFullInterpolationModelCheckingPredicateAbstractionErrorOtherInconclusiveCount{\the\numexpr \JarFullInterpolationModelCheckingPredicateAbstractionTotalCount - \JarFullInterpolationModelCheckingPredicateAbstractionCorrectCount - \JarFullInterpolationModelCheckingPredicateAbstractionWrongTrueCount - \JarFullInterpolationModelCheckingPredicateAbstractionWrongFalseCount - \JarFullInterpolationModelCheckingPredicateAbstractionErrorTimeoutCount - \JarFullInterpolationModelCheckingPredicateAbstractionErrorOutOfMemoryCount \relax}
\edef\JarFullInterpolationModelCheckingImpactErrorOtherInconclusiveCount{\the\numexpr \JarFullInterpolationModelCheckingImpactTotalCount - \JarFullInterpolationModelCheckingImpactCorrectCount - \JarFullInterpolationModelCheckingImpactWrongTrueCount - \JarFullInterpolationModelCheckingImpactWrongFalseCount - \JarFullInterpolationModelCheckingImpactErrorTimeoutCount - \JarFullInterpolationModelCheckingImpactErrorOutOfMemoryCount \relax}

%% file: evaluation/tex/plot-defs.tex
\pgfplotsset{quantile plot/.style={
    width=6cm,
    height=3.5cm,
    scale only axis,
    /pgfplots/table/x expr={\coordindex+1},
    /pgfplots/table/y index=3,
    /pgfplots/table/header=false,
    ylabel shift=-1em,
    ticklabel style={font={\smaller}},
    xmin=0,
    ymin=1,
    ymax=1000,
    legend cell align={left},
    legend style={at={(0,1)}, anchor=north west, outer xsep=5pt, outer ysep=5pt, fill=none, font={\smaller}},
    legend columns=3,
    cycle multiindex list={
      orange, red, blue, black\nextlist
      mark list*\nextlist
      solid, densely dashed, densely dashdotdotted, densely dotted},
  },
  every axis plot/.append style={thick}
}

\onlyifstandalone{
  \pgfplotsset{quantile plot/.append style={
      ylabel=CPU time (\second),
    },
  }
}

\newcommand\addgraph[2]{{
  \newcommand\csvfile{\plotpath/\detokenize{#2}}
  \IfFileExists\csvfile{
    \addplot+ table {\csvfile}; \addlegendentry{#1}
  }{
    \addplot coordinates {};
  }
}}

%% file: evaluation/tex/quantile-true.tex
\begin{tikzpicture}
\begin{semilogyaxis}[quantile plot, mark repeat=500]
\IfStandalone{\pgfplotsset{title={quantile-true}}}{\pgfplotsset{legend to name=legend:quantile-true}}
\addgraph{IMC}{../csv/IMC.quantile-true.csv}
\addgraph{PDR}{../csv/PDR.quantile-true.csv}
\addgraph{BMC}{../csv/BMC.quantile-true.csv}
\addgraph{\kInduction}{../csv/k-Induction.quantile-true.csv}
\addgraph{Predicate Abstraction}{../csv/Predicate-Abstraction.quantile-true.csv}
\addgraph{\impact}{../csv/Impact.quantile-true.csv}
\end{semilogyaxis}
\end{tikzpicture}

%% file: evaluation/tex/quantile-false.tex
\begin{tikzpicture}
\begin{semilogyaxis}[quantile plot, mark repeat=200]
\IfStandalone{\pgfplotsset{title={quantile-false}}}{\pgfplotsset{legend to name=legend:quantile-false}}
\addgraph{IMC}{../csv/IMC.quantile-false.csv}
\addgraph{PDR}{../csv/PDR.quantile-false.csv}
\addgraph{BMC}{../csv/BMC.quantile-false.csv}
\addgraph{\kInduction}{../csv/k-Induction.quantile-false.csv}
\addgraph{Predicate Abstraction}{../csv/Predicate-Abstraction.quantile-false.csv}
\addgraph{\impact}{../csv/Impact.quantile-false.csv}
\end{semilogyaxis}
\end{tikzpicture}

%% file: evaluation/tex/scatter-pdr.tex
\begin{tikzpicture}
\begin{loglogaxis}[
    xmin=1,
    xmax=1000,
    ymin=1,
    ymax=1000,
    domain=1:1001,
    clip mode=individual,
    axis equal image,
    ]
    \addplot+[blue, mark=+,only marks]
         table[
             header=false,
             skip first n=3, 
             x index=5, 
             y index=3  
             ] {evaluation/csv/PDR.scatter.table.csv};
    \addplot[gray] {x};
    \addplot[gray] {10*x};
    \addplot[gray] {x/10};
\end{loglogaxis}
\end{tikzpicture}

%% file: evaluation/tex/scatter-ki-df.tex
\begin{tikzpicture}
\begin{loglogaxis}[
    xmin=1,
    xmax=1000,
    ymin=1,
    ymax=1000,
    domain=1:1001,
    clip mode=individual,
    axis equal image,
    ]
    \addplot+[blue, mark=+,only marks]
         table[
             header=false,
             skip first n=3, 
             x index=5, 
             y index=3  
             ] {evaluation/csv/k-Induction.scatter.table.csv};
    \addplot[gray] {x};
    \addplot[gray] {10*x};
    \addplot[gray] {x/10};
\end{loglogaxis}
\end{tikzpicture}

%% file: evaluation/tex/scatter-predicate.tex
\begin{tikzpicture}
\begin{loglogaxis}[
    xmin=1,
    xmax=1000,
    ymin=1,
    ymax=1000,
    domain=1:1001,
    clip mode=individual,
    axis equal image,
    ]
    \addplot+[blue, mark=+,only marks]
         table[
             header=false,
             skip first n=3, 
             x index=5, 
             y index=3  
             ] {evaluation/csv/Predicate-Abstraction.scatter.table.csv};
    \addplot[gray] {x};
    \addplot[gray] {10*x};
    \addplot[gray] {x/10};
\end{loglogaxis}
\end{tikzpicture}

%% file: evaluation/tex/scatter-predicate-impact.tex
\begin{tikzpicture}
\begin{loglogaxis}[
    xmin=1,
    xmax=1000,
    ymin=1,
    ymax=1000,
    domain=1:1001,
    clip mode=individual,
    axis equal image,
    ]
    \addplot+[blue, mark=+,only marks]
         table[
             header=false,
             skip first n=3, 
             x index=5, 
             y index=3  
             ] {evaluation/csv/Impact.scatter.table.csv};
    \addplot[gray] {x};
    \addplot[gray] {10*x};
    \addplot[gray] {x/10};
\end{loglogaxis}
\end{tikzpicture}

%% file: evaluation/tex/quantile-impact-ECA-true.tex
\begin{tikzpicture}
\begin{semilogyaxis}[
    quantile plot,
    mark repeat=100,legend columns=1,
    ylabel=CPU time (\second)
  ]
\addgraph{IMC}{../csv/IMC.quantile-ECA-true.csv}
\addgraph{\impact}{../csv/Impact.quantile-ECA-true.csv}
\end{semilogyaxis}
\end{tikzpicture}

%% file: evaluation/tex/scatter-impact-ECA-cputime.tex
\begin{tikzpicture}
\begin{loglogaxis}[
    xmin=1,
    xmax=1000,
    ymin=1,
    ymax=1000,
    domain=1:1001,
    clip mode=individual,
    axis equal image,
    ]
    \addplot+[blue, mark=+,only marks]
         table[
             header=false,
             skip first n=3, 
             x index=8, 
             y index=3  
             ] {evaluation/csv/IMC-Impact.scatter.table.csv};
    \addplot[gray] {x};
    \addplot[gray] {10*x};
    \addplot[gray] {x/10};
\end{loglogaxis}
\end{tikzpicture}

%% file: evaluation/tex/scatter-impact-ECA-itptime.tex
\begin{tikzpicture}
\begin{loglogaxis}[
    xmin=1,
    xmax=1000,
    ymin=1,
    ymax=1000,
    domain=1:1001,
    clip mode=individual,
    axis equal image,
    ]
    \addplot+[blue, mark=+,only marks]
         table[
             header=false,
             skip first n=3, 
             x index=9, 
             y index=4  
             ] {evaluation/csv/IMC-Impact.scatter.table.csv};
    \addplot[gray] {x};
    \addplot[gray] {10*x};
    \addplot[gray] {x/10};
\end{loglogaxis}
\end{tikzpicture}

%% file: evaluation/tex/scatter-impact-ECA-itpcall.tex
\begin{tikzpicture}
    \node[inner sep=0] (image) at (0,0) {\includegraphics{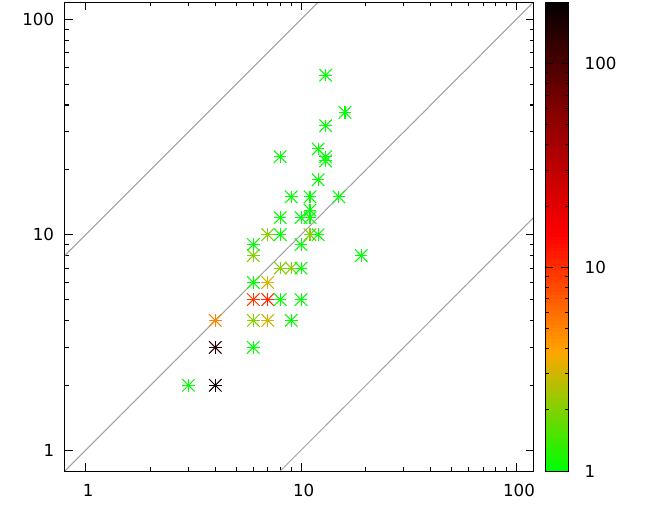}};
    \node[right=-3mm of image, rotate=90, anchor=north] {Number of tasks};
\end{tikzpicture}

%% file: conclusion.tex
\section{Conclusion}
\label{sect:conclusion}
Software verification is a hard problem, and it is imperative to leverage
as much knowledge of the verification community as possible.
\Imc (McMillan, 2003 \cite{McMillanCraig}) is a successful hardware-verification algorithm,
but in contrast to many other interpolation-based verification approaches,
this algorithm was not yet adopted to software verification,
and the characteristics of the algorithm when applied to software systems were unknown.
This paper presents the first theoretical adoption
and practical implementation of the algorithm for software verification,
providing a base-line for other researchers to build on.
Surprisingly, it has taken two decades to close this significant gap of knowledge
by investigating the applicability to software verification.
We present the novel idea of utilizing
the well established technique of large-block encoding
to extract transition relations from programs,
without encoding the control-flow structure of the program into the formulas via symbolic program counters.
The proposed adoption was implemented in the open-source software-verification framework \cpachecker and
evaluated against other state-of-the-art software-verification algorithms
on a large benchmark set of C~verification tasks for reachability properties.

Among the competing approaches, our implementation achieved a comparable performance, evaluated in terms of both effectiveness (the number of correctly solved tasks) and efficiency (the CPU time to solve tasks).
Our IMC implementation was the most effective and efficient interpolation-based approach in the evaluation.
Furthermore,
the new approach was able to solve~\nimcuniquetasks~programs for which all other approaches ran out of resources (\SI{15}{min} CPU time or \SI{15}{GB} memory usage),
which shows that the new approach \emph{improves} the state of the art and \emph{complements} the other approaches.
We hope that our promising results stimulate other researchers to further
improve the approach for software verification
and that our open-source implementation in the flexible framework \cpachecker helps
other researchers to understand the details of the algorithm.



%% file: main.bbl
\providecommand{\serysort}{}\providecommand{\svejdasort}{}
\begin{thebibliography}{10}
\providecommand{\url}[1]{{#1}}
\providecommand{\urlprefix}{URL }
\providecommand{\isbn}[1]{\mbox{ISBN:~\href{https://www.worldcat.org/isbn/#1}{#1}}}
\expandafter\ifx\csname urlstyle\endcsname\relax
  \providecommand{\doi}[1]{\url{https://doi.org/#1}}\else
  \providecommand{\doi}[1]{\url{https://doi.org/#1}}\fi

\bibitem{DragonBook}
Aho, A.V., Sethi, R., Ullman, J.D.: Compilers: Principles, Techniques, and
  Tools.
\newblock Addison-Wesley (1986).
\newblock \isbn{978-0-201-10088-4}

\bibitem{UFO}
Albarghouthi, A., Li, Y., Gurfinkel, A., Chechik, M.: \textsc{Ufo}: {A}
  framework for abstraction- and interpolation-based software verification.
\newblock In: Proc.\ CAV, LNCS~7358, pp. 672--678. Springer (2012).
\newblock \doi{10.1007/978-3-642-31424-7_48}

\bibitem{LazyAbstractionWithArrays-JOURNAL}
Alberti, F., Bruttomesso, R., Ghilardi, S., Ranise, S., Sharygina, N.: An
  extension of lazy abstraction with interpolation for programs with arrays.
\newblock Formal Methods in System Design \textbf{45}(1), 63--109 (2014).
\newblock \doi{10.1007/s10703-014-0209-9}

\bibitem{SLAMtransfer}
Ball, T., Cook, B., Levin, V., Rajamani, S.K.: \textsc{Slam} and {S}tatic
  {D}river {V}erifier: Technology transfer of formal methods inside
  {Microsoft}.
\newblock In: Proc.\ IFM, LNCS~2999, pp. 1--20. Springer (2004).
\newblock \doi{10.1007/978-3-540-24756-2_1}

\bibitem{AutomaticPredicateAbstraction2001}
Ball, T., Majumdar, R., Millstein, T., Rajamani, S.K.: Automatic predicate
  abstraction of {C} programs.
\newblock In: Proc.\ PLDI, pp. 203--213. ACM (2001).
\newblock \doi{10.1145/378795.378846}

\bibitem{SLAM}
Ball, T., Rajamani, S.K.: The \textsc{Slam} project: Debugging system software
  via static analysis.
\newblock In: Proc.\ POPL, pp. 1--3. {ACM} (2002).
\newblock \doi{10.1145/503272.503274}

\bibitem{HBMC-SMT}
Barrett, C., Tinelli, C.: Satisfiability modulo theories.
\newblock In: Handbook of Model Checking, pp. 305--343. Springer (2018).
\newblock \doi{10.1007/978-3-319-10575-8_11}

\bibitem{SVCOMP22}
Beyer, D.: Progress on software verification: {SV-COMP 2022}.
\newblock In: Proc.\ TACAS~(2), LNCS~13244, pp. 375--402. Springer (2022).
\newblock \doi{10.1007/978-3-030-99527-0_20}

\bibitem{SVCOMP22-SVBENCHMARKS-artifact}
Beyer, D.: {SV-Benchmarks}: {B}enchmark set for software verification and
  testing ({SV-COMP 2022} and {Test-Comp 2022}).
\newblock Zenodo (2022).
\newblock \doi{10.5281/zenodo.5831003}

\bibitem{CPA-DF}
Beyer, D., Chien, P.C., Lee, N.Z.: {CPA-DF}: {A} tool for configurable interval
  analysis to boost program verification.
\newblock In: Proc.\ ASE, pp. 2050--2053. IEEE (2023).
\newblock \doi{10.1109/ASE56229.2023.00213}

\bibitem{LBE}
Beyer, D., Cimatti, A., Griggio, A., Keremoglu, M.E., Sebastiani, R.: Software
  model checking via large-block encoding.
\newblock In: Proc.\ FMCAD, pp. 25--32. IEEE (2009).
\newblock \doi{10.1109/FMCAD.2009.5351147}

\bibitem{PDR}
Beyer, D., Dangl, M.: Software verification with {PDR}: {An} implementation of
  the state of the art.
\newblock In: Proc.\ TACAS~(1), LNCS~12078, pp. 3--21. Springer (2020).
\newblock \doi{10.1007/978-3-030-45190-5_1}

\bibitem{kInduction}
Beyer, D., Dangl, M., Wendler, P.: Boosting k-induction with
  continuously-refined invariants.
\newblock In: Proc.\ CAV, LNCS~9206, pp. 622--640. Springer (2015).
\newblock \doi{10.1007/978-3-319-21690-4_42}

\bibitem{AlgorithmComparison-JAR}
Beyer, D., Dangl, M., Wendler, P.: A unifying view on {SMT}-based software
  verification.
\newblock J. Autom. Reasoning \textbf{60}(3), 299--335 (2018).
\newblock \doi{10.1007/s10817-017-9432-6}

\bibitem{HBMC-dataflow}
Beyer, D., Gulwani, S., Schmidt, D.: Combining model checking and data-flow
  analysis.
\newblock In: Handbook of Model Checking, pp. 493--540. Springer (2018).
\newblock \doi{10.1007/978-3-319-10575-8_16}

\bibitem{BLAST}
Beyer, D., Henzinger, T.A., Jhala, R., Majumdar, R.: The software model checker
  \textsc{Blast}.
\newblock Int.\ J.\ Softw.\ Tools Technol.\ Transfer \textbf{9}(5-6), 505--525
  (2007).
\newblock \doi{10.1007/s10009-007-0044-z}

\bibitem{CPA}
Beyer, D., Henzinger, T.A., Th{\'e}oduloz, G.: Configurable software
  verification: Concretizing the convergence of model checking and program
  analysis.
\newblock In: Proc.\ CAV, LNCS~4590, pp. 504--518. Springer (2007).
\newblock \doi{10.1007/978-3-540-73368-3_51}

\bibitem{CPAplus}
Beyer, D., Henzinger, T.A., Th{\'e}oduloz, G.: Program analysis with dynamic
  precision adjustment.
\newblock In: Proc.\ ASE, pp. 29--38. IEEE (2008).
\newblock \doi{10.1109/ASE.2008.13}

\bibitem{CPACHECKER}
Beyer, D., Keremoglu, M.E.: \textsc{CPAchecker}: A tool for configurable
  software verification.
\newblock In: Proc.\ CAV, LNCS~6806, pp. 184--190. Springer (2011).
\newblock \doi{10.1007/978-3-642-22110-1_16}

\bibitem{ABE}
Beyer, D., Keremoglu, M.E., Wendler, P.: Predicate abstraction with
  adjustable-block encoding.
\newblock In: Proc.\ FMCAD, pp. 189--197. FMCAD (2010).
\newblock
  \href{https://ieeexplore.ieee.org/document/5770949}{\texttt{https://ieeexplore.ieee.org/document/5770949}}

\bibitem{IMC-artifact-JAR-final}
Beyer, D., Lee, N.Z., Wendler, P.: Reproduction package for article
  `{Interpolation} and {SAT}-based model checking revisited'.
\newblock Zenodo (2023).
\newblock \doi{10.5281/zenodo.8245824}

\bibitem{Benchmarking-STTT}
Beyer, D., L{\"o}we, S., Wendler, P.: Reliable benchmarking: {R}equirements and
  solutions.
\newblock Int.\ J.\ Softw.\ Tools Technol.\ Transfer \textbf{21}(1), 1--29
  (2019).
\newblock \doi{10.1007/s10009-017-0469-y}

\bibitem{LDV12}
Beyer, D., Petrenko, A.K.: {Linux} driver verification.
\newblock In: Proc.\ ISoLA, LNCS~7610, pp. 1--6. Springer (2012).
\newblock \doi{10.1007/978-3-642-34032-1_1}

\bibitem{CSIsat}
Beyer, D., Zufferey, D., Majumdar, R.: \textsc{CSIsat}: Interpolation for
  {LA+EUF}.
\newblock In: Proc.\ CAV, LNCS~5123, pp. 304--308. Springer (2008).
\newblock \doi{10.1007/978-3-540-70545-1_29}

\bibitem{BMC}
Biere, A., Cimatti, A., Clarke, E.M., Zhu, Y.: Symbolic model checking without
  {BDD}s.
\newblock In: Proc.\ TACAS, LNCS~1579, pp. 193--207. Springer (1999).
\newblock \doi{10.1007/3-540-49059-0_14}

\bibitem{CTIGAR}
Birgmeier, J., Bradley, A.R., Weissenbacher, G.: Counterexample to
  induction-guided abstraction-refinement {(CTIGAR)}.
\newblock In: Proc.\ CAV, LNCS~8559, pp. 831--848. Springer (2014).
\newblock \doi{10.1007/978-3-319-08867-9_55}

\bibitem{BlichaTACAS2022}
Blicha, M., Fedyukovich, G., Hyv{\"{a}}rinen, A.E.J., Sharygina, N.: Transition
  power abstractions for deep counterexample detection.
\newblock In: Proc.\ TACAS, LNCS~13243, pp. 524--542. Springer (2022).
\newblock \doi{10.1007/978-3-030-99524-9\_29}

\bibitem{IC3}
Bradley, A.R.: {SAT}-based model checking without unrolling.
\newblock In: Proc.\ VMCAI, LNCS~6538, pp. 70--87. Springer (2011).
\newblock \doi{10.1007/978-3-642-18275-4_7}

\bibitem{SlicingAbstractions}
Br{\"{u}}ckner, I., Dr{\"{a}}ger, K., Finkbeiner, B., Wehrheim, H.: Slicing
  abstractions.
\newblock In: Proc. FSEN, LNCS~4767, pp. 17--32. Springer (2007).
\newblock \doi{10.1007/978-3-540-75698-9_2}

\bibitem{CabodiDATE2011}
Cabodi, G., Nocco, S., Quer, S.: Interpolation sequences revisited.
\newblock In: Proc.\ DATE, pp. 1--6. {IEEE} (2011).
\newblock \doi{10.1109/DATE.2011.5763056}

\bibitem{INFER}
Calcagno, C., Distefano, D., Dubreil, J., Gabi, D., Hooimeijer, P., Luca, M.,
  O'Hearn, P.W., Papakonstantinou, I., Purbrick, J., Rodriguez, D.: Moving fast
  with software verification.
\newblock In: Proc.\ NFM, LNCS~9058, pp. 3--11. Springer (2015).
\newblock \doi{10.1007/978-3-319-17524-9_1}

\bibitem{SoftwareIC3}
Cimatti, A., Griggio, A.: Software model checking via {IC3}.
\newblock In: Proc.\ CAV, LNCS~7358, pp. 277--293. Springer (2012).
\newblock \doi{10.1007/978-3-642-31424-7_23}

\bibitem{MATHSAT5}
Cimatti, A., Griggio, A., Schaafsma, B.J., Sebastiani, R.: The
  \textsc{MathSAT5} {SMT} solver.
\newblock In: Proc.\ TACAS, LNCS~7795, pp. 93--107. Springer (2013).
\newblock \doi{10.1007/978-3-642-36742-7_7}

\bibitem{ClarkeCEGAR}
Clarke, E.M., Grumberg, O., Jha, S., Lu, Y., Veith, H.: Counterexample-guided
  abstraction refinement for symbolic model checking.
\newblock J. {ACM} \textbf{50}(5), 752--794 (2003).
\newblock \doi{10.1145/876638.876643}

\bibitem{CBMC}
Clarke, E.M., Kröning, D., Lerda, F.: A tool for checking {ANSI-C} programs.
\newblock In: Proc.\ TACAS, LNCS~2988, pp. 168--176. Springer (2004).
\newblock \doi{10.1007/978-3-540-24730-2_15}

\bibitem{AWS}
Cook, B.: Formal reasoning about the security of {Amazon} web services.
\newblock In: Proc.\ CAV~(2), LNCS~10981, pp. 38--47. Springer (2018).
\newblock \doi{10.1007/978-3-319-96145-3_3}

\bibitem{Craig57}
Craig, W.: Linear reasoning. {A} new form of the {H}erbrand-{G}entzen theorem.
\newblock J.~Symb.\ Log. \textbf{22}(3), 250--268 (1957).
\newblock \doi{10.2307/2963593}

\bibitem{k-Induction}
Donaldson, A.F., Haller, L., Kröning, D., R{\"{u}}mmer, P.: Software
  verification using k-induction.
\newblock In: Proc.\ SAS, LNCS~6887, pp. 351--368. Springer (2011).
\newblock \doi{10.1007/978-3-642-23702-7_26}

\bibitem{kIndForDMARaces}
Donaldson, A.F., Kröning, D., R{\"{u}}mmer, P.: Automatic analysis of {DMA}
  races using model checking and \emph{k}-induction.
\newblock FMSD \textbf{39}(1), 83--113 (2011).
\newblock \doi{10.1007/s10703-011-0124-2}

\bibitem{FlanaganQadeer02}
Flanagan, C., Qadeer, S.: Predicate abstraction for software verification.
\newblock In: Proc.\ POPL, pp. 191--202. ACM (2002).
\newblock \doi{10.1145/503272.503291}

\bibitem{MCMT}
Ghilardi, S., Ranise, S.: Goal-directed invariant synthesis for model checking
  modulo theories.
\newblock In: Proc.\ TABLEAUX, LNCS~5607, pp. 173--188. Springer (2009).
\newblock \doi{10.1007/978-3-642-02716-1_14}

\bibitem{GrafSaidi97}
Graf, S., Sa{\"\i}di, H.: Construction of abstract state graphs with
  \textsc{Pvs}.
\newblock In: Proc.\ CAV, LNCS~1254, pp. 72--83. Springer (1997).
\newblock \doi{10.1007/3-540-63166-6_10}

\bibitem{TraceAbstraction}
Heizmann, M., Hoenicke, J., Podelski, A.: Refinement of trace abstraction.
\newblock In: Proc.\ SAS, LNCS~5673, pp. 69--85. Springer (2009).
\newblock \doi{10.1007/978-3-642-03237-0_7}

\bibitem{UAUTOMIZER2013}
Heizmann, M., Hoenicke, J., Podelski, A.: Software model checking for people
  who love automata.
\newblock In: Proc.\ CAV, LNCS~8044, pp. 36--52. Springer (2013).
\newblock \doi{10.1007/978-3-642-39799-8_2}

\bibitem{AbstractionsFromProofs}
Henzinger, T.A., Jhala, R., Majumdar, R., McMillan, K.L.: Abstractions from
  proofs.
\newblock In: Proc.\ POPL, pp. 232--244. {ACM} (2004).
\newblock \doi{10.1145/964001.964021}

\bibitem{LazyAbstraction}
Henzinger, T.A., Jhala, R., Majumdar, R., Sutre, G.: Lazy abstraction.
\newblock In: Proc.\ POPL, pp. 58--70. {ACM} (2002).
\newblock \doi{10.1145/503272.503279}

\bibitem{RERS12}
Howar, F., Isberner, M., Merten, M., Steffen, B., Beyer, D.: The {RERS}
  grey-box challenge 2012: Analysis of event-condition-action systems.
\newblock In: Proc.\ ISoLA, LNCS~7609, pp. 608--614. Springer (2012).
\newblock \doi{10.1007/978-3-642-34026-0_45}

\bibitem{SoftwareModelChecking}
Jhala, R., Majumdar, R.: Software model checking.
\newblock ACM Computing Surveys \textbf{41}(4) (2009).
\newblock \doi{10.1145/1592434.1592438}

\bibitem{TRApproximation}
Jhala, R., McMillan, K.L.: Interpolant-based transition relation approximation.
\newblock In: Proc.\ CAV, LNCS~3576, pp. 39--51. Springer (2005).
\newblock \doi{10.1007/11513988_6}

\bibitem{PDR-kInduction}
Jovanovic, D., Dutertre, B.: Property-directed k-induction.
\newblock In: Proc.\ FMCAD, pp. 85--92. {IEEE} (2016).
\newblock \doi{10.1109/FMCAD.2016.7886665}

\bibitem{PKind}
Kahsai, T., Tinelli, C.: \textsc{PKind}: A parallel k-induction based model
  checker.
\newblock In: Proc.\ Int. Workshop on Parallel and Distributed Methods in
  Verification, EPTCS~72, pp. 55--62. EPTCS (2011).
\newblock \doi{10.4204/EPTCS.72.6}

\bibitem{LDV}
Khoroshilov, A.V., Mutilin, V.S., Petrenko, A.K., Zakharov, V.: Establishing
  {Linux} driver verification process.
\newblock In: Proc.\ Ershov Memorial Conference, LNCS~5947, pp. 165--176.
  Springer (2009).
\newblock \doi{10.1007/978-3-642-11486-1_14}

\bibitem{SPACER}
Komuravelli, A., Gurfinkel, A., Chaki, S., Clarke, E.M.: Automatic abstraction
  in {SMT}-based unbounded software model checking.
\newblock In: Proc.\ CAV, LNCS~8044, pp. 846--862. Springer (2013).
\newblock \doi{10.1007/978-3-642-39799-8_59}

\bibitem{Wolverine}
Kröning, D., Weissenbacher, G.: Interpolation-based software verification with
  \textsc{Wolverine}.
\newblock In: Proc.\ CAV, LNCS~6806, pp. 573--578. Springer (2011).
\newblock \doi{10.1007/978-3-642-22110-1_45}

\bibitem{IC3-CFA}
Lange, T., Neuh{\"{a}}u{\ss}er, M.R., Noll, T.: {IC3} software model checking
  on control flow automata.
\newblock In: Proc.\ FMCAD, pp. 97--104 (2015).
\newblock \doi{10.1109/FMCAD.2015.7542258}

\bibitem{McMillanCraig}
McMillan, K.L.: Interpolation and {SAT}-based model checking.
\newblock In: Proc.\ CAV, LNCS~2725, pp. 1--13. Springer (2003).
\newblock \doi{10.1007/978-3-540-45069-6_1}

\bibitem{IMPACT}
McMillan, K.L.: Lazy abstraction with interpolants.
\newblock In: Proc.\ CAV, LNCS~4144, pp. 123--136. Springer (2006).
\newblock \doi{10.1007/11817963_14}

\bibitem{LazyAnnotation}
McMillan, K.L.: Lazy annotation for program testing and verification.
\newblock In: Proc.\ CAV, LNCS~6174, pp. 104--118. Springer (2010).
\newblock \doi{10.1007/978-3-642-14295-6_10}

\bibitem{HBMC-interpolation}
McMillan, K.L.: Interpolation and model checking.
\newblock In: Handbook of Model Checking, pp. 421--446. Springer (2018).
\newblock \doi{10.1007/978-3-319-10575-8_14}

\bibitem{DUALITY}
McMillan, K.L., Rybalchenko, A.: Computing relational fixed points using
  interpolation.
\newblock Tech. Rep. MSR-TR-2013-6, Microsoft Research (2013).
\newblock Available online:
  \href{https://www.microsoft.com/en-us/research/publication/computing-relational-fixed-points-using-interpolation/}{MSR-TR-2013-6}

\bibitem{SeryFunFrog11}
Sery, O., Fedyukovich, G., Sharygina, N.: Interpolation-based function
  summaries in bounded model checking.
\newblock In: Proc.\ HVC, LNCS~7261, pp. 160--175. Springer (2011).
\newblock \doi{10.1007/978-3-642-34188-5_15}

\bibitem{VizelFMCAD09}
Vizel, Y., Grumberg, O.: Interpolation-sequence based model checking.
\newblock In: Proc.\ FMCAD, pp. 1--8. {IEEE} (2009).
\newblock \doi{10.1109/FMCAD.2009.5351148}

\bibitem{LDV-Toolset}
Zakharov, I.S., Mandrykin, M.U., Mutilin, V.S., Novikov, E., Petrenko, A.K.,
  Khoroshilov, A.V.: Configurable toolset for static verification of operating
  systems kernel modules.
\newblock Programming and Comp. Softw. \textbf{41}(1), 49--64 (2015).
\newblock \doi{10.1134/S0361768815010065}

\end{thebibliography}
